# HIERARCHIC THEORY OF MATTER AND ITS APPLICATION TO ANALYSIS OF WATER PERTURBATIONS BY MAGNETIC FIELD. ROLE OF WATER IN BIOSYSTEMS

**Alex Kaivarainen**

University of Turku, Department of Physics, Turku, Finland
H2O@karelia.ru
http://web.petrsu.ru/~alexk/new_articles/index.html

## CONTENTS

**REFERENCES**

## SUMMARY

 This paper contains few interrelated parts. The short version of new quantum Hierarchic theory, general for solids and liquids, created by the author is presented. Condensed matter in this theory is considered as a system of 3D standing waves (collective excitations) of different nature: thermal de Broglie waves (waves B), IR photons, related to intermolecular oscillations and thermal phonons (acoustic waves). Theory is verified by computer simulations on examples of pure water and ice, using special computer program (copyright 1997, Kaivarainen) yielding about 400 physical parameters, most of them hidden for direct experiment. The idea of new optoacoustic device with huge informational potential: Comprehensive Analyzer of Matter Properties (CAMP) with its various configurations and applications, based on Hierarchic theory is presented. The full version of this part of paper is a part of book, under preparation by Nova Publishers (New York, USA): "The Hierarchic Theory of Liquids and Solids. Computerized applications for ice, water, and Biosystems": http://arXiv.org/abs/physics/0102086. Good correlation between simulated parameters of water and proteins spatial and dynamic structure points to crucial role of water in biopolymers evolution and function.
 The comprehensive possibilities of Hierarchic theory based computer program (pCAMP, copyright, 1997, Kaivarainen) has been demonstrated in investigation of water properties perturbations under permanent magnetic fields treatment. Theoretical explanations of strongly nonlinear dependence of the different effects on the rotation frequency of test - tubes with water and the long relaxation time (memory) of these deviations from the control after magnetic treatment was off, are presented.
 A new possible mechanisms of distant and local specific interaction between ligands and proteins are discussed. The idea of infrared laser and ultrasound radiation induced cancer cells selective disintegrator, based on *biophysical* mechanism of cancer emergence, is proposed.  It is shown, that multi-fractional model of interfacial water structure, introduced by author, can be responsible for some of morphogenetic field properties, related to alternating water activity. The discovered in our work distant solvent-mediated interaction between different proteins and between proteins and cells also can be a consequence of water activity change.
 A possible role of Virtual Replica of any material object (Kaivarainen, 2006), generated by interference of Bivacuum virtual pressure waves with de Broglie waves of the objects particles, in the remote interaction between the active sites of proteins and the ligands and in the mechanism of 'homeopathic memory' is discussed.

The upgraded theory of elementary act of consciousness or 'Quantum of Mind' is described also.

The coherent IR radiation of coherent water clusters - mesoscopic Bose condensate (mBC) in microtubules can be responsible for the exchange interaction between microtubules of distant neurons, stimulating transition of mBC to noncontinuous macroscopic Bose condensation of coherent water in remote neurons. The corresponding unified wave function collapsing is accompanied by the in-phase [gel-sol] transitions in the entangled neurons bodies, their volume and shape pulsation, and synaptic reorganization. This is an important stage of elementary act of consciousness (the Quantum of Mind).

## Introduction

During many years our Laboratory of molecular biophysics was involved in study of solvent dependent large-scale dynamics of proteins determined by relative thermal mobility of their domains and subunits. The role of large-scale dynamics in the mechanism of protein function, the signal transmission, allosteric effects and other water dependent effects in protein solutions have been investigated.

The modified for this goal physical methods, like NMR, EPR (spin-label), microcalorimetry, spectroscopy, light scattering, refractometry and others were used. A number of new phenomena in physics of biopolymers have been discovered (Kaivarainen, 1985; 1989; 1995, 2001, 2003).

*The most important of them are following:*

1. The ability of proteins to change the bulk water dynamics and thermodynamic activity, as a result of large-scale pulsations of their big interdomain and intersubunit cavities, accompanied by assembly $\rightleftharpoons$ disassembly (flickering) of water clusters in these cavities and exchange of this water molecules with bulk water;

2. Solvent - mediated remote interaction between different kinds of proteins in the process of their large - scale dynamics (flexibility) change, induced, respectively, by ligand binding to the active sites, by temperature or by variation of solvent composition;

3. Solvent-mediated distant interaction between protein and cells, accompanied by cells swelling or shrinking, correlated with change of protein flexibility and water activity, enhancing or triggering the passive osmos via membranes of cells.

A new kind of interaction of water clusters, containing 30 - 70 molecules, with the open interdomain and intersubunit cavities of macromolecules/proteins, named *clusterphilic interaction,* was introduced (Kaivarainen, 1985, 1995, 2001). Such interaction can be considered, as the intermediate one between the hydrophobic and hydrophilic ones. It follows from our dynamic model of protein behavior in water, that intramolecular *clusterphilic interaction* stands for remote signal transmission, allosteric properties in multidomain and oligomeric proteins. It is a consequence of high sensitivity of clusters stability to perturbation of such protein cavities geometry, induced by the ligand binding. Stabilization or destabilization of water clusters in cavities shifts the dynamic equilibrium **B** $\rightleftharpoons$ **A** between the open (B) and closed (A) states of protein cavities to the left or right, correspondingly. As far the *assembly* $\rightleftharpoons$ *disassembly* of water clusters in cavities represent mesoscopic 1st order phase transitions, the functionally important changes of proteins configuration, accompanied the shift of **B** $\rightleftharpoons$ **A** equilibrium need very small change of free energy: $\Delta G = \Delta H - T\Delta S \simeq 0$. This change easily can be provided by binding of ligands to the active sites of proteins (Kaivarainen, 1985, 2001).

The intermolecular *clusterphilic interactions* are crucial in the interfacial effects and thixotropic structure formation in colloid systems (Kaivarainen, 1995; 2003).

The part of this research activity was summarized in book of this author: "Solvent - dependent flexibility of proteins and principles of their function", D Reidel Pub Co., 1985, ISBN: 9027715343.

The development of new Hierarchic theory of condensed matter was started by this author in 1986. This work was stimulated by the understanding that the progress in biophysics is not possible without the detailed and quantitative description of water physical properties on mesoscopic and macroscopic level. The existing theories of liquid state was not enough deep and general for this end.

One of the results of application of created computer program (pCAMP, copyrighted in USA in 1997), based on our Hierarchic theory of matter (Kaivarainen, 1995, 2001, 2003), was the discovering of molecular *mesoscopic Bose condensation (mBC)* in the ice and in liquid water at the ambient temperatures (even around $36^0C$) in form of coherent molecular clusters, named the *primary librational effectons.*

The evidences where obtained, using computer simulations, that just the dimensions and dynamics of these water clusters represent the crucial factors in evolution of biopolymer's spatial and dynamic structure.

Our Hierarchic theory of condensed matter got a lot of convincing computerized verifications on examples of water and ice from comparison of calculated and experimental physical parameters, like heat capacity, thermal conductivity, surface tension, vapor pressure, viscosity and self-diffusion. The new quantitative theories of refraction index, Brillouin light scattering, Mössbauer effect and others, based on the same hierarchical model, are also in good correspondence with experiment.

Because of numerous anomalies, water is a good system for testing of new theories of condensed matter. One may anticipate, that if the theory works well quantitatively for such complicated systems, as water and ice, it must be valid for the other liquids, glasses or crystals also.

# 1. New Hierarchic Theory of Condensed Matter and its Computerized Verification on Examples of Water & Ice

## 1.1 Basic notions and definitions of Hierarchic theory of matter

A quantum based new hierarchic quantitative theory, general for solids and liquids, has been developed.[1–3] It is assumed, that anharmonic oscillations of particles in any condensed matter lead to emergence of three-dimensional (3D) superposition of standing de Broglie waves of molecules, electromagnetic and acoustic waves. Consequently, any condensed matter could be considered as a 'gas' of 3D standing waves of corresponding nature. Our approach unifies and develops strongly the Einstein's and Debye's models and can be reduced to those after strong simplifications.

Collective excitations, like 3D standing de Broglie waves of molecules were analyzed, as a background of hierarchic model of condensed matter.

The most probable de Broglie wave (wave B) length is determined by the ratio of Plank constant to the most probable momentum of molecules, or by ratio of its most probable phase velocity to frequency. The waves B of molecules are related to their translations (tr) and librations (lb).

As the quantum dynamics of condensed matter is anharmonic and does not follow the classical Maxwell - Boltzmann distribution, the real most probable de Broglie wave length can exceed the classical thermal de Broglie wave length and the distance between centers of molecules many times. *This makes possible the atomic and molecular mesoscopic Bose condensation (mBC) in solids and liquids at temperatures, below boiling point. It is one of* *the most important results of new theory, confirmed by computer simulations on examples of water and ice.*

Four strongly interrelated new types of quasiparticles (collective excitations) were introduced in our hierarchic model:

1. *Primary effectons (tr and lb)*, existing in "acoustic" (a) and "optic" (b) states represent the coherent clusters with resulting external momentum, equal to zero. *Secondary effectons* are the result of averaging of all effectons with nonzero external momentum, using Bose-Einstein distribution;

2. *Convertons*, corresponding to interconversions between *tr* and *lb* types of the effectons (flickering clusters);

3. *Primary and secondary transitons* are the intermediate $[a \rightleftharpoons b]$ transition states of the *tr* and *lb* primary and secondary effectons;

4. *Primary and secondary deformons* represent 3D superposition of IR electromagnetic and acoustic waves, correspondingly, activated by primary and secondary *transitons* and *convertons*.

*Primary effectons* (*tr and lb*) are formed by 3D superposition of the most probable standing de Broglie waves of the oscillating ions, atoms or molecules. The volume of effectons (tr and lb) may contain from less than one, to tens and even thousands of molecules. The first condition means validity of classical approximation in description of the subsystems of the effectons. The second one points to quantum properties of coherent clusters due to mesoscopic Bose condensation (mBC), in contrast to macroscopic BC, pertinent for superfluidity and superconductivity.

The liquids are semiclassical systems because their primary (tr) effectons contain less than one molecule and primary (lb) effectons - more than one molecule. The solids are quantum systems totally because both kind of their primary effectons (tr and lb) are mesoscopic molecular Bose condensates. These consequences of our theory are confirmed by computer calculations.

The 1st order $[gas \rightarrow liquid]$ transition is accompanied by strong degeneration of rotational (librational) degrees of freedom due to emergence of primary (lb) effectons (mBC) and $[liquid \rightarrow solid]$ transition - by degeneration of translational degrees of freedom due to Bose-condensation of primary (tr) effectons.

In the general case the effecton can be approximated by parallelepiped with edges determined by de Broglie waves length in three selected directions (1, 2, 3), related to symmetry of molecular dynamics. In the case of isotropic molecular motion the effectons' shape is approximated by cube. The edge-length of primary effectons (tr and lb) is considered as the "parameter of order" in our theory of phase transitions.

The in-phase oscillations of molecules in the effectons correspond to the effecton's (a) - acoustic state and the counterphase oscillations correspond to their (b) - optic state. States (a) and (b) of the effectons differ in potential energy only, however, their kinetic energies, momentums and spatial dimensions - are the same. The b-state of the effectons has a common feature with Frölich's polar mode. The $(a \rightarrow b)$ or $(b \rightarrow a)$ transition states of the *primary effectons* (tr and lb), defined as *primary transitons*, are accompanied by a change in molecule polarizability and dipole moment without density fluctuation. At this case these transitions lead to absorption or radiation of IR photons, respectively. Superposition (interception) of three internal standing IR photons of different directions (1,2,3), normal to each other - forms primary electromagnetic deformons (tr and lb). On the other hand, the $[lb \rightleftharpoons tr]$ *convertons* and *secondary transitons* are accompanied by the density fluctuations, leading to absorption or radiation of phonons. Superposition of three standing phonons, propagating in three directions (1,2,3), normal to each other, forms *secondary acoustic deformons (tr and lb)*.

Correlated collective excitations of primary and secondary effectons and deformons (tr and lb), localized in the volume of primary tr and lb electromagnetic deformons, lead to origination of *macroeffectons, macrotransitons and macrodeformons* (tr and lb respectively).

*Macroconvertons* are the result of simultaneous transitions $[a_{lb} \rightleftharpoons a_{tr}]$ and $[b_{lb} \rightleftharpoons b_{tr}]$ between the acoustic (a) and optic (b) modes of librational and translational effectons, accompanied by disassembly $\rightleftharpoons$ assembly of coherent water clusters. This process is close to notion of 'flickering' clusters.

Correlated simultaneous excitations of tr and lb macroeffectons in the volume of superimposed *tr* and *lb* electromagnetic deformons lead to origination of *supereffectons*.

In turn, the simultaneous excitation of both*: tr and lb macrodeformons and macroconvertons* in the same volume means origination of *superdeformons*. Superdeformons are the biggest (cavitational) fluctuations, leading to microbubbles in liquids and to local defects in solids.

Total number of quasiparticles of condensed matter equal to 4!=24, reflects all of possible combinations of the four basic ones [1-4], introduced above (Table 1). This set of collective excitations in the form of 3D standing waves of three types: thermal de Broglie waves, acoustic and electromagnetic ones - is proved to be able to explain virtually all the properties of all condensed matter.

*The important positive feature of our hierarchic model of matter is that it does not need the semi-empirical intermolecular potentials for calculations, which are unavoidable in existing theories of many body systems. The potential energy of intermolecular interaction is involved indirectly in dimensions and stability of quasiparticles, introduced in our model.*

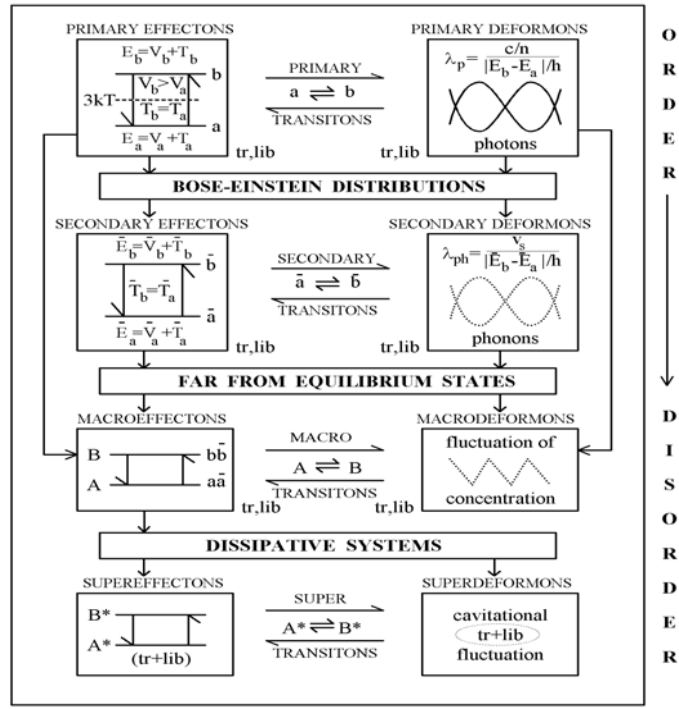

Table 1. Schematic representation of the 18 types of quasiparticles of condensed matter as a hierarchical dynamic system, based on the effectons, transitons and deformons. Total number of *quasiparticles*, introduced in Hierarchic concept is 24. Six collective excitations, related to *convertons*- interconversions between primary librational and translational effectons and their derivatives are not represented here for the end of simplicity.

The main formulae of theory are the same for liquids and solids and include following experimental parameters, which take into account their different properties: [1]- Positions of (tr) and (lb) bands in oscillatory spectra; [2]- Sound velocity; [3]- Density; [4]- Refraction index.

The knowledge of these four basic parameters at the same temperature and pressure makes it possible using our computer program, to evaluate more than 300 important characteristics of any condensed matter. Among them are such as: total internal energy, kinetic and potential energies, heat-capacity and thermal conductivity, surface tension, vapor pressure, viscosity, coefficient of self-diffusion, osmotic pressure, solvent activity, etc. Most of calculated parameters are hidden, i.e. inaccessible to direct experimental measurement.

This is the first theory able to predict all known experimental temperature anomalies for water and ice. The conformity between theory and experiment is good even without adjustable parameters. The hierarchic concept creates a bridge between micro- and macro- phenomena, dynamics and thermodynamics, liquids and solids in terms of quantum physics.

## 1.2 Total Internal Energy of Condensed Matter

The final formula for the total internal energy of ($U_{tot}$) of one mole of matter, leading from Hierarchic theory, considering condensed matter as a system of 3D standing waves is (see book, under preparation by Nova Publishers (New York, USA): "The Hierarchic Theory of Liquids and Solids. Computerized applications for ice, water, and Biosystems" http://arXiv.org/abs/physics/0102086):

$$\begin{aligned}
U_{tot} = V_0 \frac{1}{Z} \sum_{tr,lb} &\left\{ \left[ \begin{array}{c} n_{ef}\left( P^a_{ef}E^a_{ef} + P^b_{ef}E^b_{ef} + P_tE_t \right) \\ + n_d P_d E_d \end{array} \right] + \right. \\
&+ \left[ \bar{n}_{ef}\left( \bar{P}^a_{ef}\bar{E}^a_{ef} + \bar{P}^b_{ef}\bar{E}^b_{ef} + \bar{P}_t\bar{E}_t \right) + \bar{n}_d\bar{P}_d\bar{E}_d \right] + \\
&+ \left[ n_M\left( P^A_M E^A_M + P^B_M E^B_M \right) + n_D P^D_M E^D_M \right]_{tr,lb} + \\
&+ V_0 \frac{1}{Z} \left[ n_{con}\left( P_{ac}E_{ac} + P_{bc}E_{bc} + P_{cMt}E_{cMt} \right) + \right. \\
&+ \left. \left( n_{cda}P_{ac}E_{ac} + n_{cdb}P_{bc}E_{bc} + n_{cMd}P_{cMd}E_{cMd} \right) \right] + \\
&+ V_0 \frac{1}{Z} n_s \left[ \left( P^{A^*}_S E^{A^*}_S + P^{B^*}_S E^{B^*}_S \right) + n_{D^*} P^{D^*}_S E^{D^*}_S \right] \quad 1.1
\end{aligned}$$

The meaning of the variables in formulae (1), necessary for the internal energy calculations, are presented in our paper (Kaivarainen, 2001). Total potential energy of one mole of condensed matter is defined by the difference between corresponding total internal energy and total kinetic energy: $V^{tot} = U^{tot} - T^{tot}$. It is important to stress, that the same equations are valid for liquids and solids in our theory.

A lot of characteristics of condensed matter, composed from 24 quasiparticles - about 300, may be calculated, using hierarchic theory and CAMP computer program [copyright 1997, Kaivarainen]. For this end we need four basic input experimental parameters at the same temperature and pressure: 1) positions of translational and librational bands in middle/far IR spectrum of condensed matter; 2) sound velocity; 3) density or molar volume; 4) refraction index.

### 1.3 Quantitative verification of Hierarchic theory on examples of ice and water

*1.3.1. The coincidence of theoretical and experimental data for ice structure stability*

Our hierarchic theory makes it possible to calculate unprecedentedly big amount of parameters for liquids and solids. Part of them, accessible experimentally and taken from literature, are in good correspondence with CAMP - computer simulations.

For example, the calculated minimum of partition function for ice (Z) (Fig. 1a) corresponds to temperature of about $-170^0C$. For the other hand, the interval from -198 to $-173^0C$ is known, indeed, as T- anomalies one due to the fact that the heat equilibrium of ice establishes very slowly in this range (Maeno, 1988). This fact is a consequence of the less probable ice structure (minimum value of partition function Z) near $-170^0C$.

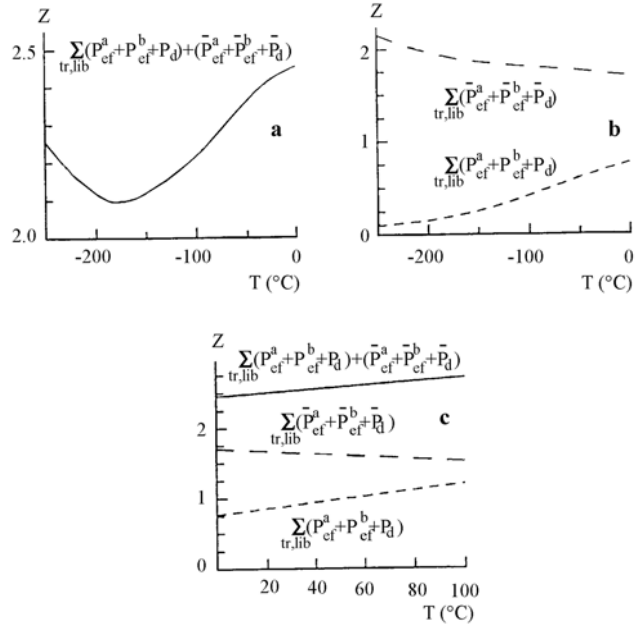

Figure 1.1 $(a, b, c)$. Temperature dependences of the total partition function $(Z)$ and contributions related to primary and secondary effectons and deformons for ice (a,b) and water (c).

*1.3.2. The coincidence of theoretical and experimental heat capacity of ice and water*

It follows from Fig. 2a that the mean theoretical value of heat capacity for ice in the interval from -75 to $0^oC$ is equal to:

$$\bar{C}_p^{ice} = \frac{\Delta U_{tot}}{\Delta T} \approx 39 J/MK = 9.3 \text{ cal}/MK \qquad 1.2$$

For water within the whole range $\Delta T = 100^0C$, the theoretical change in the internal energy is: $\Delta U = 17 - 9.7 = 7.3 kJ/M$ (Fig. 2b). This corresponds to mean value of heat capacity of water:

$$C_p^{water} = \frac{\Delta U_{tot}}{\Delta T} = 73 \ J/MK = 17.5 \ cal/MK \qquad 1.3$$

These results of calculation agree well with the experimental mean values $C_p = 18$

cal/$MK$ for water and $C_p = 9cal/MK$ for ice.[4]

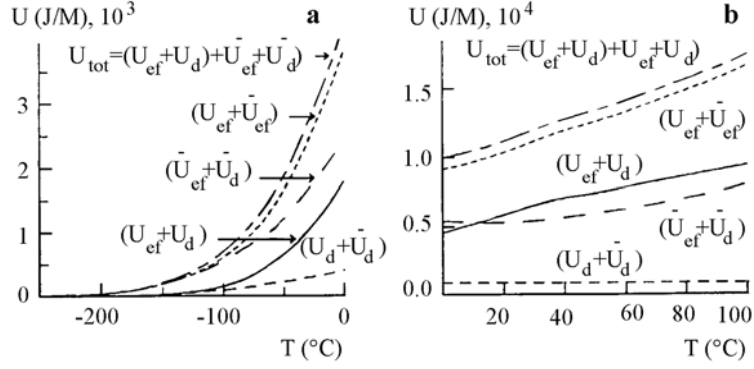

Figure. 1.2. (a,b). Temperature dependences of the total internal energy ($U_{tot}$) and different contributions for ice (a) and water (b). Following contributions to $U_{tot}$ are presented: ($U_{ef} + \bar{U}_{ef}$) is the contribution of primary and secondary effectons; ($U_d + \bar{U}_d$) is the contribution of primary and secondary deformons; ($U_{ef} + U_d$) is the contribution of primary effectons and deformons; ($\bar{U}_{ef} + \bar{U}_d$) is the contribution of secondary effectons and deformons.

### *1.3.3. New State Equation for Condensed Matter*

It was Van der Waals who choosed the first way more than a hundred years ago and derived the equation:

$$\left(P + \frac{a}{V^2}\right)\left(V - b\right) = RT \qquad 1.4$$

where the attraction forces are accounted for by the amending term ($a/V^2$), while the repulsion forces and the effects of the excluded volume accounted for the term (b).

Equation (1.4) correctly describes changes in P,V and T related to liquid-gas transitions on the qualitative level. However, the quantitative analysis of (1.4) is approximate and needs the fitting parameters. The parameters (a) and (b) are not constant for the given substance and depend on temperature. Hence, the Van der Waals equation is only some approximation describing the state of a real gas.

Using our equation for the total internal energy of condensed matter ($U_{tot}$), we can present state equation in a more general form than (1.4). For this end we introduce the notions of *internal pressure* ($P_{in}$), including *all type of interactions* between particles of matter and excluded molar volume ($V_{exc}$):

$$V_{exc} = \frac{4}{3}\pi\alpha^* N_0 = V_0\left(\frac{n^2 - 1}{n^2}\right) \qquad 1.5$$

where $\alpha^*$ is the acting polarizability of molecules in condensed matter; $N_0$ is Avogadro number, and $V_0$ is molar volume.

*The new general state equation* can be expressed as:

$$P_{tot}V_{fr} = (P_{ext} + P_{in})(V_0 - V_{exc}) = U_{ef} \qquad 1.6$$

where: $U_{ef} = U_{tot}(1 + V/T^t_{kin}) = U^2_{tot}/T_{kin}$ is the effective internal energy and:

$$(1 + V/T_{kin}) = U_{tot}/T_{kin} = S^{-1} \qquad 1.7$$

is the reciprocal value of the total structural factor; $P_{tot} = P_{ext} + P_{in}$ is total pressure, $P_{ext}$ and $P_{in}$ are external and internal pressures; $V_{fr} = V_0 - V_{exc} = V_0/n^2$ (see eq.1.5) is a free molar volume; $U_{tot} = V + T_{kin}$ is the total internal energy, V and $T_{kin}$ are total potential and kinetic energies of one mole of matter.

For the limit case of ideal gas, when $P_{in} = 0$; $V_{exc} = 0$; and the potential energy $V = 0$, we get from (1.6) the Clapeyrone - Mendeleyev equation:

$$P_{ext} V_0 = T_{kin} = RT$$

One can use equation of state (1.6) for estimation of sum of *all types of internal matter interactions*, which determines the internal pressure $P_{in}$:

$$P_{in} = \frac{U_{ef}}{V_{fr}} - P_{ext} = \frac{n^2 U_{tot}^2}{V_0 T_{kin}} - P_{ext} \qquad 1.8$$

where: the molar free volume: $V_{fr} = V_0 - V_{exc} = V_0/n^2$;

and the effective total energy: $U_{ef} = U_{tot}^2/T_{kin} = U_{tot}/S$; where $S = T_{kin}/U_{tot}$ is a total structural factor.

*1.3.4. Coincidence between calculated and experimental vapor pressure for ice and water*

There was not earlier the satisfactory quantitative theory for *vapor pressure* calculation.

Such a theory has been derived, using our notion of collective excitations: *superdeformons*, representing the biggest thermal fluctuations.[2,3] The basic idea is that the external equilibrium vapor pressure is related to internal one ($P_{in}^S$) with coefficient determined by the probability of cavitational fluctuations (superdeformons) in the **surface layer** of liquids or solids.

In other words due to excitation of superdeformons with probability ($P_D^S$), the internal pressure ($P_{in}^S$) in surface layers, determined by the total contributions of all intramolecular interactions turns to external one - vapor pressure ($P_V$). It is something like a compressed spring energy realization due to trigger switching off.

To take into account the difference between the surface and bulk internal pressure ($P_{in}$) we introduce the semi-empirical surface pressure factor ($q^S$) as:

$$P_{in}^S = q^S P_{in} = q^S \left( \frac{n^2 U_{tot}}{V_0 S} - P_{ext} \right) \qquad 1.9$$

where: $P_{in}$ corresponds to eq.(1.8); $S = T_{kin}/U_{tot}$ is a total structure factor.

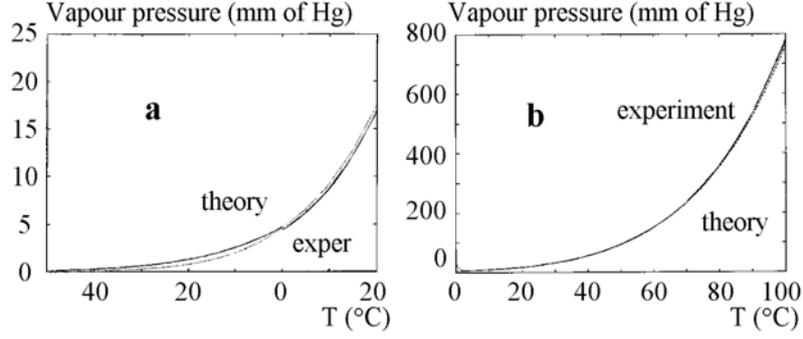

Figure. 1.3. a) Theoretical (−) and experimental (· ·) temperature dependences of vapor pressure ($P_{vap}$) for ice (a) and water (b), including phase transition region. The experimental data where taken from Handbook of Chem. & Phys. 67 ed., CRC press, 1986-1987.

Multiplying (1.9) to probability of superdeformons excitation we obtain for vapor pressure, resulting from evaporation or sublimation, the following formulae:

$$P_{vap} = P_{in}^S \cdot P_D^S = q^S \left( \frac{n^2 U_{tot}^2}{V_0 T_{kin}} - P_{ext} \right) \exp\left(-\frac{E_D^S}{kT}\right) \qquad 1.10$$

where:

$$P_D^S = \exp\left(-\frac{E_D^S}{kT}\right) \qquad 1.11$$

is a probability of superdeformons excitation (see eqs. 3.37, 3.32 and 3.33 from[3]).
The pressure surface factor ($q^S$) could be presented as:

$$q^S = P_{in}^S / P_{in}$$

Theoretical calculated temperature dependences of vapor pressure, described by (1.10) coincide very well with experimental ones for water at $q_{liq}^S = 3.1$ and for ice at $q_{sol}^S = 18$ (Fig. 1.3).

The almost five-times difference between $q_{sol}^S$ and $q_{liq}^S$ means that the *surface* properties of ice differ from *bulk* ones much more than for liquid water.

*1.3.5. Coincidence between calculated and experimental surface tension*

The resulting surface tension is introduced in our mesoscopic model as a sum:

$$\sigma = (\sigma_{tr} + \sigma_{lb}) \qquad 1.12$$

where: $\sigma_{tr}$ and $\sigma_{lb}$ are translational and librational contributions to surface tension. Each of these components can be expressed using our mesoscopic state equation (1.6), taking into account the difference between surface and bulk total energies ($q^S$), introduced in previous section:

$$\sigma_{tr,} = \frac{1}{\frac{1}{\pi}(Vlb_{ef})_{tr,lb}^{2/3}} \left[ \frac{q^S P_{tot}(P_{ef}V_{ef})_{tr,lb} - P_{tot}(P_{ef}V_{ef})_{tr,lb}}{(P_{ef} + P_t)_{tr} + (P_{ef} + P_t)_{lb} + (P_{con} + P_{cMt})} \right] \qquad 1.13$$

where $(V_{ef})_{tr,lb}$ are volumes of primary tr and lib effectons, related to their concentration

$(n_{ef})_{tr,lb}$ as:

$$(V_{ef})_{tr,lb} = (1/n_{ef})_{tr,lb};$$

$$r_{tr,lb} = \frac{1}{\pi}(V_{ef})_{tr,lb}^{2/3}$$

is an effective radius of the primary translational and librational effectons, localized on the surface of condensed matter; $q^S$ is the surface factor, equal to that used in vapor pressure calculations; $[P_{tot} = P_{in} + P_{ext}]$ is a total pressure; $(P_{ef})_{tr,lb}$ is a total probability of primary effecton excitations in the (a) and (b) states:

$$(P_{ef})_{tr} = (P_{ef}^a + P_{ef}^b)_{tr}$$

$$(P_{ef})_{lb} = (P_{ef}^a + P_{ef}^b)_{lb}$$

$(P_t)_{tr}$ and $(P_t)_{lb}$ in (13) are the probabilities of corresponding transiton excitation;
$P_{con} = P_{ac} + P_{bc}$ is the sum of probabilities of $[a]$ and $[b]$ *convertons*; $P_{cMt} = P_{ac} P_{bc}$ is a probability of Macroconvertons excitation.

The eq. (1.13) contains the ratio:

$$(V_{ef}/V_{ef}^{2/3})_{tr,lb} = l_{tr,lb} \qquad 1.14$$

where: $l_{tr} = (1/n_{ef})_{tr}^{1/3}$ and $l_{lb} = (1/n_{ef})_{lib}^{1/3}$ are the length of the ribs of the primary translational and librational effectons, approximated by cube.

The resulting surface tension can be presented as:

$$\sigma = \sigma_{tr} + \sigma_{lb} = \pi \frac{P_{tot}(q^S - 1)\left[(P_{ef})_{tr}l_{tr} + (P_{ef})l_{lb}\right]}{(P_{ef} + P_t)_{tr} + (P_{ef} + P_t)_{lb} + (P_{con} + P_{cMt})} \qquad 1.15$$

The results of computer calculations of $\sigma$ (eq. 1.15) for water and experimental data are presented at Fig.1.4.

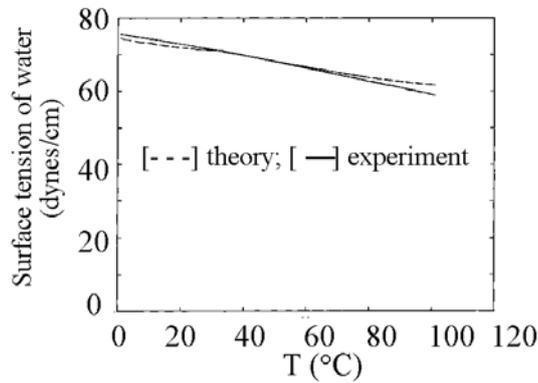

Figure 1.4. Experimental (—) and theoretical (---) temperature dependences of the surface tension for water. The experimental data where taken from Handbook of Chem. & Phys., 67 ed., CRC press, 1986-1987.

It is obvious, that the correspondence between theory and experiment is very good, confirming in such a way the correctness of our model and Hierarchic concept in general.

*1.3.6. Coincidence between calculated and experimental thermal conductivity*

Thermal conductivity may be related to phonons, photons, free electrons, holes and [electron-hole] pairs movement. We will discuss here only the main type of thermal conductivity in condensed matter, related to phonons.

Hierarchic theory introduce two contributions to thermal conductivity: related to phonons, radiated by secondary effectons and forming *secondary* translational and librational deformons $(\kappa_{sd})_{tr,lb}$ and to phonons, radiated by $a$ and $b$ convertons $[tr/lb]$, forming the convertons-induced deformons $(\kappa_{cd})_{ac,bc}$:

$$\kappa = (\kappa_{sd})_{tr,lb} + (\kappa_{cd})_{ac,bc} = \frac{1}{3} C_V v_s [(\Lambda_{sd})_{tr,lb} + (\Lambda_{cd})_{ac,bc}] \qquad 1.16$$

where: free runs of secondary phonons (tr and lb) are represented as:

$$1/(\Lambda_{sd})_{tr,lb} = 1/(\Lambda_{tr}) + 1/(\Lambda_{lb}) = (\overline{v}_d)_{tr}/v_s + (\overline{v}_d)_{lb}/v_s$$

consequently:

$$1/(\Lambda_{sd})_{tr,lb} = \frac{v_s}{(\overline{v}_d)_{tr} + (\overline{v}_d)_{lb}} \qquad 1.17$$

and free runs of convertons-induced phonons:

$$1/(\Lambda_{cd})_{ac,bc} = 1/(\Lambda_{ac}) + 1/(\Lambda_{bc}) = (v_{ac})/v_s + (v_{bc})/v_s$$

The heat capacity: $C_V = \partial U_{tot}/\partial T$ can be calculated also from our theory.

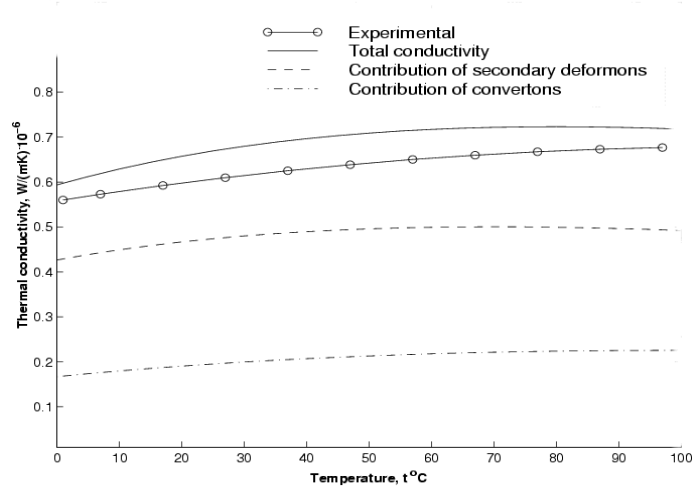

Figure 1.5. Temperature dependences of total thermal conductivity for water and contributions, related to acoustic deformons and [*lb/tr*]convertons. The experimental data were taken from Handbook of Chem. & Phys., 67 ed., CRC press, 1986-1987.

*1.3.7. Coincidence between calculated and experimental viscosity for liquids and solids*

The viscosity is determined by the energy dissipation as a result of medium (liquid or solid) structure deformation. Viscosity, corresponding to the shift deformation, is named *shear viscosity*. So-called *bulk viscosity* is related to deformation of volume parameters and corresponding dissipation. These types of viscosity have not the same values.

The new hierarchic theory of viscosity has been developed. The dissipation processes, related to $(A \rightleftharpoons B)_{tr,lb}$ cycles of translational and librational macroeffectons and (a,b)-*convertons* excitations were analyzed.

*In contrast to liquid state, the viscosity of solids* is determined by the biggest fluctuations: supereffectons and superdeformons, resulting from simultaneous excitations of translational and librational macroeffectons and macrodeformons in the same volume.[3]

The contributions of translational and librational macrodeformons to resulting viscosity are present in following way:

$$\eta_{tr,lb}^M = \left[ \frac{E_D^M}{\Delta v_f^0} \tau^M \left( \frac{T_k}{U_{tot}} \right)^3 \right]_{tr,lb} \qquad 1.18$$

where: $(\Delta v_f^0)$ is the reduced fluctuating volume; the energy of macrodeformons: $[E_D^M = -kT(\ln P_D^M)]_{tr,lb}$.

The cycle-periods of the *tr* and *lib* macroeffectons has been introduced as:

$$[\tau^M = \tau_A + \tau_B + \tau_D]_{tr,lb} \qquad 1.19$$

where: characteristic life-times of macroeffectons in A, B-states and that of transition state in the volume of primary electromagnetic deformons can be presented, correspondingly, as follows:

$$\left[ \tau_A = (\tau_a \tau_{\bar{a}})^{1/2} \right]_{tr,lb} \quad \text{and} \quad \left[ \tau_A = (\tau_a \tau_{\bar{a}})^{1/2} \right]_{tr,lb} \qquad 1.20$$

$$\left[ \tau_D = |(1/\tau_A) - (1/\tau_B)|^{-1} \right]_{tr,lb} \qquad 1.21$$

Using (1.18 - 1.21) it is possible to calculate the contributions of $(A \rightleftharpoons B)$ cycles of translational and librational macroeffectons to viscosity separately.

The averaged contribution of Macroexcitations (tr and lb) in viscosity is:

$$\eta^M = \left[ (\eta)_{tr}^M (\eta)_{lb}^M \right]^{1/2} \qquad 1.22$$

The resulting theoretical viscosity (Fig. 1.6) was calculated as a sum of the averaged contributions of macrodeformons and convertons:

$$\eta = \eta^M + \eta_c \qquad 1.23$$

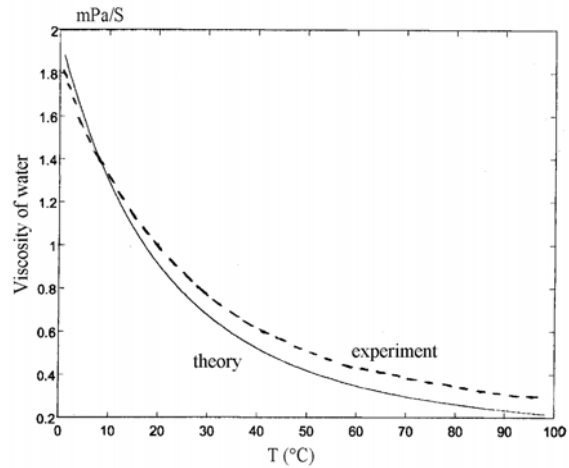

Figure 1.6. Theoretical and experimental temperature dependences of water viscosity. The experimental data where taken from Handbook of Chem. & Phys. 67 ed., CRC press,

1986-1987.

Like in the cases of thermal conductivity, viscosity and vapor pressure, the results of theoretical calculations of self-diffusion coefficient coincide well with experimental data for water in temperature interval $(0 - 100^0 C)$.[3] The coefficient of self-diffusion in solids also may be evaluated using the CAMP computer program.

The important conclusion, leading from the examples presented above, is that as far the final results of calculations are in a good accordance with experiment, it means that a lot of intermediate parameters, hidden from direct experiment, characterizing the spatial and dynamic properties of number of collective excitations of condensed matter - also correctly describe the matter properties.

### 1.4 New Optoacoustic Device, Based on Hierarchic Theory: Comprehensive Analyzer of Matter Properties (CAMP)

The new quantum Hierarchic theory of condensed matter and theory based computer program, copyrighted in 1997 (USA) by A.Kaivarainen allows the calculating of more than 300 physical parameters of any material, including water and ice. Among these parameters are: total internal energy, heat capacity, thermal conductivity, surface tension, vapor pressure, viscosity, self-diffusion, etc. Most of intermediate parameters of calculations, like dimensions of water clusters and their life-time, are hidden for direct experimental measurements. The computerized evaluation of hundreds of output parameters is possible in a second, if the following input experimental parameters are available:

1. Sound velocity;
2. Density;
3. Refraction index;
4. Positions of translational or librational bands in far and middle IR range: 30-3500) 1/cm.

These data should be obtained at the same temperature and pressure from the same sample (liquid or solid). The Hierarchic theory and program have been verified, using the listed above input parameters for water and ice from literature in temperature range: 20 – 373 K (http://arxiv.org/abs/physics/0102086). The coincidence between theory and experiment is very good. Such possibilities led this author to idea of new optoacoustic device: Comprehensive Analyzer of Matter Properties (CAMP), providing a huge amount of information of any condensed matter under study. The FT-Raman spectrometer can be used for registration of spectra in far and middle IR region. The table-top system for measurement of sound velocity, density and refraction index of the same liquid, almost at the same time, is available (DSA 5000 + RXA 156; Anton-Paar, Graz, Austria). We investigated the perturbation of water properties after permanent magnetic field treatment, using this experimental approach and CAMP computer program (Kaivarainen, 2004: http://arxiv.org/abs/physics/0207114).

One of possible configuration of CAMP may include the FT-Brillouin light scattering spectrometer, based on Fabry-Perrot interferometer or its combination with special Raman spectrometer, like produced by HORIBA Raman Research System T 64000. This configuration makes it possible a simultaneous measurement of hypersound velocity (from the Doppler shift of side bands of Brillouin spectra) and positions of intermolecular bands [translational and librational] from the Stokes/antiStokes satellite components on the central (Raman) peak of Brillouin spectra. CAMP may allow the monitoring of perturbation of very different physical properties of water, ice and other condensed matter (material) under the influence of guest molecules, temperature, pressure or external electromagnetic or acoustic fields.

Our CAMP could be the ideal instrument for analyzing of drinking water and aqueous commercial products quality, using 'fingerprints' containing more than 300 physical parameters. CAMP has the unique informational potential, making its future market of scientific equipment free.

The DEMO version of pCAMP computer program for evaluation of water and ice properties in the range: 20-373K can be downloaded from the front page of my site: web.petrsu.ru/~alexk [download pCAMP].

## 2. Quantitative Analysis of Water Perturbations by Magnetic Field, Based on Hierarchic Theory

The changes of refraction index of bi-distilled water during 30 min in rotating test-tube under magnetic field treatment with tension up to 80 kA/m has been demonstrated already (Semihina, 1981). For each magnetic field tension the optimal rotation speed, corresponding to maximum effect, has been revealed. For example, for geomagnetic field tension (46,5 A/m) the optimal number of rotation per minute (rpm) was 790 rpm.

In the works of (Semikhina and Kiselev, 1988, Kiselev et al., 1988, Berezin et al., 1988) the influence of the artificial weak magnetic field on the dielectric losses, the changes of dissociation constant, density, refraction index, light scattering and electroconductivity, the coefficient of heat transition, the depth of super-cooling for distilled water and for ice was studied. This weak alternating field was used as a modulator a geomagnetic field action.

The absorption spectra and the fluorescence of the dye (rhodamine 6G) and proteins in solutions also changed under the action of weak magnetic fields on water.

The influence of permanent geomagnetic, modulated by weak alternating magnetic fields on water and ice in the frequency range $(10^4 - 10^8) Hz$ was studied. The maximum sensitivity to field action was observed at the frequency $v_{max} = 10^5 Hz$. In accordance with our calculations, this frequency corresponds to frequency of superdeformons excitations in water (see Fig. 2.2d).

A few of physical parameters changed after the long (nearly 6 hour) influence of the variable fields ($\tilde{H}$), modulating the geomagnetic field of the tension $[\mathbf{H} \sim \mathbf{H}_{geo}]$ with the frequency in the range of $f = (1 \div 10) \times 10^2 Hz$ (Semikhina and Kiselev, 1988, Kiselev et al., 1988):

$$\tilde{\mathbf{H}} = \mathbf{H} \cos 2\pi f t \qquad 2.1$$

In the range of modulating magnetic field (**H**) tension from $0.08\ A/m$ to $212\ A/m$ the *eight maxima of dielectric losses tangent* in the above mentioned (*f*) range were observed. Dissociation constant decreases more than other parameters (by 6 times) after the incubation of ice and water in magnetic field. The relaxation time ("memory") of the changes, induced in water by fields was in the interval from 0.5 to 8 hours.

The authors explain the experimental data obtained, because of influence of magnetic field on the probability of proton transfer along the net of hydrogen bonds in water and ice, which lead to the deformation of this net.

However, this explanation is doubtful, as far the *equilibrium constant* for the reaction of dissociation:

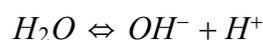

$$H_2O \Leftrightarrow OH^- + H^+$$

in ice is less by almost six orders ($\simeq 10^6$) than that for water. On the other hand the values of the *field- induced effects in ice are several times more than in water*, and the time for reaching them in ice is less.

*In the framework of our Hierarchic theory all the aforementioned phenomena could be*

*explained by the shift of the ($a \Leftrightarrow b$) equilibrium of primary translational and librational effectons to the left and decreasing the ions hydration.* In turn, this shift stimulates polyeffectons or coherent superclusters growth, under the influence of magnetic fields. Therefore, parameters such as light scattering have to enhance, while the $H_2O$ dissociation constant depending on the probability of superdeformons must decrease. The latter correlate with declined electric conductance, revealed in experiments.

*As far, the magnetic moments of molecules within the coherent superclusters or polyeffectons, formed by primary librational effectons are the additive parameters, then the values of changes induced by magnetic field must be proportional to polyeffectons sizes. These sizes are markedly higher in ice than in water and decrease with increasing temperature.*

Inasmuch the effectons and polyeffectons interact with each other by means of phonons (i.e. the subsystem of secondary deformons), and the velocity of phonons is higher in ice than in water, then the saturation of all concomitant effects and achievement of new equilibrium state in ice is faster than in water.

The frequencies of geomagnetic field modulation, at which changes in the properties of water and ice have maxima can correspond to the eigen-frequencies of the [$a \Leftrightarrow b$] equilibrium constant of primary effectons oscillations, determined by [assembly $\Leftrightarrow$ disassembly] equilibrium oscillations for coherent super clusters or polyeffectons.

The presence of dissolved molecules (ions, proteins) in water or ice can influence on the initial [$a \Leftrightarrow b$] equilibrium dimensions of polyeffectons and, consequently the interaction of solution with outer field.

Narrowing of $^1$H-NMR lines in a salt-containing water and calcium bicarbonate solution was observed after magnetic field action. This indicates that the degree of ion hydration is decreased by magnetic treatment. On the other hand, the width of the resonance line in *distilled water* remains unchanged after 30 minute treatment in the field ($135\,kA/m$) at water flow rate of $60\,cm/s$ (Klassen, 1982).

The hydration of diamagnetic ions ($Li^+$, $Mg^{2+}$, $Ca^{2+}$) decreases, while the hydration of paramagnetic ions ($Fe^{3+}$, $Ni^{2+}$, $Cu^{2+}$) increases. It leads from corresponding changes in ultrasound velocity in ion solutions (Duhanin and Kluchnikov, 1975).

There are numerous data which pointing to increasing of coagulation of different particles and their sedimentation velocity after magnetic field treatment. These phenomena point to dehydration of particles in treated by magnetic field aqueous solutions, stimulating particles association. In turn, this provide a reducing the scale formation in heating systems, widely used in practice. Crystallization and polymerization also increase in magnetic field. It points to decrease of water activity, deterioration of its properties as a solvent.

*For better understanding the character of perturbations, induced by magnetic fields on water, and the mechanism of water 'memory' we perform a systematic quantitative investigation, using our Hierarchic theory based computer program (copyright, 1997, Kaivarainen).*

In order to keep pH stable, the 1 mM phosphate buffer, pH 7.0 has been used as a test system. The influence of permanent magnetic field with tension:

$$\mathbf{H} = 200\,mT = 2000\,G = 160\,kA/m \qquad 2.1$$

with space between the two pairs of North and South poles of permanent magnets, equal to 17.5 mm on buffer, has been studied.

The volume of liquid of about 6 ml in standard glass test-tubes with diameter 10 mm was rotated during 21 hours and sometimes 70 hours at room temperature: $t = (23 \pm 0.5)^0 C$.

In our work for study the non-monotonic, nonlinear character of **H**- field interaction with water, we used few rotational frequencies (rotations per minute, rpm) of test-tube in magnetic field (2.1) with weak buffer: [50] rpm; [125] rpm and [200] rpm. For the control measurements, the same volume of buffer in similar test-tube, similar temperature and light conditions were used in the absence of external magnetic field (2.1) and static conditions. The possible influence of test-tube rotation on water properties inside without magnetic field action has been controlled also.

Two degassing procedures of buffer samples in test-tubes in vacuum chamber were performed: the 1st - before starting the magnetic field treatment of test-tubes and 2nd - after treatment just before measurements. The relaxation time (memory) of the induced by magnetic field changes in water at 23 0 C was at least 48 hours. This time decreases with temperature of treated buffer increasing.

Three parameters has been checked for evaluation of magnetic field influence on water structure in buffer, using the same samples during 1.5-2 hours after finishing of H -treatment, much less than relaxation time at $25^0C$:

1. Density;
2. Sound velocity;
3. Refraction index;
4. Position of translational (tr) and librational (lb) bands in IR spectra.

The density and sound velocity were measured simultaneously in the same cuvette at $(25.000 \pm 0.001)^0C$, using device from Anton - Paar company (Austria): DSA- 5000. The molar volume of substance is equal to ratio of its molecular mass (gr./M) to density $(gr/m^3)$:

$$\mathbf{V}_M(\mathbf{m}^3/\mathbf{M}) = \frac{\text{Mol. mass (gr/M)}}{\text{density (gr/m}^3)} \qquad 2.2$$

It means that the bigger is density of substance, the less is its molar volume ($\mathbf{V}_M$).

The refraction index was measured, using differential refractometer (Knouffer, Germany) of high sensitivity at temperature $25^0C$. The IR spectra of attenuated total reflection (ATR) in the middle IR range of the control and **H** -treated samples were registered by FR-IR spectrometer (Perkin -Elmer).

The results of density, sound velocity, refraction index and positions of translational and librational bands in FT- IR spectra are presented at the different frequencies of test-tube rotation during 21 hours are presented in Table 2-1.

**TABLE 2-1.**
**The basic experimental parameters of 1 mM phosphate buffer, pH 7.0 ($25^0C$), treated by H-field at different frequencies of test-tube rotation and the control - untreated buffer**

| Frequency of test-tube rotation (rpm) | 50 | 125 | 200 | Untreated by H-field buffer |
|---|---|---|---|---|
| Molar volume (m³/M) | $1,80541 \cdot 10^{-5}$ | $1,80473 \cdot 10^{-5}$ | $1,80540 \cdot 10^{-5}$ | $1,80538 \cdot 10^{-5}$ |
| Sound velocity (m/sec.) | 1496,714 | 1495,014 | 1496,874 | 1496,614 |
| Refraction index | 1,333968 | 1,333928 | 1,333970 | 1,333978 |
| Positions of tr. bands (cm$^{-1}$) | 194 | 194 | 194 | 194 |
| Positions of lb. bands (cm$^{-1}$) | 682,5 | 682,5 | 682,5 | 682,5 |

All the experiments were repeated 3-5 times with close results. The described experiments was repeated 5 times with qualitatively same results.

The strongest effects of H-field treatment of buffer were obtained at test-tube rotating velocity 125 rpm during 22 hours at 23 C. The control buffer was incubated in similar test-tube without H-field and rotation.

The molar volume and sound velocity in water after magnetic treatment at 50 rpm and 200 rpm increase as respect to control and are smaller and opposite to these values, obtained at 125 rpm. The date obtained point to complicated nonlinear polyextremal dependence of treated buffer effects on rotation frequency of test-tube in permanent magnetic field.

The difference between treated by magnetic field 1mM phosphate buffer, pH 7.0 (sample) and control buffer (reference) was measured on differential refractometer just after evaluation of density and sound velocity of the same liquids. In all test tube rotation frequencies, the refraction index decreases. Like for density and sound velocity, the strongest change in refraction index were obtained at rotating velocity 125 rpm during 22 hours at 23 C.

The reversibility and relaxation time (memory) after incubation of the treated at 125 rpm buffer at 23 C during 22 h and degassing was studied. The starting values of differences with control of density, sound velocity and refraction index decreases about 5 times. However, the relaxation even over 70 h at 23 C of the effects in water, induced by magnetic field was not total. After incubation of the magnetized and reference/control buffer in water bath at $100^0$C during 5 minutes, the differences between them decreases about 10 times and sound velocity - about 4 times.

The change of refraction index of buffer at 50 rpm and 200 rpm in H-field is few times smaller, than at (125) rpm, however, has the same sign, in contrast to changes of density and sound velocity. The control experiment on the influence of rotation of tube with buffer in magnetic field in conditions, when rotation is absent, shows, that the changes in density and sound velocity are negligible, however, the change of refraction index has almost the same value as at 50 and 200 rpm. These results points that different physical parameters of water have different sensibility to magnetic field and rotation effects and could be related to different levels of water structure changes (microscopic / mesoscopic / macroscopic) in magnetic field. Pure rotation of the test-tube at 125 rpm in the absence of H-field do not influence water properties.

The opposite changes of primary input parameters of buffer at different test-tube rotation frequencies, presented at Table 1, are accompanied by the opposite deviations of big number of simulated by our computer program output parameters of quantum excitations of buffer (see Table 2).

The clear correlation has been revealed between the sign of changes of primary (input) parameters and most of calculated (output) ones.

We can see from Table 2-2, that the molar volume and sound velocity of buffer has been increased after treatment at 50 rpm and 200 rpm, in contrast to decreasing of these parameters at 125 rpm. The refraction index decreases in all three cases.

The full description of parameters, presented in tables above could be found in book: "New Hierarchic Theory of Condensed Matter and its Computerized Application to Water, Ice and Biosystems" on line: http://arxiv.org/abs/physics/0102086.

**TABLE 2-2**.
**The primary experimental and a part of calculated parameters for control buffer ($p_c$) and their relative deviations: $\Delta p/p = (p_c - p_H)/p_c$ in treated by magnetic field buffer ($p_H$) at different frequencies of test-tube rotation at $25^0 C$**

| Experimental parameters of buffer | Control: untreated by H-field buffer | Δp/p 50 rpm | Δp/p 125 rpm | Δp/p 200 rpm |
|---|---|---|---|---|
| Molar volume ($m^3$/M) | $1,805384 \cdot 10^{-5}$ | $-1.60077 \cdot 10^{-5}$ | $3.60976 \cdot 10^{-4}$ | $-4.98509 \cdot 1$ |
| Sound velocity (m/sec) | $1496,614$ | $-6.68175 \cdot 10^{-5}$ | $1.06908 \cdot 10^{-3}$ | $-1.73726 \cdot 1$ |
| Refraction index | $1,3339779$ | $1 \cdot 10^{-5}$ | $3.74819 \cdot 10^{-5}$ | $5.9971 \cdot 10$ |
| Positions of translational bands ($cm^{-1}$) | 194 | 0 | 0 | 0 |
| Positions of librational bands ($cm^{-1}$) | 682,5 | 0 | 0 | 0 |
| **Calculated parameters of buffer** | | | | |
| 1. Contribution of translations to the total internal energy, J/M | $6456.831$ | $1.84378 \cdot 10^{-4}$ | $-2.852 \cdot 10^{-3}$ | $5.15981 \cdot 1$ |
| 2. Contribution of librations to the total internal energy, J/M | $4743.6564$ | $1.69679 \cdot 10^{-4}$ | $-2.624 \cdot 10^{-3}$ | $4.74845 \cdot 1$ |
| 3. The total internal energy, J/M | $11256.701$ | $1.78027 \cdot 10^{-4}$ | $-2.753 \cdot 10^{-3}$ | $4.98192 \cdot 1$ |
| | **Untreated by H-field buffer** | **Δp/p 50 rpm** | **Δp/p 125 rpm** | **Δp/p 200 rpm** |
| 4. Total kinetic energy, J/M | $320.643$ | $3.17206 \cdot 10^{-4}$ | $-4.988 \cdot 10^{-3}$ | $8.6102 \cdot 10^{-4}$ |
| 5. Total potential energy, J/M | $10936.058$ | $1.7392 \cdot 10^{-4}$ | $-2.688 \cdot 10^{-3}$ | $4.87561 \cdot 10^{-4}$ |
| 6. The length of edges of libr. primary effectons, Å | $13.636054$ | $-6.6808 \cdot 10^{-5}$ | $1.069 \cdot 10^{-3}$ | $-1.73731 \cdot 10^{-4}$ |
| 7. Number of molecules in the edge of primary libr. effectons | $4.270908$ | $-6.68242 \cdot 10^{-5}$ | $1.06909 \cdot 10^{-3}$ | $-1.737 \cdot 10^{-4}$ |
| 8. The ratio of total potential energy to total kinetic energy | $35.1066$ | $-1.3923 \cdot 10^{-4}$ | $2.223 \cdot 10^{-3}$ | $-3.63151 \cdot 10^{-4}$ |
| 9. Number of mol. in primary librational effectons | $84.5759$ | $-1.8441 \cdot 10^{-4}$ | $2.844 \cdot 10^{-3}$ | $-5.16258 \cdot 10^{-4}$ |
| 10. Aver. dist. between primary librational effectons, Å | $46.057$ | $-6.6808 \cdot 10^{-5}$ | $1.069 \cdot 10^{-3}$ | $-1.7372 \cdot 10^{-4}$ |
| 11. The internal pressure, Pa | $3.8951763 \cdot 10^{10}$ | $6.983 \cdot 10^{-5}$ | $-8.1 \cdot 10^{-4}$ | $1.52188 \cdot 10^{-4}$ |
| 12. Total thermal conductivity, $10^{-6}$ W/(mK) | $0.75014597$ | $2.8661 \cdot 10^{-6}$ | $6.1415 \cdot 10^{-5}$ | $4.5058 \cdot 10^{-5}$ |
| 13. Vapor pressure, Pa | 3255.4087 | $6.98223 \cdot 10^{-5}$ | $-8.101 \cdot 10^{-4}$ | $1.5218 \cdot 10^{-4}$ |
| 14. Viscosity of liquids, Pa·s | $7.570872 \cdot 10^{-4}$ | $4.48641 \cdot 10^{-4}$ | $-6.988 \cdot 10^{-3}$ | $1.1057 \cdot 10^{-3}$ |
| 15. Total coeff. of self-diffusion in liquids, m$^2$/s | $1.9779126 \cdot 10^{-9}$ | $-2.06784 \cdot 10^{-5}$ | $1.907 \cdot 10^{-4}$ | $-1.1325 \cdot 10^{-5}$ |
| 16. Surface tension, dynes/cm | 71.2318 | $3.01831 \cdot 10^{-6}$ | $2.598 \cdot 10^{-4}$ | $-2.1507 \cdot 10^{-5}$ |
| 17. Acting polarizability of molecules, cm$^3$ | $3.13507 \cdot 10^{-30}$ | $3.15782 \cdot 10^{-6}$ | $4.57 \cdot 10^{-4}$ | $1.0366 \cdot 10^{-5}$ |
| 18. The acting field energy, J | $1.76833 \cdot 10^{-19}$ | $1.459 \cdot 10^{-5}$ | $2.0917 \cdot 10^{-3}$ | $4.745 \cdot 10^{-5}$ |
| 19. Coeff. of total light scattering, m$^{-1}$ | $1.3686059 \cdot 10^{-4}$ | $2.24316 \cdot 10^{-5}$ | $5.532 \cdot 10^{-4}$ | $2.5793 \cdot 10^{-5}$ |
| 20. Contrib of primary libr. effect in (a) state to the total reduced inform, bit | 11.483 | $-1.844 \cdot 10^{-4}$ | $2.844 \cdot 10^{-3}$ | $-5.1632 \cdot 10^{-4}$ |
| 21. Contrib. of primary libr. effect in (b) state to the total reduced inform, bit | 0.25 | $-1.844 \cdot 10^{-4}$ | $2.844 \cdot 10^{-3}$ | $-5.1625 \cdot 10^{-4}$ |

## 2.1 Possible mechanism of water properties perturbations under magnetic field treatment

We can propose few possible targets and corresponding mechanisms of interaction of external permanent magnetic field with pure water and weak buffer, moving in this field:

1) Influence on the coherent molecular clusters (molecular Bose condensate), oriented by laminar flow of water, stimulated their polymerization via Josephson junctions; 2) Influence of Lorentz force on trajectory (orbit) of ions, like $H_3^+O$ and hydroxyl ions $HO^-$. These ions originate as a result of dissociation of water molecules to proton and hydroxyl in the process of cavitational fluctuations, accompanied by high temperature fluctuations:

$$H_2O \rightleftharpoons H^+ + HO^- \rightleftharpoons H_3^+O + HO^- \qquad 2.3$$

In buffers, containing inorganic ions, i.e. phosphate ions $PO_4^{2-}$ and others, these ions are also under the Lorentz force influence.

The stability of closed orbits of ions and their influence on stability of polyeffectons and other properties of water is related to conditions of standing de Broglie waves of corresponding ions in flowing liquid. Other interesting form of self-organization of water systems in permanent magnetic field is a possibility of interaction of internal EM fields with radio-frequency, corresponding to frequency of certain fluctuations and $H_2O$ dipoles reorientations (collective excitations) in water at conditions, close to cyclotronic resonance:

$$\omega_c = 2\pi v_c = \frac{ZeB_z}{m^*c} \qquad 2.4$$

The permanent magnetic field, inducing cyclotronic orbits of ions and the alternating EM field, accompanied cyclotronic resonance at certain conditions, could be external as respect to liquid or internal, induced by resulting magnetic field of polyeffectons and correlated in volume density/symmetry of water dipoles fluctuations.

*Let us discuss at first the interaction of magnetic field with macroscopically diamagnetic matter*, like water, as an example. The magnetic susceptibility ($\chi$) of water is a sum of two opposite contributions (Eizenberg and Kauzmann, 1969):

1) average negative diamagnetic part, induced by external magnetic field:

$$\bar{\chi}^d = \frac{1}{2}(\chi_{xx} + \chi_{yy} + \chi_{zz}) \cong -14.6(\pm 1.9) \cdot 10^{-6} \qquad 2.5$$

2) positive paramagnetism related to the polarization of water molecule due to asymmetry of electron density distribution, existing without external magnetic field. Paramagnetic susceptibility ($\chi^p$) of $H_2O$ is a tensor with the following components:

$$\chi^p_{xx} = 2.46 \times 10^{-6}; \quad \chi^p_{yy} = 0.77 \cdot 10^{-6}; \quad \chi^p_{zz} = 1.42 \times 10^{-6} \qquad 2.6$$

The resulting susceptibility:

$$\chi_{H_2} = \bar{\chi}^d + \bar{\chi}^p \cong -13 \times 10^{-6} \qquad 2.7$$

The second contribution in the magnetic susceptibility of water is about 10 times lesser than the first one. But the first contribution to the magnetic moment of water depends on external magnetic field and must disappear when it is switched out in contrast to second one.

The coherent primary librational effectons of water even in liquid state contain about 100 molecules $\left[(n_M^{ef})_{lb} \simeq 100\right]$ at room temperature (Fig. 3.4a of this paper). In ice $(n_M^{ef})_{lb} \geq 10^4$. In (a)-state the vibrations of all these molecules are synchronized in the same phase, and in (b)-state - in counterphase. Correlation of $H_2O$ forming effectons means that the energies of interaction of water molecules with external magnetic field are additive: $\epsilon^{ef} = n_M^{ef} \mu_p H$.

In such a case this total energy of effecton interaction with field may exceed thermal energy: $\epsilon^{ef} > kT$

In the case of polyeffectons formation, as a result of association of primary librational effectons via Josephson junctions, this inequality becomes much stronger. Corresponding chains, formed by mesoscopic magnets may self-organize to long-life metastable 3D structures, influencing macroscopic properties of water, like viscosity, self-diffusion, surface tension, thermal conductivity, light scattering. These macroscopic changes are interrelated with changes on microscopic molecular level, like polarizabilities and related parameters, like refraction index, acting field energy and light scattering (Kaivarainen,1995; 2007).

The life-time of closed ionic orbits, corresponding to standing waves conditions and 3D polyeffectons determines the relaxation time of changes in water properties (memory) after magnetic treatment.

The energy of interaction of primary librational effectons and other collective excitations in water with external magnetic field is dependent on few factors:
1. Dimensions of primary librational effectons;
2. Stability (life-time) of primary librational effectons;
3) Interaction with $H_3^+O$ and hydroxyl ions (see eq. 2.3) and the guest molecules (inorganic ions or organic molecules, polymers, including proteins, lipids, etc.).

The effectons dimensions are determined by the most probable momentums of water molecules in selected directions: $\lambda_{1,2,3} = h/p_{1,2,3}$

The stability of the effectons is dependent on number of molecules $[n_M(L_{lb})]$ in the edge of primary librational effectons, approximated by cube.

The closer this number is to integer one, the more stable are effectons and more favorable condition for their polymerization are existing. For the other hand, the closer this number is to semi-integer values, the less stable are primary effectons and smaller is probability of their 'head' to 'tail' polymerization and closed cycles/rings of polyeffectons formation. Such nonlinearity of stability is a result of competition between sterical and quantum-mechnical factors, responsible for the primary effectons formation.

If water is flowing in a tube, it increases the relative orientations of all effectons in volume and stimulate the coherent superclusters formation. All the above discussed effects must increase. Similar ordering phenomena happen in a rotating test-tube with liquid. After switching off the external magnetic field, the relaxation of *induced ferromagnetism* in water begins. It may be accompanied by the oscillatory behavior of water properties.

Remnant ferromagnetism in water was experimentally established using a SQUID superconducting magnetometer by Kaivarainen et al. in 1992 (unpublished data). Water was treated in permanent magnetic field $500\,G$ for two hours. Then it was frozen and after switching off the external magnetic field the remnant ferromagnetism was registered at helium temperature. Even at this low temperature a slow relaxation in form of decreasing of ferromagnetic signal was revealed.

## 2.2 Cyclotronic frequency of ions in rotating tube with water and possible mechanism of water structure stabilization/destabilization under magnetic field treatment

The frequency of rotation of charged particle with charge (Ze) and effective mass ($\mathbf{m}^*$), moving in permanent magnetic field **H** with velocity **v** along circle orbit with radius (**r**), is determined by the equality of the Lorentz force (**F**) and centripetal force:

$$\mathbf{F} = e\mathbf{E} + \frac{|Ze|}{c}[\mathbf{vB}] = \frac{\mathbf{m}^*\mathbf{v}^2}{\mathbf{r}} \qquad 2.11$$

In the absence of electric field (**E** = **0**), we get from (2.11) the known formula for cyclotronic frequency:

$$\omega_c = 2\pi\nu_c = \frac{1}{\mathbf{r}/\mathbf{v}} = \frac{|Ze|\mathbf{H}}{\mathbf{m}^*\mathbf{c}} \qquad 2.12$$

The radius of cyclotronic orbit is dependent on velocity of charged particle and its de Broglie wave length radius ($\mathbf{L}_B = \hbar/\mathbf{m}^*\mathbf{v}$) as:

$$\mathbf{r} = \frac{\mathbf{c}}{|Ze|\mathbf{H}}\mathbf{m}^*\mathbf{v} = \frac{\hbar\mathbf{c}}{|Ze|\mathbf{H}}\frac{1}{\mathbf{L}_B} \qquad 2.13$$

The condition of orbit stability, corresponds to that of de Broglie standing waves condition: $\mathbf{r} = \mathbf{1L}_B;\ \mathbf{2L}_B;\ 3\mathbf{L}_B,\ldots n\mathbf{L}_B$

From (2.13) we get these conditions of stability in form:

$$n^2 = 1, 4, 9\ldots = \frac{\hbar\mathbf{c}}{|Ze|\mathbf{H}}\frac{1}{\mathbf{L}_B^2} = \frac{\mathbf{c}}{|Ze|\mathbf{H}}\frac{(\mathbf{m}^*\mathbf{v})^2}{\hbar} \qquad 2.14$$

The velocity of particle, moving in the volume of liquid in rotating test-tube with internal radius 0.5 cm on distance **R** = 0.3 cm from the center is:

$$\mathbf{v} = \frac{2\pi\mathbf{R}}{T_{rot}} = 2\pi\nu_{rot}\mathbf{R} = \omega_{rot}\mathbf{R} \qquad 2.15$$

Putting (2.15) in (2.14), we get the formula for rotation frequency of test-tube in permanent magnetic field, corresponding to stable cyclotronic orbits of charge (Ze) with effective mass ($\mathbf{m}^*$) at given distance (**R**) from center of rotation:

$$\omega_{rot} = \frac{\mathbf{n}}{\mathbf{R}\,\mathbf{m}^*}\left(\frac{\hbar\,|Ze|\mathbf{H}}{c}\right)^{1/2} \qquad 2.16$$

or in dimensionless mode:

$$\mathbf{n} = \omega_{rot}\frac{\mathbf{R}\,\mathbf{m}^*}{\left(\hbar\,|Ze|\mathbf{H}/\mathbf{c}\right)^{1/2}} = \text{integer number} \qquad 2.17$$

Taking into account formula for cyclotronic resonance (2.12), we get from (2.16) the relation between the test-tube rotation angle frequency and cyclotron frequencies:

$$\omega_{rot} = \frac{\mathbf{n}}{\mathbf{R}}\left(\frac{\hbar}{\mathbf{m}^*}\omega_c\right)^{1/2} \qquad 2.18$$

One of the consequence of our formulas (2.16-2.18) is that the stability of cyclotronic orbit and related stability of water structure in test-tube should have the polyextremal dependence on rotation frequency ($v_{rot}$) of the test tube. A stable water structure would corresponds to integer values of (**n**) and unstable water structure to its half-integer values. We have to keep in mind that formulas (2.16 - 2.18) are approximate, because the component of Lorentz force: $\frac{Ze}{c}[\mathbf{vB}]$ is true only for particles, **moving in the plane, normal** to direction of **H** vector. Obviously, in the case of rotating test-tube with water and ions this condition do not take a place. However, for qualitative explanation of nonlinear, polyextremal dependence of the water parameters changes on rotation frequency of test-tube, the formulas obtained are good enough. They confirm the proposed mechanism of water structure stabilization or destabilization after magnetic treatment, as a consequence of stable or unstable cyclotronic orbits of ions origination in water and aqueous solutions.

The cyclotronic resonance represents the absorption of alternating electromagnetic field energy with frequencies ($\omega_{EM}$), equal or multiple to cyclotron frequency:

$$\hbar\omega_{EM} = n\omega_c, \quad \text{where } n = 1, 2, 3, \ldots \qquad 2.19$$

The integer **n** correspond to condition of maximum of EM energy absorption and half-integer value - to minimum absorption. If the frequency $\omega_{EM}$ is close to one of the radio-frequencies of correlated fluctuations in water, presented on Fig.3.2, then the auto-cyclotron resonance effect in liquid is possible. Such new phenomena may represent a new kind of macroscopic self-organization in water and other, ions-containing liquids at selected conditions.

**2.3 Possible mechanism of distant specific attraction between ligands and proteins**

The another kind of "memory of water", discovered by Jacques Benveniste in 1988, includes the ability of water to carry information about biologically active guest molecules and possibility to record, transmit and amplify this information. This phenomenon involves the successive diluting and shaking of [water + guest] system to a degree where the final solution contains no guest molecules more at all. However, using hypersensitive biological cells-containing test systems, he observed that this highly diluted solution initiated a reaction in similar way, as if the active guest molecules were still present in water. From the first high dilution experiments in 1984 to the present, thousands of experiments have been made in DigiBio company (Paris), enriching and considerably consolidating the initial knowledge of such kind of water memory. It was demonstrated also by DigiBio team, that low frequency ($< 20$ kHz $= 2 \times 10^4$ s$^{-1}$) electromagnetic waves are able to activate specific biological cells. This prompted J. Benveniste to hypothesize that the molecular signal is composed of such low frequency waves and that the ligand coresonates with the receptor on these frequencies, stimulating specific attraction between them.

The perturbation of water properties, induced by haptens and antibodies, concurrent inhibitors and enzymes, viruses and cells in separate and mixed solutions can be studied, using our Hierarchic theory of condensed matter and this theory based Comprehensive Analyzer of Matter Properties (CAMP). The ways for specific water treatment by EM fields, corresponding to activation or inhibition of concrete biological processes, can be found out. The limiting stages of relaxation of water perturbations, induced by 'guest' molecules after successive dilution and shaking, responsible for 'memory' of water can be investigated also. Such investigations could turn the homeopathy from the art to quantitative science.

The trivial Brownian collisions between interacting molecules can not explain a high rate of specific complex formation. One of possible explanation of specific distant interaction/attraction between the ligand and its receptor is the exchange resonant EM interaction, proposed by DigiBio team. The another explanation of such interaction between – sterically and dynamically complementary molecules in water, like in reaction: [antibody + hapten] or [enzyme + substrate] can be based on our Wave-Particle duality, Bivacuum and Virtual Replicas Unified theory (Kaivarainen, 2006).

This basically new Bivacuum and water-mediated interaction between Virtual Replicas (VR) of [S]ender (ligand) and [R]eceiver (protein) is a result of superposition: [$\mathbf{VR}^S \rightleftharpoons \mathbf{VR}^R$]. The carrying high frequency of VR of [S] and [R], equal to frequency of their [Corpuscle ⇌ Wave] pulsation and fundamental frequency of Bivacuum ($\omega_0 = m_0 c^2/\hbar$), is modulated by thermal vibrations of molecules (i.e. their thermal de Broglie waves), providing possibility of resonant distant interaction between molecules with close internal and external dynamics. The Virtual Replicas has a properties of 3-dimensinal Virtual/Quantum Holograms. Such kind of Bivacuum - mediated interaction (Kaivarainen, 2003) should be accompanied by increase of dielectric permittivity between interacting molecules, decreasing the Van-der-Waals interactions between water molecules and enhancing the coefficient of diffusion of ligand in this selected directions. The probability of cavitational fluctuations in water with average frequency of about $10^4$ Hz (see Fig 3.2d below), like revealed by DigiBio group, also should increase in the volume of $\mathbf{VR}^S$ and $\mathbf{VR}^R$ superposition, providing spatially selective way of ligand diffusion.

It's possible, that both mechanisms: the electromagnetic one and basically new Bivacuum mediated interaction - are interrelated with each other and contribute to specific distant interaction between ligands and the active sites of protein receptors *in vitro* and *in vivo*. This is a intriguing subject of future study.

## 3. Water as a Regulating Factor of Biopolymers Dynamic Structure Properties and Evolution

The dynamic model of proteins (Kaivarainen, 1985, 1995, 2007) leads to the following classification of dynamics in the native globular proteins (see also http://arXiv.org/abs/physics/0003093).

*1. Small-scale (SS) dynamics:* low amplitude (less than 1Å) thermal fluctuations of atoms, aminoacids residues, and displacements of alpha - helixes and beta - structures within domains and subunits, at which the effective Stokes radius of domains does not change. This type of motion can differ in the content of A and B conformers, corresponding to closed and open to water interdomain and intersubunit cavities. The range of characteristic times at SS dynamics of the surface aminoacids residues is around ($10^{-10} - 10^{-12}$) sec., determined by activation energies of conformational transitions and microviscosity. It corresponds to calculated frequency of ($a \rightleftharpoons b$) transitions of primary translational [tr] effectons (Fig. 3.2). The SS internal vibrations of aminoacids in rigid core of domains may be very slow $\sim 10^4 \, s^{-1}$;

2. *Large-scale (LS) dynamics* is subdivided into *LS-pulsation and LS-librations* (see Fig.3.1) in form of limited diffusion of domains and subunits of proteins:

a) *LS-pulsations* are represented by relative translational-rotational displacements of domains and subunits at distances about 3Å or more. Thus, big cavities of proteins, fluctuate between states of less (A) and more (B) water-accessibility. The life-times of these states depending on protein structure and external conditions are in the range of ($10^{-4}$ to $10^{-7}$) s. The [A - B] pulsations are accompanied by reversible sorption-desorption of (20 - 50) water molecules from the protein's cavities;

b) LS-*librations* represent the relative rotational - translational motions of domains and subunits in composition of A and B conformers with correlation times $\tau_M \simeq (1-5) \times 10^{-8}$s without [A - B] transitions.

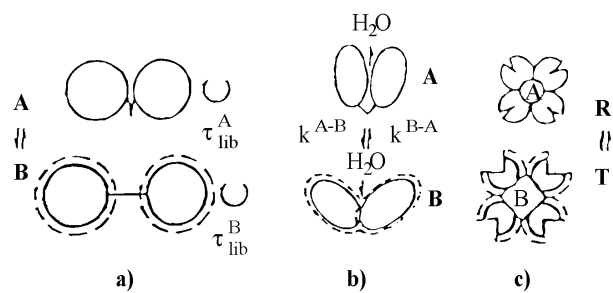

Figure 3.1 Examples of large-scale (LS) protein dynamics: $A \Leftrightarrow B$ pulsations and librations with correlation times ($\tau_{lb}^B < \tau_{lb}^A$) (Kaivarainen, 1985, 1995, 2007): a) mobility of domains connected by flexible hinge or contact region, like in the light chains of immunoglobulins; b) mobility of domains that form the active sites of proteins, like in hexokinase, papain, pepsin, lysozyme etc. due to flexibility of contacts; c) mobility of subunits forming the oligomeric proteins like hemoglobin. Besides transitions of the active sites of each subunit, the ($A \Leftrightarrow B$) pulsations with frequencies of ($10^4 - 10^6$) $s^{-1}$ are pertinent to the common central cavity.

The librational mobility of domains and subunits is revealed by the fact that the experimental value of $\tau_M$ is less than the theoretical one ($\tau_M^t$) calculated on the Stokes-Einstein formula:

$$\tau_M^t = (V/k) \eta/T \qquad 3.1$$

This formula is based on the assumption that the whole protein can be approximated by a rigid sphere. It means, that the large-scale dynamics can be characterized by the "flexibility factor", in the absence of aggregation equal to ratio:

$$fl = (\tau_M/\tau_M^t) \leq 1 \qquad 3.2$$

LS - librations of domains are accompanied by "flickering" of water cluster in the open cavity between domains or subunits. The process of water cluster "flickering", i.e. [disassembly ⇌ assembly] is close to the reversible first-order phase transition, when:

$$\Delta G_{H_2O} = \Delta H_{H_2O} - T\Delta S_{H_2O} \approx 0 \qquad 3.3$$

Such type of transitions in water-macromolecular systems could be responsible for so called "enthalpy-entropy compensation effects".

### 3.1 Role of water in dynamics of proteins

The "flickering clusters" means excitation of [*lb/tr*] conversions between librational and translational primary water effectons, accompanied by [association/dissociation] of coherent water. The water cluster (primary lb effecton) association and dissociation in protein cavities in terms of mesoscopic model represent the (*ac*) - convertons or (*bc*) - convertons. These excitations stimulate the LS- librations of domains in composition of B-conformer. The frequencies of (ac) and (bc) convertons, has the order of about $10^8 c^{-1}$, like the frequency of primary librational effectons excitation. This value coincides well with experimental characteristic times for protein domains librations Fig 3.2 b. The (ac) and (bc) convertons represent transitions between similar states of primary librational and translational effectons: $[a_{lb} \rightleftharpoons a_{tr}]$ and $[b_{lb} \rightleftharpoons b_{tr}]$ (see Section 1).

For the other hand, the Macroconvertons, representing simultaneous excitation of (*ac* + *bc*) convertons, are responsible for $[B \rightleftharpoons A]$ large-scale pulsations of proteins. The frequency of macroconvertons excitation is about $5 \cdot 10^6$ s$^{-1}$ at physiological temperatures (Fig.3.2c).

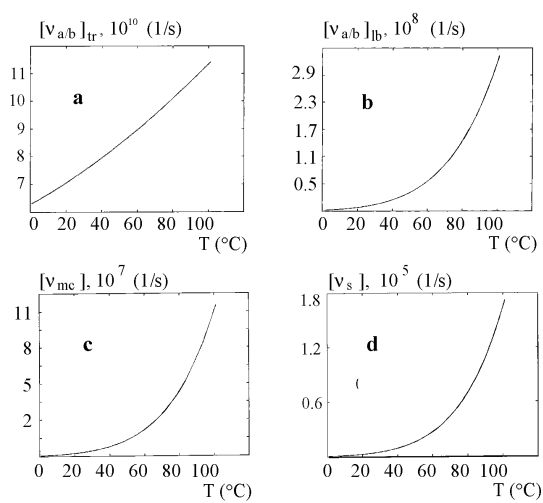

Figure 3.2. (a) - Frequency of primary [tr] effectons excitations; (b) - Frequency of primary [lb] effectons excitations; (c) - Frequency of [*lb/tr*] Macroconvertons (flickering clusters) excitations; (d) - Frequency of Superdeformons excitations.

At the temperature interval $(0\text{-}100)^0 C$ the frequencies of translational and librational macrodeformons (tr and lb) are in the interval of $(1.3\text{-}2.8) \times 10^9 s^{-1}$ and $(0.2\text{-}13) \times 10^6 s^{-1}$ correspondingly. The calculated frequency of primary translational effectons $[a \Leftrightarrow b]_{tr}$ excitations at $20^0 C$ (Fig. 3.2 a) is $v \sim 7 \times 10^{10} (1/s)$. It corresponds to electromagnetic wave length in water $\lambda = (cn)/v \sim 6mm$ with refraction index ($n = 1.33$). For the other hand, there are a lot of evidence, that irradiation of very different biological systems with such coherent electromagnetic field exert great influences on their properties.

The frequency of Superdeformons excitation (Fig.3.2d) is much lower, than that of macroconvertons: $v_s \sim (10^4 - 10^5) \, s^{-1}$. Superdeformons are responsible for cavitational fluctuations in water and disassembly of protein filaments. The pulsation frequency of oligomeric proteins, like hemoglobin or disassembly (peptization) of actin and microtubules could be also related with such big fluctuations. The life-times of (A) and (B) conformer markedly exceeds the transition-time between them ($10^{-9}$ to $10^{-11}$) s.

*The (A - B) pulsations of various cavities in allosteric proteins are correlated.* The corresponding A and B conformers have different Stokes radii and effective volume. The geometrical deformation of the inter-subunits large central cavity of oligomeric proteins and the destabilization of the water cluster located in it, lead to relaxational change of (A - B) equilibrium constant, providing their cooperative properties. At the temperature interval (0-100 C$^0$) the frequencies of translational and librational macrodeformons (tr and lb) are in the interval of (1.3-2.8) $\times 10^9$ 1/s and (0.2-13)$\times 10^6$ s$^{-1}$ correspondingly. It is obvious, that between the dynamics/function of proteins, membranes, etc. and dynamics of their aqueous environment the strong correlation exists.

### 3.2 The role of water in mechanism of protein-ligand specific complex formation and signal transmission between domains and subunits

According to our model of specific complexes formation the following order of events is assumed (Fig. 3.3):

1. Ligand (L) collides with the active site (AS), formed usually by two domains, in its open (b) state: the structure of water cluster in AS is being perturbed and water is forced out of AS cavity totally or partially;

2. Transition of AS from the open (b) to the closed (a) state occurs due to strong shift of [$a \Leftrightarrow b$] equilibrium to the left, i.e. to the AS domains large scale dynamics;

3. A process of dynamic adaptation of complex [L + AS] begins, accompanied by the directed ligand diffusion in AS cavity due to its domains small-scale dynamics and deformation of their tertiary structure;

4. If the protein is oligomeric with few AS, then the above events cause changes in the geometry of the central cavity between subunits in the open state leading to the destabilization of the large central water cluster and the shift of the $A \rightleftharpoons B$, corresponding to $R \rightleftharpoons T$ equilibrium of quaternary structure leftward. Water is partially forced out from central cavity. Due to the feedback mechanism this shift can influence the [$a \Leftrightarrow b$] equilibrium of the remaining free AS and promotes its reaction with the next ligand. Every new ligand stimulates this process, promoting the positive cooperativity. The negative cooperativity also could be resulted from the interaction between central cavity and active sites;

5. The terminal [*protein – ligand*] complex is formed as a consequence of the relaxation process, representing deformation of domains and subunits tertiary structure. This stage could be much slower than the initial ones [1-3]. As a result of it, the stability of the complex grows up.

*Dissociation of specific complex* is a set of reverse processes to that described above which starts from the [$a^* \to b$] fluctuation of the AS cavity.

In multidomain proteins like antibodies, which consist of 12 domains, and in oligomeric proteins, the cooperative properties of $H_2O$ clusters in the cavities can determine the mechanism of signal transmission from AS to the remote effector regions and allosteric protein properties.

The stability of a librational water effecton as coherent cluster strongly depends on its sizes and geometry. This means that very small deformations of protein cavity, which violate the [cavity-cluster] complementary condition, induce a cooperative shift of [$A \Leftrightarrow B$] equilibrium leftward. The clusterphilic interaction, introduced by this author earlier, turns to hydrophobic one due to [*lb/tr*] conversion.

This process can be developed step by step. For example, the reorientation of variable domains, which form the antibodies active site (AS) after reaction with the antigen determinant or hapten deforms the next cavity between pairs of variable and constant domains forming F$_{ab}$ subunits (Fig.2.3). The leftward shift of [$A \Leftrightarrow B$] equilibrium of this cavity, in turn, changes the geometry of the big central cavity between $F_{ab}$ and $F_c$ subunits, perturbing the structure of the latter. Therefore, the signal transmission from the AS to the effector sites of $F_c$ subunits occurs due to the balance shift between clusterphilic and hydrophobic interactions. This signal may be responsible for complement- binding sites activation and triggering the receptors function on the lymphocyte membranes.

The leftward shift of [$A \Leftrightarrow B$] equilibrium in a number of cavities in the elongated multidomain proteins can lead to the significant decrease of their linear size and dehydration. The mechanism of muscular contraction is probably based on such phenomena and clusterphilic interactions. The clusterphilic interactions means that interaction of the open protein cavity with water cluster is energetically more preferable, than with the same number of molecules after cluster disassembly (Kaivarainen, 1985, 2001, 2007). For such a nonlinear system the energy is necessary for reorientation of the first couple of domains only. The process then goes on spontaneously with decreasing the averaged protein chemical potential. The chemical potential of the A- conformer is usually lower than that of B- conformer ($\bar{G}_A < \bar{G}_B$) and the relaxation of protein is accompanied by the leftward $A \Leftrightarrow B$ equilibrium shift of cavities, accompanied by decreasing of the averaged protein dimensions. The shift of [A – B] equilibrium of central cavity of oligomeric proteins determines their cooperative properties during consecutive ligand binding in the active sites. Signal transmission from the active sites to the remote regions of macromolecules is also dependent of [A – B] equilibrium (Fig. 3.3). The evolution of the ideas of the protein-ligand complex formation proceeded in the following sequence:

1. "Key-lock" or the rigid conformity between the geometry of an active site and that of a ligand (Fisher, 1894);

2. "Hand-glove" or the so-called principle of induced conformity (Koshland, 1962);

3. Our model allows to put forward a new "Principle of Stabilized Conformity (PSC)" instead that of "induced conformity" in protein-ligand specific reaction (Kaivarainen, 1985, 2007). *Principle of Stabilized Conformity (PSC)* means that the geometry of the active site (AS), optimal from energetic and stereochemical conditions, is already existing *before* reaction with ligand. The optimal geometry of AS is to be the only one selected among the number of others and stabilized by ligand, but not induced "de nova";

4. New additional mechanism of specific distant coresonant interaction between ligands and protein, based the electromagnetic and superposition of Virtual Replicas of interacting molecules, is proposed in section 2.3 of this paper.

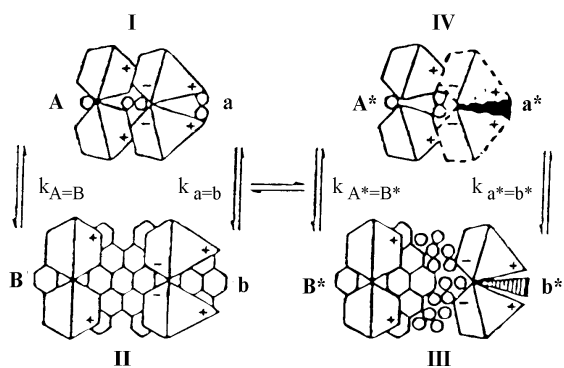

Figure 3.3. The schematic picture of the protein association (Fab subunits of antibody with a ligand), which is accompanied by destabilization of water clusters in cavities, according to dynamic model (Kaivarainen, 1985). The dotted line denotes the perturbation of the tertiary structure of the domains forming the active site. Antibodies of IgG type contain usually two such Fab subunit and one Fc subunit, conjugated with 2Fab by flexible hinge, forming the general Y-like structure.

### 3.3 The role of water in spatial parameters of proteins

The number of water molecules within the primary librational effectons of water, which could be approximated by a cube, decreases from 280 at ($0^0$C) to 3 at ($100^0$C) (Fig. 2.4). It should be noted that at physiological temperatures (35-40 C) such quasiparticles contain nearly 40 water molecules. This number is close to that of water molecules that can be placed in the open interdomain protein cavities judging from X-ray data[5]. Structural domains are space-separated units with a mass of $(1-2)\times 10^{-3}$ D. Protein subunits, as a rule consist of two or more domains.

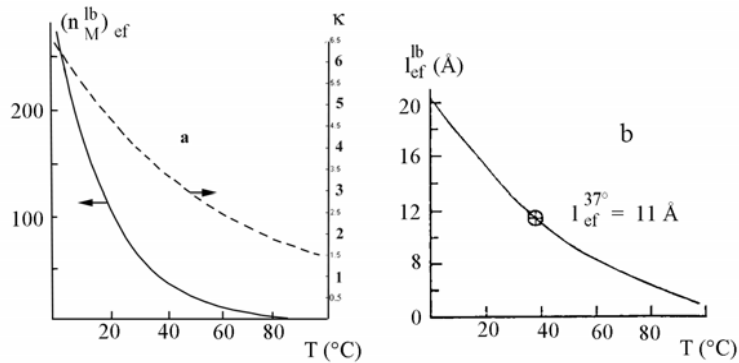

Figure 3.4. (*a*) : The temperature dependencies of the number of $H_2O$ molecules in the volume of primary librational effecton ($(n_M^{lb})_{ef}$, left axis) and the number of $H_2O$ per length of this effecton edge ($\kappa$, right axis); (b): the temperature dependence of the water primary librational effecton (approximated by cube) edge length $[l_{ef}^{lib} = \kappa (V_0/N_0)^{1/3}]$.

The number of $H_2O$ molecules within the *primary libration effectons* of water, which could be approximated by a cube, decreases from $n_M = 280$ at $0^0$ to $n_M \simeq 3$ at $100^0$ (Fig. 3a). It should be noted that at physiological temperatures $(35 - 40^0)$ such quasiparticles contain nearly 40 water molecules. Similar by order dimensions of heavy water clusters (about 10 Å) with saturated hydrogen bonds were revealed using inelastic neutron scattering method by Texeira et al., in 1987. The dimensions of water clusters are close to dimensions of the open interdomain protein cavities judging from X-ray data (Kaivarainen, 1985). The flickering of these clusters, i.e. their [disassembly ⇌ assembly] due to [lb\tr] conversions in accordance to our model is directly related to the large-scale dynamics of proteins, presented on Fig. 3.1). It is important finding that the linear dimensions of the interdomain water clusters (about 11 Å) in 'open' states of protein cavities at physiological temperature, calculated using our software, are close to common ones for protein domains. Such spatial correlation indicate that the properties of water exerted a strong influence on the biological evolution of macromolecules, namely, their dimensions and allosteric properties due to cooperativity of intersubunit water clusters. The correlation between dimensions of microtubules (about 10 microns) and wave-length of standing librational IR photons, composing primary electromagnetic deformons in water, points that not only spatial characteristics of biopolymers, but also the cell's dimensions are determined by water properties [1, 2]. Consequently, the calculations, based on our hierarchic theory, give a strong evidence, that water was one of the most important factors in evolution of biopolymers and cells. The new mechanism of ATP hydrolysis energy utilization in muscle contraction and role of water in cancer also is presented in full paper http://arXiv.org/abs/physics/0003093.

## 4 Multi-Fractional Model of Interfacial Water Structure. Its Contribution to Morphogenetic Field

Our model of multi-fractional water structure, formed on solid-liquid interface, is based on Hierarchic theory of condensed matter (Kaivarainen, 1995; 2001; 2007). Classification and description of *four* interfacial water fractions, in accordance to our Multi-fractional model of interfacial (solid-liquid) water structure:

*1. The first fraction is Primary hydration Shell (PS)* with maximum energy of interaction with surface. The structure and dynamics of this 1st fraction can differ strongly from those of bulk water. Its thickness: corresponds to 1-3 solvent molecule. In accordance to generally accepted and experimentally proved models of hydration of macromolecules and colloid particles, we assume that PS of strongly bound water molecules, serves like intermediate shell, neutralizing the specific chemical properties of surface (charged, polar, nonpolar, etc.). Such strongly bound water can remain partially untouched even after strong dehydration of samples in vacuum. This 1st fraction of interfacial water serves like a matrix for second fraction - *vicinal water* shell formation. The properties of vicinal water are independent on specific chemical structure of the surface - from quartz plates, mineral grains and membranes to large macromolecules (Clegg and Drost-Hansen, 1991). This can be a result of "buffering" effect of primary hydration shell;[8]

   2. *This author suggest, that the second fraction - vicinal water (VW)* is formed by primary librational [lb] effectons - coherent molecular clusters with properties of mesoscopic Bose-condensate (mBC) with saturated hydrogen bonds and less density than the average one of bulk water. It is a result of [lb] effecton adsorption from the bulk volume on the primary hydration shell (PS). Vicinal water (VW) can be formed in the volume of pores, near the curved and plate interfaces as a result of interaction with strongly surface-bound water of PS.

   The decreasing of most probable [lb] thermal momentums of water molecules, especially in direction, normal to the surface of macromolecule or colloid particle as a consequence of interaction with primary shell (PS), should lead to increasing of corresponding edge's length of primary [lb] effectons, forming VW, as compared to the bulk water effectons. This selected immobilization of water molecules change the cube-like shape of effectons of the bulk water to shape of elongated parallelepiped for the effectons of VW. It is a result of increasing of corresponding wave B length of water molecules as a ratio of Plank constant to their most probable momentum.

   The increasing of life-time of these enlarged primary [lb] effectons in the (a) state - means the increasing of their stability and concentration in the volume of VW. As far we assume, that VW is a result of adsorption of primary librational effectons on primary hydration shell and their elongation in direction, normal to surface, we can make some predictions:

   a) The thickness of VW can be about 30-75 Å, depending on properties of surface (geometry, polarity), temperature, pressure and presence the perturbing solvent structure agents (the linear dimension of primary librational effecton of bulk water at $25^0 C$ is only about 15 Å);

   b) The elongation of primary [lb] effectons in direction, normal to the surface, should be resulted in increasing the intensity of librational IR photons superradiation in the same direction (Dicke, 1954);

   c) The vicinal water (VW) of second interfacial fraction should differ by number of physical parameters from the bulk water. For example, VW should have: lower density; bigger heat capacity; bigger sound velocity; bigger viscosity; smaller dielectric relaxation frequency, etc. The lower mobility of water molecules of vicinal water is confirmed directly by almost 10 times difference of dielectric relaxation frequency ($2 \times 10^9$ Hz) as respect to bulk one ($19 \times 10^9$ Hz) (Clegg and Drost-Hansen, 1991). The increasing of temperature should lead to decreasing the vicinal librational effectons dimensions and thickness of VW shell;

   3. *The third fraction of interfacial water:* surface-stimulated Bose-condensate (SS-BC) is represented by 3D network of primary [lb] effectons (mBC) with a thickness of (50-300 Å), stabilized by Josephson contacts. It is a next hierarchical level of interfacial water self-organization, using the second vicinal fraction (VW) as a matrix of nucleation centers for SS-BS. The time of gradual formation of this 3D net of linked to each other coherent clusters (strings of polyeffectons), can be much longer than that of VW and more sensitive to temperature and mechanical perturbations. The second and third fractions of interfacial water can play an important role in biological cells activity regulation;

4. The biggest and most fragile *forth fraction* of interfacial water is a result of slow orchestration of bulk primary effectons in the volume of primary (electromagnetic) [lb] deformons. The primary deformons appears as a result of superposition of three standing IR librational photons, normal to each other. Corresponding IR photons are radiated by the enlarged primary [lb] effectons of vicinal water and those of SS-BC. The linear dimension of librational IR deformons is about half of librational IR photons wave length, i.e. 5 microns.

This "electromagnetically orchestrated water (EM-OW)" fraction easily can be destroyed not only by temperature increasing, ultrasound and Browning movement, but even by mechanical shaking. The time of spontaneous reassemble of this fraction after destruction has an order of hours and is dependent strongly on temperature, viscosity and dimensions of colloid particles.

### 4.1 Possible role of interfacial water near cell's microfilaments in morphogenetic field formation

In biosystems, like living cells, the IR radiation of second (VW) and third (SS-BC) fractions of interfacial water, orchestrated in the internal core of microtubules (MT) and around MTs and actin filaments, may contribute to "morphogenetic field", revealed by A. Gurwich in form of EM field. Later it was confirmed in different laboratories, that EM radiation is accompanied the cells division and differentiation. The directed IR *superradiation* of interfacial water in cytoplasm should be dependent on orientation of microtubules and actin filaments. The known non-linear optical effect - "superradiation" (Dicke, 1957) is a part of our Hierarchic theory of condensed matter (Kaivarainen, 1991, 1995, 2007). It is a consequence of water coherent clusters - mesoscopic Bose condensate (mBC) ability to quantum beats between their optic and acoustic modes. Superradiation should be maximum from the ends of microtubules, in accordance with theory of this effect. Superposition of corresponding coherent IR photons can be responsible for formation of primary deformons, stimulating cavitational fluctuations in certain volumes of cytoplasmic water and reversible disassembly of MTs and actin filaments.

In accordance to our Hierarchic model of consciousness, described in chapter 6, the intensity of IR coherent photons superradiation is maximum from the ends of microtubules (MTs). Their superposition leads to formation of hologram like system of primary deformons, responsible for distant cell-cell interaction and regulation of their cytoplasmic dynamics.

Each $\alpha\beta$ tubulins dimer of MT is a dipole with negative charges, shifted towards $\alpha$ subunit. Consequently, microtubules, as an oriented elongated structure of dipoles system, have the piezoelectric properties (Athestaedt, 1974).

Hollow core of MT is has a diameter of 140 Å. All the internal water of MT may represent the 1st, 2nd and 3d fractions of interfacial water, described above. The spatial orientation of two bundles of MTs, containing about 2×25 = 50 MTs is determined by orientation of two centrioles, forming as a rule the right angle (Fig 4.1). Consequently, the directions of penetration of IR librational photons, superradiated from the ends of MTs, usually fixed on cell's membrane, also have the almost right angle relative orientation. Interception of these coherent IR photons with those, superradiated by other cells leads to formation of 3D standing waves, i.e. IR primary [lb] deformons with linear dimension of 5 microns ( 1/2 [lb] photon wave-length).

Distribution of density of inorganic ions, especially bivalent like $Ca^{2+}$, and probability of their fluctuations, affecting the water activity, should be regulated by anisotropy of the electric field tension in the volume of 3D electromagnetic standing waves. We suppose, that the corresponding spatial distribution of water activity ($a_{H_2O}$) plays the important modulation role in proteins dynamics/function and dynamic equilibrium of [assembly ⇌ disassembly] of microtubules and actin filaments, responsible for cell's shape.

The process of cavitational fluctuations 'collapsing' with frequency of superdeformons excitation ($\sim 10^4\, Hz$) is accompanied by high-frequency (UV and visible) "biophotons" radiation due to recombination of dissociated to hydroxyl and proton water molecules. These biophotons may be responsible for short range morphogenetic field in contrast to coherent IR photons, standing for long-range morphogenetic field. This could be one of the possible mechanism of morphogenetic field action. The another component of morphogenetic field may be related with superposition of virtual replicas (VR) of DNA, microtubules, actin filaments, responsible for cells 3D structure.

The notion of VR leads from our Unified model of Bivacuum and wave - corpuscle duality, as base for quantum entanglement, discussed in paper: "Unified Theory of Bivacuum, Particles Duality, Fields & Time. New Bivacuum Mediated Interaction, Overunity Devices, Cold Fusion & Nucleosynthesis": http://arxiv.org/abs/physics/0207027.

## 5. Distant Solvent-Mediated Interaction Between Different Proteins and Between Proteins and Cells

### 5.1. Distant solvent-mediated interaction between macromolecules

The most of macromolecules, including proteins, are existing in dynamic equilibrium between two conformers (A and B) with different hydration ($n_{H_2O}$) and flexibility:

$$A + n_{H_2O} \Leftrightarrow B \quad\quad 5.1$$

Usually the correlation time of more hydrated B - conformer ($\tau_B$), dependent on its *effective* volume ($V_B$) and solvent viscosity ($\eta$) are less, than that of more rigid A-conformer ($\tau_A \sim V_A$) in accordance to Stokes - Einstein law:

$$\tau_{A,B} = \frac{V_{A,B}}{k}(\eta/T)$$

$k$ is the Boltzmann constant $\quad\quad 5.2$

$$\tau_A > \tau_B \quad \text{and} \quad V_A > V_B$$

This means that flexibility of more hydrated B-conformer, determined by large-scale dynamics, is higher, than that of A-conformer.

For such a case, the change of the bulk water activity ($a_{H_2O}$) in solution by addition of other macromolecules or inorganic ions induce the change of the equilibrium constant: $K_{A \Leftrightarrow B} = (K_{B \Leftrightarrow A})^{-1}$ and the dynamic behavior of macromolecules (Kaivarainen, 1985, 1995) :

$$\Delta \ln K_{B \Leftrightarrow A} = n_{H_2O} \Delta \ln a_{H_2O} \quad\quad 1.3$$

The developed by this author experimental approach, based on quite different dependencies of microdynamics and macrodynamics (Brownian rotation) on viscosity of solvent in solution of spin-labeled macromolecules (see section 3.6 in book by Kaivarainen, 1985) makes it possible to evaluate *separately*:

a) the frequency of spin-label rotation as respect to protein surface ($\nu_R \sim 1/\tau_R$), where $\tau_R$ is a correlation time of spin-label itself, depending on microviscosity of protein matrix (small-scale dynamics) and

b) the effective frequency of rotation of protein, as a whole (large-scale dynamics), i.e.

the averaged ($v_M \sim 1/\tau_M$), where $\tau_M$ is the effective correlation time of the mixture of A and B - conformers, depending on $A \Leftrightarrow B$ equilibrium.

The isothermal viscosity $(\eta)_T$ dependences of resulting experimental correlation time of spin-label, conjugated with protein ($\tau_{R+M}$), makes it possible to get a separate information about the protein small-scale and large-scale dynamics:

$$\left(\frac{1}{\tau_{R+M}}\right)_T = \frac{1}{\tau_R} + \frac{1}{\tau_M} \qquad 5.3a$$

or using (1.2) in conditions $T = const$, we get:

$$\left(\frac{1}{\tau_{R+M}}\right)_T = \left(\frac{1}{\tau_R}\right)_T + \frac{(\eta/T)_{st}}{(\tau_M)_{st}}\left(\frac{T}{\eta}\right)_T = \left(\frac{1}{\tau_R}\right)_T + \frac{k}{(V_M)_{st}}\left(\frac{T}{\eta}\right)_T \qquad 5.3b$$

where: $(\tau_M)_{st}$ and $(V_M)_{st}$ are the effective correlation time and effective volume of protein, reduced to standard conditions, when the ratio $(\eta/T)_{st} = 3 \times 10^{-5}$ P/K, corresponding to this value for pure water at $25^0 C$.

The slopes of dependence of $\left(\frac{1}{\tau_{R+M}}\right)_T$ on $\left(\frac{T}{\eta}\right)_T$ give us the parameters of large-scale dynamics: $(\tau_M)_{st} \sim (V_M)_{st}$. Usually these parameters are dependent on temperature and ligand binding, shifting the $A \Leftrightarrow B$ equilibrium to the left or right. The interceptions of isotherms with ordinate ($1/\tau_{R+M}$), extrapolated to infinitive solvent viscosity ($T/\eta$) → 0, give the parameter of small-scale dynamics in spin-label environment ($1/\tau_R$) (Kaivarainen, 1975, 1985).

Similar viscosity approach for separate investigation of the large scale and small-scale dynamics of proteins has been developed also, using microcalorimetry method (Kaivarainen et.al., 1993).

In mixed systems: [PEG + spin-labeled antibody] the dependence of large-scale (LS) dynamics of antibody on the molecular mass of polyethylenglycol (PEG) is similar to dependence of water activity and freezing temperature ($T_f$) of PEG solution (see book by Kaivarainen, 1985, Fig. 82).

The presence of PEG with molecular mass and concentration, increasing ($a_{H_2O}$) and ($T_f$), stimulate LS-dynamics of proteins decreasing their effective volume $V$ and correlation time $\tau_M$ in accordance to eqs.(5.3 and 5.2).

If $\Delta T_f = T_f^0 - T_f$ is the difference between the freezing point of a solvent ($T_f^0$) and solution ($T_f$), then the relation between water activity in solution and $\Delta T_f$ is given by known relation:

$$\ln a_{H_2O} = -\left[\frac{\Delta H}{R(T_f^0)^2}\right]\Delta T_f \qquad 5.4$$

where $\Delta H$ is the enthalpy of solvent melting; R is the gas constant.

In our experiments with polymer solutions the 0.1 M phosphate buffer
$pH\ 7.3 + 0.3M$ NaCl was used as a solvent.

One can see from (5.3) that the negative values of $\Delta T_f$ in the presence of certain polymers means the increasing of water activity in three component [water - ions - polymer] system ($\Delta \ln a_{H_2O} > 0$). In turn, it follows from eqs. (5.1 - 5.3) that the increasing of $a_{H_2O}$ shifts the $[A \Leftrightarrow B]$ equilibrium of proteins in solutions to the right. Consequently, the flexibility of the proteins will increase as far $\tau_B > \tau_A$.

The correlation between freezing temperature $T_f$, the water activity ($a_{H_2O}$) and immunoglobulin flexibility ($\tau_M$), corresponding to (5.4 and 5.2) was confirmed in our experiments (Kaivarainen, 1985, Table 13).

It was shown that *protein-protein* distant interaction depends on their LS dynamics and conformational changes induced by ligand binding or temperature (Kaivarainen, 1985).

The temperature dependencies of correlation time of spin labeled human serum albumin (HSA-SL), characterizing the rigidness of macromolecule (the effective volume) in regular solvent and in presence of 3% $D_2O$ is presented at Fig.5.1.

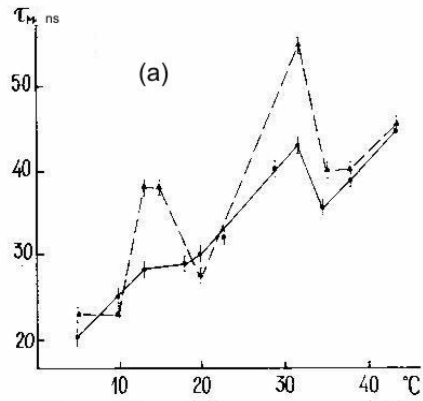

**Fig. 5.1**. The temperature dependence of resulting correlation time ($\tau_M$) of spin-labeled human serum albumin (HSA-SL) and similar dependence in presence of 3% $D_2O$ (dotted line). Concentration of HSA was 25 mg/cm$^3$ in 0.01 M phosphate buffer (pH 7.3) + 0.15 M NaCl.

Fig.5.2 demonstrates a distant, solvent-mediated interaction between human serum albumin (HSA) and spin labeled hemoglobin (Hb-SL). We can see, that in presence of HSA its temperature - induced changes of flexibility/rigidity (correlation time, fig. 5.2) influence the flexibility of Hb-SL. The peak of correlation time of HSA around 15$^0$C, stimulated by presence of 3% $D_2O$ also induce a corresponding transition in large-scale dynamics of Hb-SL.

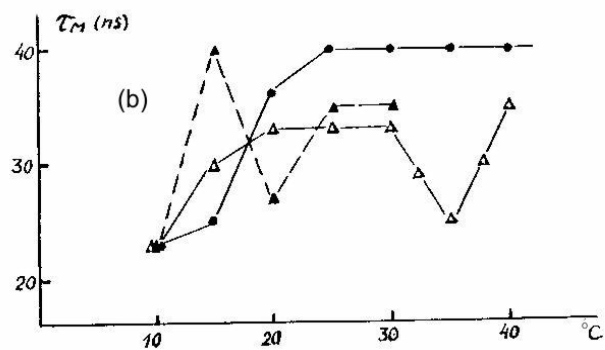

**Fig.5.2**. The temperature dependence of resulting correlation time ($\tau_M$) of:
a) spin-labeled oxyhemoglobin (Hb-SL) - black dots;
b) similar dependence in presence of human serum albumin (HSA);
c) similar dependence in presence of human serum albumin (HSA) and 3% $D_2O$ (dotted line).
The concentration of oxyhemoglobin was 20 mg/cm$^3$, the concentration of HSA was 10 mg/cm$^3$ in solvent: 0.01 M phosphate buffer (pH 7.3) + 0.15 M NaCl.

Between the clusterphilic interaction, discussed in previous section, and protein flexibility or rigidity, characterized by $\tau_B$, the positive correlation exists. It means that the increasing of dimensions *of librational bulk water effectons is accompanied by enhancement of the water clusters dimensions in protein cavities, following, in turn, by shift of A ⇔ B* equilibrium of the cavities to the right - to the more flexible conformer ($\tau_B < \tau_A$).

Our interpretation is confirmed by the fact, that lowering of the temperature and corresponding increasing of the dimensions of bulk librational effectons (coherent water clusters) and stabilizing the open states of interdomain and intersubunit cavities - increases the flexibility of protein.

At the *low concentration of macromolecules* ($C_M$), when the average distance ($r$) between them, dependent on molar concentration ($C_M$):

$$r = \frac{11.8}{C_M^{1/3}} \, \text{Å} \qquad 5.5$$

is much bigger, than dimensions of primary librational effecton (cluster in state of mesoscopic Bose condensate) of bulk water ($r \gg \lambda_{lb}$) the large-scale [$A \Leftrightarrow B$] pulsations of proteins, accompanied by exiting of acoustic waves in solvent, induce the enhancement of *water activity ($a_{H_2O}$)*.

Such influence of pulsing proteins on solvent can be responsible for the distant solvent-mediated interaction between macromolecules at low concentration, described above.

The acoustic momentums in protein solutions are the result of the jump-way $B \to A$ transitions of interdomain or intersubunit cavities with characteristic transition time about $10^{-10}$ sec. This rapid transition follows the cavitational fluctuation of a water cluster formed by 30 - 70 water molecules in space between domains and subunits of oligomeric proteins. This fluctuation of density is a result of conversion of librational primary effecton to number translational ones (disassembly of coherent molecular cluster).

In concentrated solutions of macromolecules, when the distance between macromolecules starts to be less than linear dimension of primary librational effecton of water: $r \leq \lambda_{lb}$, the formation of thixotropic structures and trivial aggregation begin dominate. This is a consequence of decreasing of water activity in solutions.

### 5.2. **Distant solvent-mediated interaction between proteins and cell**

The unknown earlier phenomena, like distant, solvent - mediated interaction between different proteins (fig.2), was revealed in our experiments, using modified method of spin-label (Kaivarainen, 1985, section 8.5). The similar kind of interaction between proteins and cells, modulated by temperature and protein ligand state, was discovered on examples of mixed systems: erythrocyte suspension + human serum albumin (HSA), using turbidimetry (light scattering) method (Fig 5.3).

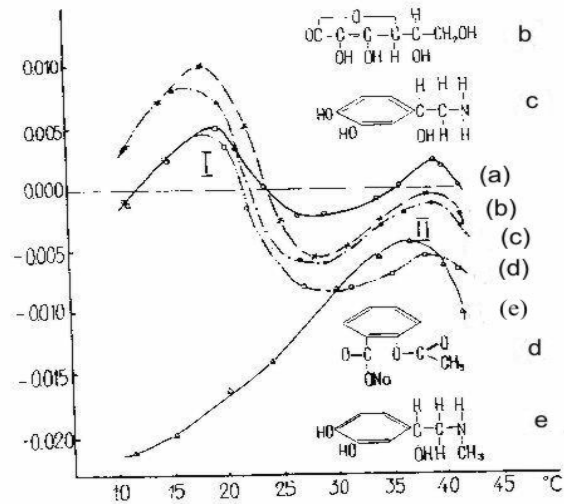

**Fig**.**5**.**3**. The differential temperature dependence of turbidity of human erythrocytes suspension at the light wave length $\lambda$ = 600 nm in the presence and absence of human serum albumin (HSA) in different ligand state:

$$\Delta D^* = D^*_{ER+HSA} - D^*_{ER}$$

a) the interaction with intact HSA; b) HSA + ascorbic acid; c) HSA + noradrenaline; d) HSA + sodium acetyl-salicylate; e) HSA + adrenaline. The experiments where performed in modified physiological Henx solution (pH7.3), where the glucose was excluded to avoid possible artefacts on osmotic processes (Kaivarainen, 1985, Fig.84).

The concentration of erythrocytes was $2 \times 10^5$ cm$^{-1}$ and the concentration of HSA was 15mg/cm$^3$. All ligands had fivefold molar excess over HSA. The erythrocytes suspension was incubated at least 24 hours before measurements for exhaustion of cellular ATF and elimination of the active osmos. A number of special experimental controls where performed to prove the absence of absorption of HSA on the membranes of cells and direct effect of ligands on turbidity of erythrocyte suspension (Kaivarainen, 1985, section 8.5).

The *swelling* and *shrinking* of erythrocytes, reflected by increasing and decreasing of their suspension turbidity, is a consequence of increasing and decreasing of the external water activity (Fig.5.3). The latter factors are responsible for the direction of water diffusion across the membranes of cells ('in' or 'out') as a result of passive osmos. They are in-phase with corresponding changes of HSA large-scale dynamics (flexibility). The latter was revealed in special experiments with spin-labeled HSA-SL (see Fig.5.1). The same kind of temperature dependence of HSA - SL flexibility was confirmed in modified Henx solution.

We got the modified Boyle van Hoff equation, pointing to direct correlation between the increments of cell's volume and the external water activity (Kaivarainen, 1985): $\Delta \mathbf{V} \sim \Delta \mathbf{a}_{H_2O}$.

Both thermoinduced conformational transitions of serum albumin occur in physiological region of surface tissues of body. Consequently, presented data point to possible role of albumin (the biggest protein fraction in blood) in thermal adaptation of many animals, including human, using the discovered *solvent - mediated mechanism of protein - protein and protein - cell interaction.*

*5.2.1 Possible mechanism of water activity increasing in solutions of macromolecules*

Let us analyze in more detail the new effect, discovered in our work: the increasing of water activity ($a_{H_2O}$) in presence of macromolecules in three component [water - salt - macromolecules] system. The Gibbs-Duhem law for this case can be presented as (Kaivarainen, 1988):

$$X_{H_2O}\Delta \ln a_{H_2O} + X_M \frac{\Delta \mu_M}{RT} + X_i \Delta \ln a_i = 0 \qquad 5.6$$

where $X_{H_2O}$, $X_M$, $X_i$ are the molar fractions of water, macromolecules and ions in the system;

$$a_j = y_j X_j = \exp\left(-\frac{\mu_0 - \mu_j}{RT}\right) = \exp\left(-\frac{\Delta \mu_j}{RT}\right) \qquad 5.7$$

is the activity of each component related to its molar fraction ($X_j$) and coefficients of activity ($y_j$);

$$\mu_M = f_B G_B + f_A G_A \simeq f_B(G_B - G_A) + G_A \qquad 5.8$$
$$\Delta \mu_M \simeq (G_B - G_A)\Delta f_B \qquad 5.8a$$

is the mean chemical potential ($\mu_M$) of a macromolecule (protein), pulsing between A and B conformers with corresponding partial free energies $G_A$ and $G_B$ and its change ($\Delta \mu_M$), as a result of $A \rightleftharpoons B$ equilibrium shift, taking into account that $f_B + f_A \cong 1$ and $\Delta f_B = -\Delta f_A$ and

$$\Delta \ln a_i = (\Delta a_i / a_i) \simeq -\Delta \kappa_i \qquad 5.9$$

where the fraction of thermodynamically excluded ions (for example, due to ionic pair formation):

$$\kappa_i = 1 - y_i \qquad 5.10$$

One can see from (5.6) that when $a_{H_2O} < 1$, it means that

$$\mu^0_{H_2O} > \mu^S_{H_2O} = H^S_{H_2O} - TS^S_{H_2O} \qquad 5.11$$

It follows from eq. (5.11) that the decreasing of water entropy ($\bar{S}$) in solution related to hydrophobic and clusterphilic interactions may lead to increasing of $\mu^S_{H_2O}$ and water activity.

It is easy to see from (5.6) that the elevation the concentration and $X_M$ of macromolecules in a system at constant temperature and $\Delta \bar{\mu}_M$ may induce a rise in water activity ($a_{H_2O}$) only if the activity of ions ($a_i = y_i X_i$) is decreased. The latter could happen due to increasing of fraction of thermodynamically excluded ions ($\kappa$) (eqs. 5.9 and 5.10).

There are *two processes* which may lead to increasing the probability of ionic pair formation and fraction $\kappa$ elevation.

The *first* one is the forcing out of the ions from the ice-like structure of enlarged librational effectons, stimulated by the presence relatively high concentration of macromolecules and strong interfacial effects (2d and 3d fractions of hydration shell ( Kaivarainen 2003). This effect of excluded volume increases the effective concentration of inorganic ions and probability of association, accompanied by their dehydration.

The *second* process dominates at the low concentration of $[A \Leftrightarrow B]$ pulsing macromolecules, when the thixotropic structure fail to form ($r = 11.8/C_M^{1/3} \gg \lambda_{lb}$). The acoustic waves in solvent, generated by pulsing proteins stimulate the fluctuation of ion concentration (Kaivarainen, 1995) also increasing the probability of neutral ionic pairs formation and their dehydration.

In accordance to Gibbs-Duhem law (5.6), the decreasing of ionic activity: $\Delta \ln a_i < 0$,

should be accompanied by increasing of water activity: $\Delta \ln a_{H_2O} > 0$ and increasing of $\Delta \mu_M$, meaning shift of $[A \Leftrightarrow B]$ equilibrium to the right - toward more hydrated and flexible B-conformer of proteins, when $\Delta f_B > 0$ (see eq.5.8a).

### 5.3 Possible mechanisms of the remote interaction between the active sites of proteins and the ligands

The kind of "memory of water", discovered by Jacques Benveniste team in 1988 (Davenas, et.al. 1988), includes the ability of water to carry information about biologically active guest molecules and possibility to record, transmit and amplify this information. This phenomenon involves the successive diluting and shaking of [water + guest] system to a degree where the final solution contains no guest molecules more at all. However, using hypersensitive biological cells-containing test systems, he observed that this highly diluted solution initiated a reaction in similar way, as if the active guest molecules were still present in water. From the first high dilution experiments in 1984 to the present, thousands of experiments have been made in DigiBio company (Paris), enriching and considerably consolidating the initial knowledge of such kind of water memory. It was demonstrated also by DigiBio team, that low frequency ( 20 kHz $= 2 \times 10^4$ s$^{-1}$) electromagnetic waves are able to activate biological cells. These results indicate that the molecular signal is composed of waveforms in the $(10 - 44)$ kHz range which are specific to each molecular entity. This prompted J. Benveniste to hypothesize that the molecular signal is composed of such low frequency waves and that the ligand coresonates with the cell receptor on these frequencies, stimulating specific attraction between them.

*However, we have to note, that just in this frequency range the resonant cavitational fluctuations in water can be excited by EM or acoustic fields (Kaivarainen, 1995, 2001, 2007). This effects, following by excitation of acoustic waves, also can be responsible for the membranes perturbations and cells metabolic processes activation.*

The perturbation of water properties, induced by haptens and antibodies, concurrent inhibitors and enzymes, viruses and cells in separate and mixed solutions can be studied, using our Hierarchic theory of condensed matter and this theory based Comprehensive Analyzer of Matter Properties (CAMP) (Kaivarainen, 2003). The ways for specific water treatment by EM fields, corresponding to activation or inhibition of concrete biological processes, can be found out. The limiting stages of relaxation of water perturbations, induced by 'guest' molecules after successive dilution and shaking, responsible for 'memory' of water can be investigated also. Such investigations in case of success could turn the homeopathy from the art to quantitative science.

*It can be calculated, that the trivial Brownian collisions between interacting molecules in aqueous medium can not explain a high rate of specific complex formation.* One of possible explanation of specific distant interaction/attraction between the ligand and its receptor, is the exchange resonant EM interaction, proposed by DigiBio team.

The another possible explanation of the specific attraction between sterically and dynamically complementary macromolecules and molecules in water, like in system: [antibody + hapten] or [enzyme + substrate] is based on our Unified theory of Bivacuum, particles, duality and theory of Virtual Replicas of material objects (Kaivarainen, 2006; http://arxiv.org/abs/physics/0207027).

This new kind of Bivacuum mediated interaction between Sender [S] (ligand) and Receiver [R] (protein) can be a result of superposition of two their *virtual replicas*: $[\mathbf{VR}^S \rightleftharpoons \mathbf{VR}^R]$. The carrying frequency of Virtual Pressure Waves (VPW$^\pm$) is equal to basic frequency of [S] and [R] elementary particles [Corpuscle $\rightleftharpoons$ Wave] pulsation $(\omega_0 = m_0 c^2/\hbar)^i$, different for electrons and protons and responsible for their rest mass and charge origination. These basic frequencies are modulated by thermal vibrations of atoms and molecules. Just this modulation phenomena explains the nature of de Broglie waves of particles and provide a possibility of resonant distant interaction between molecules with close internal (quantum) and external (thermal) dynamics/frequencies via Virtual guides of spin, momentum and energy ($VirG_{S,M,E}$). A three-dimensional superposition of modulated by elementary particles of the object standing $VPW_m^\pm$ compose the internal and external Virtual Replicas ($VR_{in}$ and $VR_{ext}$). The Virtual Replicas (VR) has a properties of 3-dimensinal Virtual Quantum Holograms (Kaivarainen, 2006). The $VR = VR_{in} + VR_{ext}$ of atoms, molecules and macroscopic objects reflect:

a) the internal properties of the object ($VR_{in}$), including its dynamics, inhomogeneity, asymmetry, etc.;

b) the external surface properties and shape of macroscopic objects ($VR_{ext}$).

*The virtual replicas (VR) of the atoms and their coherent groups in state of mesoscopic Bose condensation as well as their superpositions may exist in the 'empty' Bivacuum and in any gas or condensed matter. It latter case the properties of matter can be slightly changed.*

Our Unified theory, including notion of VR, is in-line with Bohm and Pribram holographic paradigm and make it more detailed and concrete.

The proposed kind of Bivacuum mediated interaction (Kaivarainen, 2006) should be accompanied by the increase of dielectric permittivity between interacting molecules, decreasing the Van-der-Waals interactions between water molecules and enhancing the coefficient of diffusion in selected space between active site of protein and specific to this site ligand. The probability of cavitational fluctuations in water at room temperature with average frequency around $3 \times 10^4$ Hz (like revealed by DigiBio group), also should increase in the volume of $VR^S$ and $VR^R$ superposition, i.e. in the space between active site and the ligand.

The following three mechanisms of specific complex formation can be provided by:

a) the thermal fluctuations and diffusion in solution of protein and specific ligand;

b) the electromagnetic resonance exchange interaction between oscillating dipoles of protein active site and ligand (Benveniste hypothesis) and

c) the Bivacuum-mediated remote attraction between the ligand and active site, as a result of their Virtual replicas superposition (Kaivarainen, 2003-2007).

It is possible, that all three listed mechanisms of specific complex - formation are interrelated and enhance each other. The elucidation of the role of each of them in specific distant interaction/attraction between ligands and protein's active sites *in vitro* and *in vivo* - is a intriguing subject of future research.

### 5.4 Virtual replica of drugs in water and possible mechanism of 'homeopathic memory'

The memory of water, as a long relaxation time from nonequilibrium to equilibrium state, may have two explanations. One of them, for the case of magnetically treated water was suggested by this author in paper: "New Hierarchic Theory of Water & its Application to Analysis of Water Perturbations by Magnetic Field. Role of Water in Biosystems", placed to the arXiv: http://arxiv.org/abs/physics/0207114 and described in previous section.

The another explanation of water memory, described below, may explain the mechanism of imprinting of guest molecules properties in water after multiple dilution and shaking, used in homeopathy. The hypothesis of 'homeopathic memory' is based on two consequences of our Unified theory (UT), including theory of Virtual Replicas of the actual objects (Kaivarainen, 2006; http://arxiv.org/abs/physics/0207027):

1. It follows from UT, that between any actual object (**AO**), like guest molecule in water, and its virtual replica (**VR**), - the *direct* and *back* reaction is existing: (**AO** ⇌ **VR**). For example, when the drug molecule interact with binding site of receptor or substrate with enzyme both of reagents are already 'tuned' by their virtual replicas superposition. We remind, that **VR** of the guest molecules may exist in pure Bivacuum and in gas, liquid or solid phase of matter;

2. It follows from UT, that a stable virtual replica of the guest molecule, as a system of 3D standing virtual pressure waves ($\textbf{VPW}_m^\pm$): $VR = VR_{in} + VR_{ext}$ and its ability to infinitive spatial multiplication [**VRM(r, t)**], may exist after replacement of the object from the primary location to very remote position or even after its total destruction/disintegration.

If we consider a solution of any biologically active guest molecule in water (or in other liquid, in general case), then it follows from the above consequences of Unified theory, that the guest $\textbf{VR}_{guest}$ may retain its ability to affect the target via its superposition with virtual replica of the target $\textbf{VR}_{target}$ (i.e. the active site of cell's receptor, antibody, or enzyme). This can be a result of mentioned above back reaction of modulated virtual replica of target on the actual target (object): $[\textbf{VR}_{guest} \bowtie \textbf{VR}_{target}] \to \textbf{AO}$ even after super-dilution, when no one guest molecule (ligand) is no longer present in solution.

The liquid shaking (potentiation) after each step of homeopathic drug dilution is important in 'memorizing' of peculiar information about drug properties. However, the role of this procedure remains obscure.

Our approach suggest the following explanation of such 'imprinting' phenomena, accompanied the shaking or vigorous stirring, involving two stages:

**1**. Each *act of shaking* of water after dilution is accompanied by the collective motion of huge number of water in the test vessel and activation of the collective de Broglie wave of water molecules (modulation wave), participating in all 24 quantum excitations of water ($i = 24$) in accordance to our Hierarchic theory. The most probable de Broglie wave length of big number of water molecules, created in the moment of shaking or vigorous stirring can be expressed as:

$$\vec{\lambda}_s = \frac{h}{m_{H_2O} \vec{\textbf{v}}_{col}} \qquad 15.17$$

where: $m_{H_2O}$ is a mass of one water molecule; $\vec{\textbf{v}}$ is a velocity of collective motion of water molecules in stirred volume, determined by velocity and direction of stirring or shaking.

Similar formula is valid for de Broglie wave length of the solute/guest molecules in the process of shaking. However instead of mass of water molecules, the mass of guest molecule ($m_{guest}$) should be used.

As a result of shaking/stirring and creation of collective de Broglie wave of water, the degree of entanglement between coherent water clusters in state of mesoscopic Bose condensation (mBC) can be increased by two different mechanisms, depending on velocity of shaking or stirring ($\vec{\textbf{v}}_{col}$), in accordance to our theory of turbulence (section 12.2).

Theory predicts that at $30^0C$, when the most probable group velocity, related to water librations is $(\textbf{v}_{gr})_{lb} \simeq 2 \times 10^3 cm/s$, the critical flow velocity $\textbf{v}^1(r)$, necessary for *mechanical boiling* of water, corresponding to transition from the laminar flow to turbulent one (conditions 12.13 and 12.14) should be about $2.6 \times 10^3 cm/s = 26$ m/s.

At relatively low $\vec{\textbf{v}}_{col} < 10$ m/s, when the collective $\vec{\lambda}_s$ is bigger, than the average separation between primary librational effectons: about 55 Å at 25 $^0$C (see Fig. 51), this makes possible their unification on this base, representing partial transition from mesoscopic to macroscopic Bose condensation.

At higher velocity of shaking/stirring, when $\vec{\textbf{v}}_{col} \gtrsim 25$ m/s and the short-time turbulence originates in the whole volume of water under treatment, this means synhronization of conversions between primary librational and translational effectons or synhronized macro-convertons excitation in this volume. This process is accompanied by the enhancement of the exchange interaction between remote clusters in state of mBC by means of phonons and librational photons. This is followed by unification of mBC to fragile nonuniform macroscopic Bose condensation of all water volume in the vessel.

**2**. The water properties modulation by shaking or stirring, increasing the correlation between water clusters in state of mesoscopic BC. It makes the superposition of virtual replica of drag/guest molecule (**VR**$_{guest}$ - also related to its de Broglie wave length, atomic and elementary particles composition) with virtual replicas of coherent water molecules [$\sum \mathbf{VR}^i_{H_2O}$] more effective. This superposition represents the act of **VR**$_{guest}$ imprinting in the ordered fraction of water in the treated volume.

The process of imprinting is accompanied the guest molecule Virtual Replica spatial multiplication [**VRM**(**r**,**t**)] in new aqueous environment (Kaivarainen, 2006, http://arxiv.org/abs/physics/0207027, sections 13.4-13.5). We assume in our explanation of homeopathic memory, that the **VR**$_{guest}$ of guest/drug molecule retains some physical properties of this molecule itself. The proposed mechanism of homeopathic memorization of VR in space and time [**VRM**(**r**,**t**)] may explain the homeopathic drugs action in super-dilute solutions, below Avogadro number, prepared by successive shaking or stirring.

The mechanism proposed has already some experimental confirmation. Louis Rey (2003) in Switzerland, has published a paper in the mainstream journal, Physica A, describing the experiments that suggest water does have a memory of molecules that have been diluted away. With good reproducibility this fact was demonstrated by thermoluminescence method.

In this technique, the material is 'activated' by irradiation at low temperature, with UV, X-rays, electron beams, or other high-energy elementary particles. This causes electrons to come loose from the atoms and molecules, creating 'electron-hole pairs' that become separated and trapped at different energy levels. When the irradiated material is warmed up, it releases the absorbed energy and the trapped electrons and holes come together and recombine. This is accompanied by release of a characteristic glow of light, peaking at different temperatures depending on the magnitude of the separation between electron and hole.

As a general rule, the phenomenon is observed in crystals with an ordered arrangement of atoms and molecules, but it is also seen in disordered materials such as glasses. In this mechanism, imperfections in the atomic/molecular lattice are considered to be the sites at which luminescence appears.

Rey used this technique to investigate water, starting with heavy water or deuterium oxide that's been frozen into ice at a temperature of 77K.

As the ice warms up, a first peak (I) of luminescence appears near 120K, and a second peak (II) near 166 K. Heavy water gives a much stronger signal than regular water. In both cases, samples that were not irradiated gave no signals at all.

*It was shown, that the peak II comes from the hydrogen-bonded network within ice, whereas peak I comes from the individual molecules.* Rey then investigated what would happen when he dissolved some chemicals in the water and diluted it in steps of one hundred fold with vigorous stirring (as in the preparation of homeopathic remedies), until he reached a concentration of $10^{-30}$ g/cm$^3$ and compare that to the control that has not had any chemical dissolved in it *and diluted in the same way*. The samples were frozen and activated with irradiation as usual.

When lithium chloride (LiCl) was added, and then diluted away, the thermoluminescent glow became reduced, but the reduction of peak II was greater relative to peak I. Sodium chloride (NaCl) had the same effect albeit to a lesser degree. This is in accordance with our conjecture about important role of modulation of de Broglie waves of coherent water molecules by de Broglie wave of the guest particles, as far the molecular mass of $Li^+$ is almost 3 times less than that of $Na^+$. Consequently, the de Broglie wave length of the former is correspondingly bigger than that of latter at the same collective velocity ($\vec{v}_{col}$) of shaking:

$$\vec{\lambda}_{Li} = \frac{h}{m_{Li}\vec{v}_{col}} \sim 4\vec{\lambda}_{Na} = \frac{h}{m_{Na}\vec{v}_{col}}$$

It appears, therefore, that substances like LiCl and NaCl can modify the hydrogen-bonded network of water, and that this modification remains even when the molecules have been diluted away. The fact that this 'memory' remains because of vigorous stirring or shaking at successive dilutions, indicates that the 'memory' depends on a dynamic process, like introduced in our explanation collective de Broglie wave of water molecules or instant turbulence. Corresponding dynamization increases the interaction between water clusters in state of mesoscopic Bose condensate and their possible unification.

### 5.5. **Water and selective cancer cells destructor**

The hypothesis was proposed by this author (Kaivarainen, 1995; 2001), that one of the reasons of unlimited cancer cell division is related to partial disassembly of cytoskeleton's actin-like filaments and microtubules due to some genetically controlled mistakes in biosynthesis of ionic pumps, water channels and increasing the osmotic diffusion of water into transformed cell.

Decreasing of the intra-cell concentration of any types of ions ($Na^+$, $K^+$, $H^+$, $Mg^{2+}$ etc.), as a result of corresponding ionic pump malfunction, incorporated in biomembranes, also may lead to disassembly of filaments.

The shift of equilibrium: [assembly ⇔ disassembly] of microtubules (MTs) and actin filaments to the right increases the amount of intra-cell water, involved in hydration shells of protein and decreases water activity. As a consequence of concomitant osmotic process, cells tend to swell and acquire a ball-like shape. The number of direct contacts between transformed cells decrease and the water activity in the intercell space increases also.

Certain decline in the external inter-cell water activity, because of dense intercell contacts deterioration leads to the absence of triggering signal for inhibition of cells division/proliferation. The shape of normal cells under control of cell's filament is a specific one, providing good dense intercell contacts with limited amount of water in contacts, in contrast to situation with transformed cells. The activity of water in latter case is simply not low enough to stop the cells division.

In accordance to mechanism of cancer development proposed here, the *absence of contact inhibition* in the case of cancer cells, is a result of loose [cell-cell] contacts and loosing the interaction between cells because of disassembly of microtubules and actin filaments, connecting cells in normal conditions.

If our model of cancer emergency is correct, then the problem of limitless transformed cells proliferation, at least partly, is related to the problem of intercell water activity decreasing by incorporation, for example, of special intercell water soluble macro/poly-ions. The cells membranes should be nontransparent for such macroions.

Another approach to cancer healing, based on the mechanism described above, is the IR laser ($CO_2$) treatment of transformed cells with IR photons frequencies (about $3\times10^{13}$ s$^{-1}$), stimulating excitation of cavitational fluctuations in water (emergency and collapsing of microbubbles), inducing a collective disassembly of MTs, actin filaments and *gel → sol* transition. The corresponding centrioles disintegration will prevent cells division/proliferation and should give a good therapeutic effect. The IR laser based selective cancer cells destructor, proposed by this author, can be combined with ultrasound treatment of tissues and blood. Such treatment with frequency of about 40 kHz, in accordance to calculations, based on our Hierarchic theory of water, also should stimulate cavitational fluctuations in the swallowed cancer cells and their disintegration.

The method of cancer cells destruction proposed here, is based on the assumption that stability of MTs in transformed cells is weaker than that of normal cells fraction of 'free' water higher, than in normal cells. This difference should provide the selectivity of destructive action of the ultrasound and laser beam on the cancer and normal cells, remaining the normal cells undamaged.

Our Hierarchic theory of water (Kaivarainen, 1995, 2001; 2003) predicts besides the mentioned above frequency of ultrasound radiation around 40 kHz, the one more resonant to water fluctuations frequency around $10^7$ Hz. The combination of such kind of treatment of blood and organs should increase the selective destruction of cancer cells. For example, the carrying signal with frequency $10^7$ Hz can be modulated by frequency $4 \times 10^4$ Hz.

# 6. Elementary Act of Consciousness or the Quantum of Mind

Our approach to elementary act of consciousness has some common features with well-known Penrose - Hameroff model, interrelated act of consciousness with the wave function of microtubules collapsing. So we start from description of their Orchestrated objective reduction (Orch OR) model.

## 6.1 The Orchestrated objective reduction (Orch OR) model of Penrose and Hameroff

For biological qubits, Penrose and Hameroff chose open and closed clefts between the pair of tubulin subunits in microtubules. Tubulin qubits would interact and compute by entanglement with other tubulin qubits in microtubules in the same and different neurons.

It was known that the pair of alpha and beta tubulin subunits flexes 30 degrees, giving two different conformational shapes. Could such different states exist as superpositions ? The authors considered three possible types of tubulin superpositions: separation at the level of the entire protein, separation at the level of the atomic nuclei of the individual atoms within the proteins, and separation at the level of the protons and neutrons (nucleons) within the protein.

The calculated gravitational energy (E) at the level of atomic nuclei of tubulins had the highest energy, and would be the dominant factor in the wave function collapsing.

The best electrophysiological correlate of consciousness is gamma EEG, synchronized oscillations in the range of 30 to 90 Hz (also known as "coherent 40 Hz") mediated by dendritic membrane depolarizations (not axonal action potentials). This means that roughly 40 times per second (every 25 milliseconds) neuronal dendrites depolarize synchronously throughout wide regions of brain.

Using the indeterminacy principle $E = \hbar/t$ for OR, the authors take 25 ms for (t), and calculated (E) in terms of number of tubulins (since E was known for one tubulin). It can be asked: how many tubulins would be required to be in isolated superposition to reach OR threshold in $t = 25$ ms ? The answer turned out to be $2 \times 10^{11}$ tubulins.

Each brain neuron is estimated to contain about $10^7$ tubulins (Yu and Bass, 1994). If, say, 10 percent of each neuron's tubulins became coherent, then Orch OR of tubulins within roughly 20,000 (gap-junction connected) neurons would be required for a 25 ms conscious event, 5,000 neurons for a 100 ms event, or 1,000 neurons for a 500 ms event, etc.

These estimates (20,000 to 200,000 neurons) fit very well with others from more conventional approaches suggesting tens to hundreds of thousands of neurons are involved in consciousness at any one time.

How would microtubule quantum superpositions avoid environmental decoherence? Cell interiors are known to alternate between liquid phases (solution: "sol") and quasi-solid (gelatinous: "gel") phases due to polymerization states of the ubiquitous protein actin. In the actin-polymerized gel phase, cell water and ions are ordered on actin surfaces, so microtubules are embedded in a highly structured (i.e. non-random) medium. Tubulins are also known to have C termini "tails", negatively charged peptide sequences extending string-like from the tubulin body into the cytoplasm, attracting positive ions and forming a plasma-like Debye layer which can also shield microtubule quantum states. Finally, tubulins in microtubules were suggested to be coherently pumped laser-like into quantum states by biochemical energy (as proposed by H. Fröhlich).

Actin gelation cycling with 40 Hz events permits input to, and output from isolated microtubule quantum states. Thus during classical, liquid phases of actin depolymerization, inputs from membrane/synaptic inputs could "orchestrate" microtubule states. When actin gelation occurs, quantum isolation and computation ensues until OR threshold is reached, and actin depolymerizes. The result of each OR event (in terms of patterns of tubulin states) would proceed to organize neuronal activities including axonal firing and synaptic modulation/learning. Each OR event (e.g. 40 per second) is proposed to be a conscious event.

One implication of the Orch OR model is that consciousness is a sequence of discrete events, related to collapsing of general for these states wave function.

However, it is not clear in the above approach, what the mechanism is responsible for the coherence of all these remote proteins and their entanglement. Why they are unified by single wave function, like in case of macroscopic Bose condensation, i.e. superfluidity or superconductivity?

Our model, described below, has the answers to this crucial questions.

### 6.2 The basis of Hierarchic model of consciousness, including the distant and nonlocal interactions

In accordance to our Hierarchic model of elementary act of consciousness (HMC) or Quantum of Mind, each specific kind of neuronal ensembles excitation, accompanied by jump-way reorganization of big number of *dendrites and synaptic contacts* (elementary actual act of consciousness - Quantum of Mind) - corresponds to certain change of hierarchical system of three - dimensional (3D) standing waves of following kinds:

- thermal de Broglie waves (waves B), produced by anharmonic translations and librations of molecules;
- electromagnetic (IR) waves;
- acoustic waves;
- virtual pressure waves

The changes of de Broglie waves of atoms and molecules, participating in elementary act of consciousness modulate virtual pressure waves of Bivacuum ($\mathbf{VPW}^+ \bowtie \mathbf{VPW}^-$). These modulated standing virtual waves form the Virtual Replica (VR) of 'tuned' neuronal ensembles and especially [microtubules (MTs) + internal coherent water clusters] systems. The notion of VR of any material object was introduced by this author in Unified theory of Bivacuum, duality of particles, time and fields (see Appendix and Kaivarainen, 2006, 2007). It is subdivided on the surface and the volume Virtual Replicas.

The surface VR have the resemblance with regular optical hologram and contains information only about 3D shape of the object. The volume VR is a consequence of penetration of ($\mathbf{VPW}^+ \bowtie \mathbf{VPW}^-$) throw the object and scattering on its internal de Broglie waves of particles. Consequently the volume VR contains info about the dynamic interior of the object. This means that the total VR is much more informative than regular hologram and contain the whole information about the shape of the object and its internal structure.

The jump-way transition of the VR, as a result of elementary act of consciousness, representing gel-sol transition in cytoplasm of the neuron bodies and the tuned neurons ensembles in-phase pulsation and axonal firing, accompanied by redistribution of synaptic contacts between the starting and final states of dendrites, can be named: *virtual act of consciousness or Quantum of Mind*.

The *quantum act of consciousness*, corresponds to collapsing of unified wave function of big number of coherent water clusters (mBC) in microtubules from state of nonuniform macroscopic Bose condensation to mesoscopic one, representing less organized dissipative state.

The Penrose-Hameroff model considers only the coherent conformational transition of cavities between pairs of tubulins between open to closed states of big number of MTs as the act of wave function collapsing.

In our model this transition is only the triggering act, stimulating quantum transition of the big number of entangled water clusters in state of coherent macroscopic Bose condensation to decoherent state of mesoscopic Bose condensation (mBC). Consequently, the cycles of consciousness can be considered as a reversible transitions of certain part of brain between *coherence* and *decoherence*.

The nonuniform coherent state of macroscopic BC of water in microtubules system in time and space, e.g. periodically entangled, but spatially separated flickering clusters of mBC is different of continuous macroscopic BC, pertinent for superfluidity and superconductivity.

Consequently, the most important collective excitations providing the entanglement and quantum background of consciousness and can be the quantum integrity of the whole organism are coherent water clusters (primary librational effectons), representing mBC of water in microtubules.

Due to rigid core of MTs and stabilization of water librations (decreasing of most probable librational velocity $\mathbf{v}_{lb}$) the dimensions of mBC inside the MTs ($\lambda_B^{lb} = h/m\mathbf{v}_{lb}$) are bigger, than in bulk water and cytoplasm. The dimensions and stability of mBC is dependent on relative position of nonpolar cavities between $\alpha$ and $\beta$ tubulins, forming MTs and cavities dynamics.

In the open state of cavities the water clusters (mBC) are assembled and stable, making possible the macroscopic BC via entanglement in a big number of neurons MTs and in closed state of protein cavities the clusters are disassembled and macroscopic entanglement and nonuniform BC are destroyed.

The quantum beats between the ground - acoustic (a) and excited - optic (b) states of primary librational effectons (mBC) of water are accompanied by super-radiation of coherent librational IR photons and their absorption (see Introduction and Kaivarainen, 1992). The similar idea for water in microtubules was proposed later by Jibu at al. (1994, p.199).

The process of coherent IR photons radiation ⇌ absorption is interrelated with dynamic equilibrium between open (B) and closed (A) states of nonpolar clefts between α and β tubulins. These IR photons exchange interaction between 'tuned' systems of MTs stands for *distant* interaction between neurons in contrast to nonlocal one provided by conversion of mesoscopic BC to macroscopic BC.

The collective shift in geometry of nonpolar clefts/pockets equilibrium from the open to closed state is accompanied by the shrinkage of MTs is a result of turning of clusterphilic interaction to hydrophobic ones and *dissociation of water clusters* (see section 13.4.1). This process induce the disjoining of the MTs ends from the membranes of nerve cell bodies and gel → sol transition in cytoplasm, accompanied by disassembly of actin filaments.

Strong abrupt increasing of the actin monomers free surface and the fraction of water, involved in hydration shells of these proteins, decreases the internal water activity and initiate the water passive osmosis into the nerve cell from the external space. The cell swallows and its volume increases. Corresponding change of cell's body volume and shape of dendrites is followed by synaptic contacts reorganization. This is a final stage of multistage act of consciousness.

The destabilization of water clusters and conversion of clusterphilic interaction to hydrophobic one can be a consequence of such known quantum optic phenomena, as *bistability*. It represents the water polarization change as a result of $a \rightleftharpoons b$ equilibrium shift in librational primary effectons to the right. In turn, this is a consequence of librational IR photons pumping and the excited b-state of librational effectons saturation.

The related to above phenomena: the *self-induced transparency* is a consequence of the light absorption saturation by primary librational effectons (Andreev, et al., 1988). This saturation can be followed by the pike regime (light emission pulsation, after $b$-state saturation of librational effectons and subsequent super-radiation of big number of entangled water clusters in state of mBC in the process of their correlated $\sum (\mathbf{b} \to \mathbf{a})$ transitions.

The entanglement between coherent nucleons of opposite spins of H and O of remote water clusters in a big number of MTs, in accordance to our theory of nonlocality, can be realized via bundles of Bivacuum virtual guides $\mathbf{VirG}_{SME}$ of spin, momentum and energy (Kaivarainen, 2006 a,b).

*6.2.1 The mechanism of the Entanglement channels formation between remote coherent de Broglie waves of the nucleons*

The bundles of $\mathbf{VirG}_{SME}$, connecting pairs of protons and neutrons of opposite spins of remote coherent molecules in state of mesoscopic Bose condensation (mBC) were named the *Entanglement channels* (Kaivarainen, 2006, 2007):

$$\textbf{Entanglement channel} = \left[ \mathbf{N(t,r)} \times \sum_{}^{\mathbf{n}} \mathbf{VirG}_{SME}(\mathbf{S} <=> \mathbf{R}) \right]^{i}_{x,y,z} \quad 6.1$$

where: ($\mathbf{n}$) is a number of pairs of similar tuned elementary particles (protons, neutrons and electrons) of opposite spins of the remote entangled atoms and molecules; $N(t,r)$ is a number of coherent atoms/molecules in the entangled molecular (e.g. water) clusters in state of mBC.

The Virtual Guides (microtubules), connecting the remote elementary particles have a properties of quasi- one- dimensional virtual Bose condensate.

A *single* Virtual Guides of spin, momentum and energy (see Fig. 50) are assembled from 'head-to-tail' polymerized Bivacuum bosons of opposite polarization: $\mathbf{BVB}^{+} = [\mathbf{V}^{+} \uparrow\downarrow \mathbf{V}^{-}]$ and $\mathbf{BVB}^{-} = [\mathbf{V}^{+} \downarrow\uparrow \mathbf{V}^{-}]$:

$$Single \; \mathbf{VirG}_{SME}^{\mathbf{BVB}^{+}} = \mathbf{D(r,t)} \times \mathbf{BVB}^{+}; \qquad \mathbf{VirG}_{SME}^{\mathbf{BVB}^{-}} = \mathbf{D(r,t)} \times \mathbf{BVB}^{-} \quad 6.2$$

A *double* Virtual Guides are composed from Cooper pairs of Bivacuum fermions of opposite spins ($\mathbf{BVF}^{\uparrow} \bowtie \mathbf{BVF}^{\downarrow}$):

$$Double \; \mathbf{VirG}_{SME}^{\mathbf{BVF}^{\uparrow} \bowtie \mathbf{BVF}^{\downarrow}} = \mathbf{D(r,t)} \times [\mathbf{BVF}^{\uparrow}_{+} \bowtie \mathbf{BVF}^{\downarrow}_{-}]^{s}_{S=0} \quad 6.3$$

where: $\mathbf{D(r,t)}$ is a number of Bivacuum dipoles in Virtual guides, dependent on the distance (**r**) between remote but tuned de Broglie waves of elementary particles of opposite spins. The diameter of these dipoles and spatial gap between their torus and antitorus (see Appendix) are pulsing in-phase.

Just the *Entanglement channels* are responsible for nonlocal Bivacuum mediated interaction between the mesoscopic BC, turning them to macroscopic BC. For the entanglement channels activation the interacting mBC systems should be in nonequilibrium state.

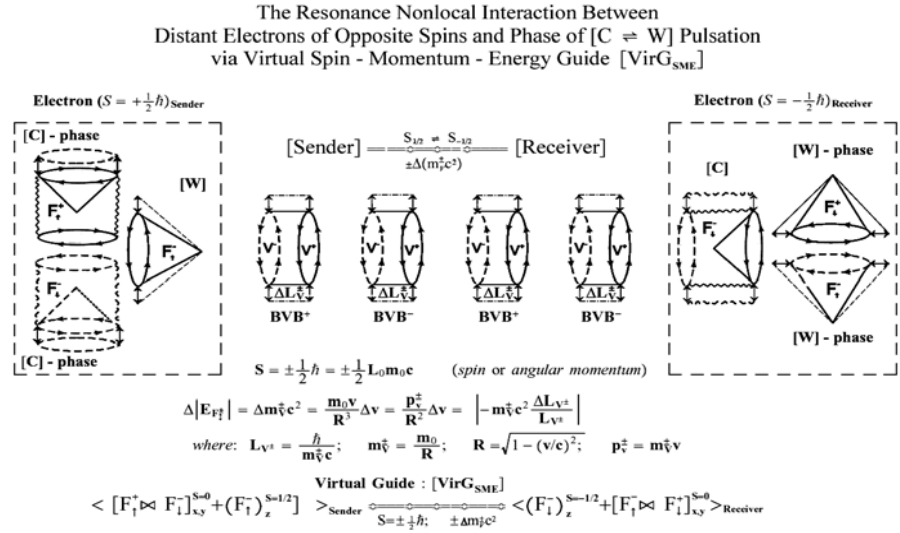

**Figure 6.1**. The mechanism of nonlocal Bivacuum mediated interaction (entanglement) between two distant unpaired sub-elementary fermions of 'tuned' elementary triplets (particles) of the opposite spins $< [\mathbf{F}_\uparrow^+ \bowtie \mathbf{F}_\downarrow^-] + \mathbf{F}_\uparrow^- >_{Sender}^i$ and $< [\mathbf{F}_\downarrow^+ \bowtie \mathbf{F}_\uparrow^-] + \mathbf{F}_\downarrow^- >_{Receiver}^i$, with close frequency of $[\mathbf{C} \rightleftharpoons \mathbf{W}]$ pulsation and close de Broglie wave length ($\lambda_B = h/m_V^+ v$) of particles. The tunnelling of momentum and energy increments: $\Delta|m_V^\pm c^2| \sim \Delta|\mathbf{VirP}^+| \pm \Delta|\mathbf{VirP}^-|$ from Sender to Receiver and vice-verse via Virtual spin-momentum-energy Guide [$\mathbf{VirG}_{SME}^i$] is accompanied by instantaneous pulsation of diameter ($2\Delta L_V^\pm$) of this virtual guide, formed by Bivacuum bosons $\mathbf{BVB}^\pm$ or double microtubule, formed by Cooper pairs of Bivacuum fermions: $[\mathbf{BVF}^\uparrow \bowtie \mathbf{BVF}^\downarrow]$. The nonlocal spin state exchange between [S] and [R] can be induced by the change of polarization of Cooper pairs: $[\mathbf{BVF}^\uparrow \bowtie \mathbf{BVF}^\downarrow] \rightleftharpoons [\mathbf{BVF}^\downarrow \bowtie \mathbf{BVF}^\uparrow]$ and Bivacuum bosons: $\mathbf{BVB}^+ \rightleftharpoons \mathbf{BVB}^-$, composing the double or single $\mathbf{VirG}_{SME}(\mathbf{S} <=> \mathbf{R})^i$, correspondingly (Kaivarainen, 2005; 2006 and 2007).

The assembly of huge number of bundles of virtual microtubules of Bivacuum, like Virtual Channels side-by-side can compose virtual multilayer membranes. Each of this layer, pulsing in counterphase with the next one between the excited and ground states are interacting with each other via dynamic exchange by pairs of virtual pressure waves $[\mathbf{VPW}^+ \bowtie \mathbf{VPW}^-]$. This process occur without violation of the energy conservation law and is accompanied by nonlocal Bivacuum gap oscillation over the space of virtual BC of Bivacuum dipoles. The value of spatial gap between the actual and complementary torus and antitorus of Bivacuum fermions is dependent on their excitation state quantum number (**n** = **0, 1, 2, 3**...):

$$[\mathbf{d}_{V^+ \Updownarrow V^-}]_n = \frac{h}{\mathbf{m}_0 \mathbf{c}(1 + 2\mathbf{n})} \qquad 6.4$$

The Bivacuum gap oscillations can be responsible for the lateral or transversal nonlocality of Bivacuum in contrast to longitudinal one, connecting the nucleons with opposite spin, realized via **VirG**$_{SME}$ (Kaivarainen, 2006a, b).

The gel-sol transition in the number of entangled neurons is accompanied by decreasing of viscosity of cytoplasm. The tuned - parallel orientation of MTs in tuned remote cells change as a result of Brownian motion, accompanied by decoherence and loosing the entanglement between water clusters in MTs.

This is followed by relaxation of the internal water + microtubulins to normal dynamics and grows of (+) ends of MTs up to new contacts formation with cells membranes, stabilizing cells dendrites new geometry and synaptic contacts distribution. *This new configuration and state of the nerve system and brain, represents the transition of the Virtual Replica of the brain from the former state to the new one.*

We assume in our model the existence of back reaction between the properties of Virtual Replica of the nerve system of living organisms, with individual properties, generated by systems:

$$[microtubules\ of\ neurons + DNA\ of\ chromosomes] \qquad 6.5$$

and the actual object - the organism itself. The corresponding subsystems can be entangled with each other by the described above Virtual Channels.

The interference of such individual (self) virtual replica VR{self} with virtual replicas of other organisms and inorganic macroscopic system may modulate the properties of VR{self}.

Because of back reaction of VR{self} on corresponding organism, the interaction of this organism with resulting/Global virtual replica of the external macroscopic world may be realized.

The twisting of centrioles in cells to parallel orientation, corresponding to maximum energy of the MTs interaction of remote neurons, is a first stage of the *next* elementary act of consciousness.

The superradiated photons from enlarged in MTs water clusters have a higher frequency than $(\mathbf{a} \rightleftharpoons \mathbf{b})_{lb}$ transitions of water primary librational effectons of cytoplasmic and inter-cell water. This feature provides the regular *transparency* of medium between 'tuned' microtubules of remote cells for librational photons.

The [gel→sol] transitions in cells is interrelated with tuned nerve cells (ensembles) coherent excitation, their membranes depolarization and the *axonal firing*.

### 6.2.2 Two triggering mechanisms of elementary act of consciousness

It is possible in some cases, that the excitation/depolarization of the nerve cells by the external factors (sound, vision, smell, tactical feeling) are triggering - *primary* events and [gel→sol] transitions in nerve cells are the *secondary* events.

However, the opposite mechanism, when the tuning of remote cells and [gel → sol] transitions are the *primary* events, for example, as a result of thinking/meditation and the nerve cells depolarization of cells are *secondary* events, is possible also.

The 1st mechanism, describing the case, when depolarization of nerve membranes due to external factors is a *primary* event and *gel → sol* transition a *secondary* one, includes the following stages of elementary act of consciousness:

a) simultaneous depolarization of big enough number of neurons, forming ensemble, accompanied by opening the potential-dependent channels and increasing the concentration of $Ca^{2+}$ in cytoplasm of neurons body;

b) collective disassembly of actin filaments, accompanied by [gel → sol] transition of

big group of depolarized neurons stimulated by $Ca^{2+}$ − activated proteins like gelsolin and villin. Before depolarization the concentration of $Ca^{2+}$ outside of cell is about $10^{-3} M$ and inside about $10^{-7} M$. Such strong gradient provide fast increasing of these ions concentration in cell till $10^{-5} M$ after depolarization.

c) strong decreasing of cytoplasm viscosity and disjoining of the (+) ends of MTs from membranes, making possible the spatial fluctuations of MTs orientations inducing decoherence switching off the entanglement between mBC;

d) volume/shape pulsation of neuron's body and dendrites, inducing reorganization of ionic channels activity and synaptic contacts in the excited neuron ensembles. These volume/shape pulsations occur due to reversible decrease of the intra-cell water activity and corresponding swallow of cell as a result of increasing of passive osmotic diffusion of water from the external space into the cell.

In the opposite case, accompanied process of braining, the depolarization of nerve membranes, axonal firing is a *secondary* event and *gel* → *sol* transition a *primary* one, stimulated in turn by simultaneous dissociation of big number of water clusters to independent molecules. The latter process represents the conversion of primary librational effectons to translational ones, following from our theory (see 'convertons' in the Introduction).

The frequency of electromagnetic field, related to change of ionic flux in excitable tissues usually does not exceed $10^3$ Hz (Kneppo and Titomir, 1989).

The electrical recording of human brain activity demonstrate a coherent (40 to 70 Hz) firing among widely distributed and distant brain neurons (Singer, 1993). Such synchronization in a big population of groups of cells points to possibility of not only the regular axon-mediated interaction, but also to fields-mediated interaction and quantum entanglement between remote neurons bodies.

The dynamic virtual replicas (VR) of all hierarchical sub-systems of brain and its space-time multiplication VRM(r,t) contain information about all kind of processes in condensed matter on the level of coherent elementary particles (Kaivarainen, 2006 a,b). Consequently, our model agrees in some points with ideas of Karl Pribram (1977), David Bohm and Basil Hiley (1993) of holographic mind, incorporated in the hologram of the Universe.

### 6.3 The comparison of Hierarchic model of consciousness and Quantum brain dynamics model

Our approach to Quantum Mind problem has some common features with model of Quantum Brain Dynamics (QBD), proposed by L.Riccardi and H.Umezawa in 1967 and developed by C.I.Stuart, Y.Takahashi, H.Umezava (1978, 1979), M.Jibu and K.Yasue (1992, 1995).

In addition to traditional electrical and chemical factors in the nerve tissue function, this group introduced two new types of *quantum* excitations (ingredients), responsible for the overall control of electrical and chemical signal transfer: *corticons and exchange bosons* (dipolar phonons).

The *corticons* has a definite spatial localization and can be described by Pauli spin matrices. The *exchange bosons*, like phonons are delocalized and follow Bose-Einstein statistics. "By absorbing and emitting bosons coherently, corticons manifest global collective dynamics…, providing systematized brain functioning" (Jibu and Yasue,1993). In other paper (1992) these authors gave more concrete definitions:

"*Corticons* are nothing but quanta of the molecular vibrational field of biomolecules (quanta of electric polarization, confined in protein filaments). *Exchange bosons* are nothing but quanta of the vibrational field of water molecules…".

It is easy to find analogy between spatially localized "corticons" and our primary effectons as well as between "exchange bosons" and our secondary (acoustic) deformons. It is evident also, that our Hierarchic theory is more developed. It is based on detailed description of all possible collective excitations, making possible the quantitative analysis of any condensed matter, including water and biological systems.

Jibu, Yasue, Hagan and others (1994) discussed a possible role of quantum optical coherence in cytoskeleton microtubules: implications for brain function. They considered MTs as a *wave guides* for coherent superradiation. They also supposed that coherent photons, penetrating in MTs, lead to *"self-induced transparency"*. Both of these phenomena are well known in fiber and quantum optics. We also use these phenomena in our model for one of explanation of transition from mesoscopic entanglement between water clusters in MTs to macroscopic one, as a result of IR photons exchange between coherent clusters, representing mesoscopic Bose condensate (mBC). However, we have to note, that the unification of mBC to macroscopic BC in 'tuned' MTs is possible even without self-induced transparency.

It follows also from our approach that the mechanism of macroscopic BC of water clusters do not need the hypothesis of Frölich that the proteins (tubulins of MTs) can be coherently pumped into macroscopic quantum states by biochemical energy.

We also do not use the idea of Jibu et al. that the MTs works like the photons wave - guides without possibility of side radiation throw the walls of MTs, increasing the probability of macroscopic entanglement between remote cells of the organism.

### 6.4 The Properties of the Actin Filaments, Microtubules and Internal Water

There are six main kind of actin existing. Most general F-actin is a polymer, constructed from globular protein G-actin with molecular mass 41800. Each G-actin subunit is stabilized by one ion $Ca^{2+}$ and is in noncovalent complex with one ATP molecule. Polymerization of G-actin is accompanied by splitting of the last phosphate group. The velocity of F-actin polymerization is enhanced strongly by hydrolysis of ATP. However, polymerization itself do not needs energy. Simple increasing of salt concentration (decreasing of water activity), approximately till to physiological one - induce polymerization and strong increasing of viscosity.

The actin filaments are composed from two chains of G-actin with diameter of 40 Å and forming double helix. The actin filaments are the polar structure with different properties of two ends.

*Let us consider the properties of microtubules (MT) as one of the most important component of cytoskeleton, responsible for spatial organization and dynamic behavior of the cells.*

The [assembly ⇔ disassembly] equilibrium of microtubules composed of $\alpha$ and $\beta$ tubulins is strongly dependent on internal and external water activity (*a*), concentration of $Ca^{2+}$ and on the electric field gradient change due to MTs piezoelectric properties.

The $\alpha$ and $\beta$ tubulins are globular proteins with equal molecular mass (*MM* = 55.000), usually forming $\alpha\beta$ dimers with linear dimension 8 nm. Polymerization of microtubules can be stimulated by NaCl, $Mg^{2+}$ and GTP (1:1 tubulin monomer) (Alberts *et al.*, 1983). The presence of heavy water (deuterium oxide) also stimulates polymerization of MTs.

In contrast to that the presence of ions of $Ca^{2+}$ even in micromolar concentrations, action of colhicine and lowering the temperature till $4^0C$ induce disassembly of MT.

Due to multigenic composition, $\alpha$ and $\beta$ tubulins have a number of isoforms. For example, two-dimensional gel electrophoresis revealed 17 varieties of $\beta$ tubulin in mammalian brain (Lee *et al.*, 1986). Tubulin structure may also be altered by enzymatic modification: addition or removal of amino acids, glycosylation, etc.

*Microtubules* are hollow cylinders, filled with water. Their internal diameter about $d_{in}$ =140Å and external diameter $d_{ext}$ = 280 Å (Figure 6.2). These data, including the dimensions of $\alpha\beta$ dimers were obtained from x-ray crystallography (Amos and Klug, 1974). However we must keep in mind that under the conditions of crystallization the multiglobular proteins and their assemblies tends to more compact structure than in solutions due to lower water activity. This means that in natural conditions the above dimensions could be a bit bigger.

The length of microtubules (MT) can vary in the interval:

$$l_t = (1 - 20) \times 10^5 Å \qquad 6.6$$

The spacing between the tubulin monomers in MT is about 40 Å and that between $\alpha\beta$ dimers: 80 Å are the same in longitudinal and transversal directions of MT.

Microtubules sometimes can be as long as axons of nerve cells, *i.e.* tenth of centimeters long. Microtubules (MT) in axons are usually parallel and are arranged in bundles. Microtubules associated proteins (MAP) form a "bridges", linking MT and are responsible for their interaction and cooperative system formation. Brain contains a big amount of microtubules. *Their most probable length is about $10^5 Å$, i.e. close to librational photon wave length.*

The viscosity of ordered water in such narrow microtubules seems to be too high for transport of ions or metabolites at normal conditions.

All 24 types of quasi-particles, introduced in the Hierarchic Theory of matter (Table 1), also can be pertinent for ordered water in the microtubules (MT). However, the dynamic equilibrium between populations of different quasi-particles of water in MT must be shifted towards primary librational effectons, comparing to bulk water due to increased clusterphilic interactions (see section 13.4.1 of this book). The dimensions of internal primary librational effectons have to be bigger than in bulk water as a consequence of stabilization of MT walls the mobility of water molecules, increasing their most probable de Broglie wave length.

The interrelation must exist between properties of internal water in MT and structure and dynamics of their walls, depending on [$\alpha - \beta$] tubulins interaction. Especially important can be a quantum transitions like convertons [$tr \Leftrightarrow lb$]. The convertons in are accompanied by [dissociation/association] of primary librational effectons, i.e. flickering of coherent water clusters, followed by the change of angle between $\alpha$ and $\beta$ subunits in tubulin dimers.

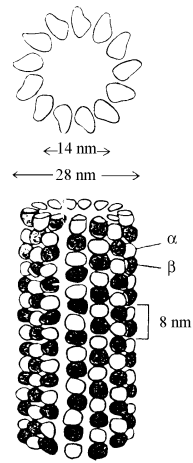

53**Figure 6**.**2**. Construction of microtubule from *α* and *β* tubulins, globular proteins with molecular mass 55 kD, existing in form of dimers (*αβ*). Each *αβ* dimer is a dipole with negative charges, shifted towards *α* subunit (De Brabander, 1982). Consequently, microtubules, as an oriented elongated structure of dipoles system, have the piezoelectric properties (Athestaedt, 1974; Mascarennas, 1974).

Intra-microtubular *clusterphilic interactions* stimulate the growth of tubules from *αβ* tubulin dimers. The structural physical-chemical asymmetry of *αβ* dimers composing microtubules determines their different rates of growth from the opposite ends ([+] and [-]).

The equilibrium of "closed" (A) and "open"(B) states of nonpolar cavities between *α* and *β* tubulins in (*αβ*) dimers can be shifted to the (B) one under the change of external electric field in a course of membrane depolarization. It is a consequence of piezoelectric properties of MTs and stimulate the formation of coherent water clusters in the open cavities of (*αβ*) dimers. The open cavities serve as a centers of water cluster formation and molecular Bose condensation.

The parallel orientation of MT in different cells, optimal for maximum [MT-MT] resonance interaction could be achieved due to twisting of centrioles, changing spatial orientation of MT. However, it looks that the normal orientation of MT as respect to each other corresponds to the most stable condition, *i.e.* minimum of potential energy of interaction (see Albreht-Buehner, 1990).

It is important to stress here that the orientation of two centrioles as a source of MT bundles in each cell are always normal to each other.

The linear dimensions of the primary librational effectons edge ($l_{ef}^{lb}$) in pure water at physiological temperature ($36^0 C$) is about 11 Å and in the ice at $0^0 C$ it is equal to 45 Å.

We assume that in the rigid internal core of MT, the linear dimension (edge length) of librational effecton, approximated by cube is between 11Å and 45 Å *i.e.* about $l_{ef}^{lb} \sim 23$Å. It will be shown below, that this assumption fits the spatial and symmetry properties of MT very well.

The most probable group velocity of water molecules forming primary *lb* effectons is:

$$\mathbf{v}_{gr}^{lb} \sim h/(m_{H_2O} \times l_{ef}^{lb}) \qquad 6.7$$

The librational mobility of internal water molecules in MT, which determines ($\mathbf{v}_{gr}^{lb}$) should be about 2 times less than in bulk water at $37^0 C$, if we assume for water in microtubules: $l_{ef}^{lb} \sim 23$Å.

Results of our computer simulations for pure bulk water shows, that the distance between centers of primary [lb] effectons, approximated by cube exceed their linear dimension to about 3.5 times (Fig 51 b). For our case it means that the average distance between the effectons centers is about:

$$d = l_{ef}^{lb} \times 3.5 = 23 \times 3.5 \sim 80 \text{Å} \qquad 6.8$$

This result of our theory points to the equidistant (80 Å) localization of the primary *lb* effectons in clefts between *α* and *β* tubulins of each (*αβ*) dimer in the internal core of MTs.

In the case, if the dimensions of librational effectons in MTs are quite the same as in bulk water, i.e. 11 Å, the separation between them should be:
$d = l_{ef}^{lb} \times 3.5 = 11 \times 3.5 \sim 40$ Å.

This result points that the coherent water clusters can naturally exist not only between *α* and *β* subunits of each pair, but also between pairs of (*αβ*) dimers.

In the both cases the spatial distribution symmetry of the internal flickering clusters in MT may serve as an important factor for realization of the signal propagation along the MT (conformational wave), accompanied by alternating process of closing and opening the clefts between neighboring $\alpha$ and $\beta$ tubulins pairs.

This large-scale protein dynamics is regulated by dissociation $\rightleftharpoons$ association of water clusters in the clefts between ($\alpha\beta$) dimers of MT (Fig.6.2) due to [$lb/tr$] convertons excitation and librational photons and phonons exchange between primary and secondary effectons, correspondingly.

The dynamic equilibrium between $tr$ and $lb$ types of the intra MT water effectons must to be very sensitive to $\alpha - \beta$ tubulins interactions, dependent on nerve cells excitation and their membranes polarization.

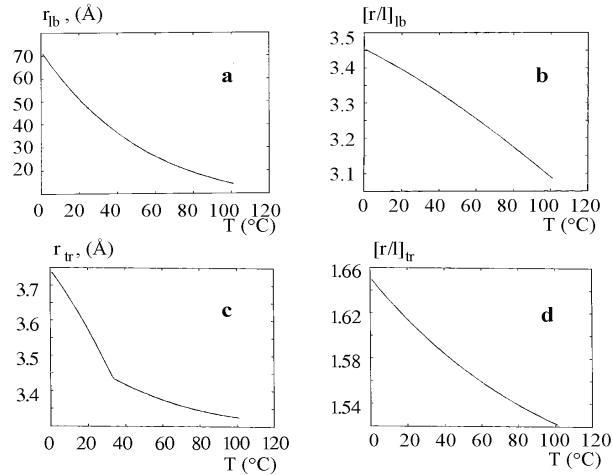

**Figure 6.3**. Theoretical temperature dependencies of:
   (a) - the space between centers of primary [lb] effectons (calculated in accordance to eq.4.62);
   (b) - the ratio of space between primary [lb] effectons to their length (calculated, using eq.4.63);
   (c) - the space between centers of primary [tr] effectons (in accordance to eq.4.62);
   (d) - the ratio of space between primary [tr] effectons to their length (eq.4.63).

**Two statements of our Hierarchic model of consciousness are important**:
   1. The ability of intra-MT primary water effectons (tr and lb) for superradiation of six coherent IR photons from each of the effectons side, approximated by parallelepiped:
   two identical - "longitudinal" IR photons, penetrating along the core of microtubule, forming the longitudinal standing waves inside it and two pairs of identical - "transverse" IR photons, also responsible for the distant, nonlocal interaction between microtubules. *In accordance to superradiation mechanism the intensity of longitudinal radiation of MTs is much bigger than that of transverse one;*
   2. The parameters of the water clusters radiation (frequency of librational photons, coherency, intensity) are regulated by the interaction of the internal water with MT walls, dependent on the [open $\Leftrightarrow$ closed] states equilibrium of cavity between $\alpha$ and $\beta$ tubulins.

### 6.5 The system of librational and translational IR standing waves in the microtubules

We found out that the average length of microtubules (*l*) correlates with length of standing electromagnetic waves of librational and translational IR photons, radiated by corresponding primary effectons:

$$l_{lb} = \kappa \frac{\lambda_p^{lb}}{2} = \frac{\kappa}{2n\tilde{\nu}_p^{lb}} \qquad 6.9$$

and

$$l_{tr} = \kappa \frac{\lambda_p^{tr}}{2} = \frac{\kappa}{2n\tilde{\nu}_p^{tr}} \qquad 6.10$$

here $\kappa$ is the integer number; $\lambda_p^{lb,tr}$ is a librational or translational IR photon wave length equal to:

$$\lambda_p^{lb} = (n\tilde{\nu}_p^{lb})^{-1} \simeq 10^5 \text{Å} = 10\mu \qquad 6.11$$

$$\lambda_p^{tr} = (n\tilde{\nu}_p^{tr})^{-1} \simeq 3.5 \times 10^5 \text{Å} = 30\mu \qquad 6.11a$$

where: $n \simeq 1.33$ is an approximate refraction index of water in the microtubule; $\tilde{\nu}_p^{lb} \simeq (700 - 750)\, cm^{-1}$ is wave number of librational photons and $\tilde{\nu}_p^{tr} \simeq (200 - 180)\, cm^{-1}$ is wave number of translational photons.

It is important that the most probable length of MTs in normal cells is about $10\mu$ indeed. So, just the librational photons and the corresponding primary effectons play the crucial role in the entanglement inside the microtubules and between MTs. The necessary for this quantum phenomena tuning of molecular dynamics of water is provided by the electromagnetic interaction between separated coherent water clusters in state of mBC.

### 6.6 The role of electromagnetic waves in the nerve cells

In the normal animal-cells, microtubules grow from pair of centriole in center to the cell's periphery. In the center of plant-cells the centrioles are absent. Two centrioles in cells of animals are always oriented at the right angle with respect to each other. The centrioles represent a construction of 9 triplets of microtubules (Fig. 52), i.e. two centriole are a source of: $(2 \times 27 = 54)$ microtubules. The centriole length is about $3000\,\text{Å}$ and its diameter is $1000\,\text{Å}$.

These dimensions mean that all 27 microtubules of each centrioles can be orchestrated in the volume ($\mathbf{v}_d$) of one translational or librational electromagnetic deformon:

$$\left[\mathbf{v}_d = \frac{9}{4\pi}\lambda_p^3\right]_{tr,lb} \qquad 6.12$$

where: $(\lambda_p)_{lb} \sim 10^5 \text{Å}$ and $(\lambda_p)_{tr} \sim 3.5 \times 10^5 \text{Å}$

Two centrioles with normal orientation as respect to each other and a lot of microtubules, growing from them, contain the internal orchestrated system of librational water effectons. It represent a quantum system with correlated $(a \rightleftharpoons b)_{lb}^{1,2,3}$ transitions of the effectons. The resonance superradiation or absorption of a *number* of librational photons ($3q$) in the process of above transitions, is dependent on the number of primary *lb* effectons ($q$) in the internal hollow core of a microtubule:

$$q = \frac{\pi L_{MT}^2 \times l}{V_{ef}^{lb}} \qquad 6.13$$

The value of $q$ - determines the intensity (amplitude) of coherent longitudinal librational IR photons radiation from microtubule with internal radius $L_{MT} = 7nm$ and length ($l$), for the case, when condition of standing waves (6.9) is violated.

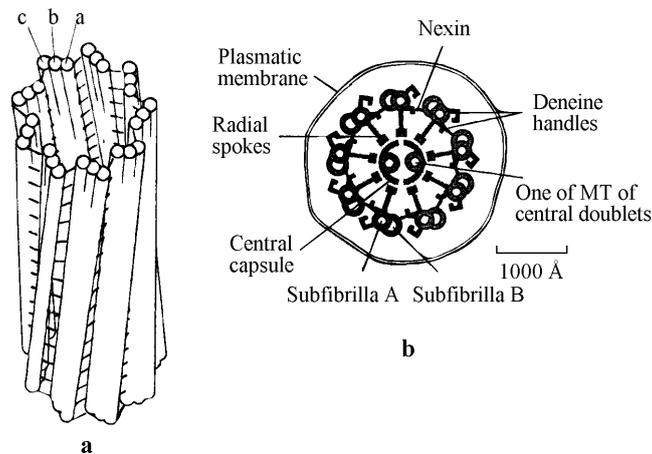

**Figure 6.4** (*a*) : the scheme of centriole construction from nine triplets of microtubules. The length and diameter of cylinder are 3000 Å and 1000 Å, correspondingly. Each of triplets contain one complete microtubule and two noncomplete MT;
(b): the scheme of cross-section of cilia with number of MT doublets and MT-associated proteins (MAP): [2 × 9 + 2] = 20. One of MT of periphery doublets is complete and another is noncomplete (subfibrilles A and B).

It is important that the probabilities of pair of longitudinal and two pairs of *transversal* photons, emission as a result of superradiance by primary librational effectons are equal, being the consequence of the same collective $(b \to a)_{lb}$ transition. These probabilities can be "tuned" by the electric component of electromagnetic signals, accompanied axon polarization and nerve cell excitation due to piezoelectric properties of MT.

Coherent *longitudinal* emission of IR photons from the ends of each *pair* of microtubules of two perpendicular centrioles of the *same cell* and from ends of *one* microtubule of *other cell* can form a 3*D* superposition of standing photons (primary deformons) as a result of 3 photons pairs interception.

The system of such longitudinal electromagnetic deformons, as well as those formed by transversal photons, have a properties of pilotless 3D hologram. Such an electromagnetic hologram can be responsible for the following physico-chemical phenomena:

 -Nonmonotonic distribution of intra-cell water viscosity and diffusion processes in cytoplasm due to corresponding nonmonotonic spatial distribution of macro-deformons;

 -Regulation of spatial distribution of water activity ($a_{H_2O}$) in cytoplasm as a result of corresponding distribution of inorganic ions (especially bivalent such as $Ca^{2+}$) in the field of standing electromagnetic waves. Concentration of ions in the nodes of standing waves should be higher than that between them. Water activity ($a_{H_2O}$) varies in the opposite manner than ions concentration.

The spatial variation of ($a_{H_2O}$) means the modulation of [assembly ⇔ disassembly] equilibrium of filaments of the actin and partly MTs. As a consequence, the volume and shape of cell compartments will be modulated also. The activity of numerous oligomeric allosteric enzymes can be regulated by the water activity also.

The following properties of microtubules can affect the properties of 3D standing waves, radiated by them:
 a) total number of microtubules in the cell;
 b) spatial distribution of microtubules in the volume of cytoplasm;
 c) distribution of microtubules by their length.

The constant of ($a \Leftrightarrow b$) equilibrium of primary librational effectons

$$(K_{a \Leftrightarrow b})_{lb} = \exp[-(E_a - E_b)/kT]_{lb} \qquad 6.14$$

and that of ($A^* \Leftrightarrow B^*$) equilibrium of super-effectons are dependent on the structure and dynamics of $\alpha\beta$ tubulin pairs forming MT walls.

This equilibrium is interrelated, in turn, with librational photons frequency $(\nu_{lb})^{1,2,3}$:

$$[\nu_{lb} = c(\tilde{\nu})_{lb} = (V_b - V_a)_{lb}/h]^{1,2,3} \qquad 6.15$$

which is determined by the difference of potential and total energies between (b) and (a) states of primary effectons in the hollow core of microtubules, as far the kinetic energies of these states are equal $T_b = T_a$:

$$[V_b - V_a = E_b - E_a]_{lb}^{1,2,3} \qquad 6.16$$

$(\tilde{\nu})_{lb}^{1,2,3}$ is the librational band wave number.

The refraction index ($n$) and dielectric constant of the internal water in MT depends on $[a \Leftrightarrow b]$ equilibrium of the effectons because the polarizability of water and their interaction in (a) state are higher, than that in (b) state.

### 6.7 The reactions accompanied nerve excitation

The normal nerve cell contains few dendrites, increasing the surface of cell's body. It is enable to form synaptic contacts for reception the information from thousands of other cells. Each neuron has one axon for transmitting the "resulting" signal in form of the electric impulses from the ends of axons of cells-transmitters to neuron-receptor.

The synaptic contacts, representing narrow gaps (about hundreds of angstrom wide) could be subdivided on two kinds: the *electric* and *chemical* ones. In chemical synapsis the signal from the end of axon - is transmitted by *neuromediator, i.e. acetylholine.* The neuromediator molecules are stored in *synaptic bubbles* near *presynaptic membrane.* The releasing of mediators is stimulated by ions of $Ca^{2+}$. After diffusion throw the synaptic gap mediator form a specific complexes with receptors of post synaptic membranes on the surface of neurons body or its dendrites. Often the receptors are the ionic channels like $(Na^+, K^+)$ - ATP pump. Complex - formation of different channels with mediators opens them for one kind of ions and close for the other. Two kind of mediators interacting with channels: small molecules like acetylholine., monoamines, aminoacids and big ones like set of neuropeptides are existing..

The quite different mechanism of synaptic transmission, related to stimulation of production of secondary mediator is existing also. For example, activation of adenilatcyclase by first mediator increases the concentration of intra-cell cyclic adenozin-mono-phosphate (cAMP). In turn, cAMP can activate enzymatic phosphorylation of ionic channels, changing the electric properties of cell. This secondary mediator can participate in a lot of regulative processes, including the genes expression.

In the normal state of dynamic equilibrium the ionic concentration gradient producing by ionic pumps activity is compensated by the electric tension gradient. The *electrochemical gradient* is equal to zero at this state.

The equilibrium concentration of $Na^+$ and $Cl^+$ in space out of cell is bigger than in cell, the gradient of $K^+$ concentration has an opposite sign. The external concentration of very important for regulative processes $Ca^{2+}$ (about $10^{-3}M$) is much higher than in cytosol (about $10^{-7}M$). Such a big gradient provide fast and strong increasing of $Ca^{2+}$ internal concentration after activation of corresponding channels.

At the "rest" condition of equilibrium the resulting concentration of internal anions of neurons is bigger than that of external ones, providing the difference of potentials equal to 50-100mV. As far the thickness of membrane is only about 5nm or 50Å it means that the gradient of electric tension is about:

$$100.000 \; V/sm$$

i.e. it is extremely high.

Depolarization of membrane usually is related to penetration of $Na^+$ ions into the cell. This process of depolarization could be inhibited by selected diffusion of $Cl^-$ into the cell. Such diffusion can produce even *hyperpolarization* of membrane.

*The potential of action and nerve impulse can be excited in neuron - receptor only if the effect of depolarization exceeds certain threshold.*

In accordance to our Hierarchic model of elementary act of consciousness (HMC) three most important consequences of neuron's body polarization can occur:

- reorganization of MTs system and change of the ionic channels activity, accompanied by short-term memorization;

-reorganization of synaptic contacts on the surface of neuron and its dendrites, leading to long-term memory;

- generation of the nerve impulse, transferring the signal to another nerve cells via axon.

The propagation of nerve signal in axons may be related to intra-cellular water activity ($a_{H_2O}$) decreasing due to polarization of membrane. As a result of feedback reaction the variation of $a_{H_2O}$ induce the [*opening/closing*] of the ionic channels, thereby stimulating signal propagation along the axons.

We put forward the hypothesis, that the periodic transition of clusterphilic interaction of the ordered water between inter-lipid tails in nonpolar central regions of biomembranes to hydrophobic one, following by water clusters disassembly and vice verse, could be responsible for lateral nerve signal propagation/firing via axons (Kaivarainen, 1985, 1995, 2001). The anesthetic action can be explained by its violation of the ordered water structure in the interior of axonal membranes, thus preventing the nerve signal propagation. The excessive stabilization of the internal clusters by nonpolar atoms, like Ar, Kr and Xe, also prevent the axonal firing.

The change of the ionic conductivity of the axonal membranes of the axons in the process of signal propagation is a secondary effect in this explanation.

The proposed mechanism, like sound propagation, can provide distant cooperative interaction between different membrane receptors on the same cell and between remote neurons bodies without strong heat producing because of compensation processes, absorbing the excessive heat energy. The latter phenomena is in total accordance with experiments.

As far the $\alpha\beta$ pairs of tubulins have the properties of "electrets" (Debrabander, 1982), the *piezoelectric properties* of core of microtubules can be predicted (Athenstaedt, 1974; Mascarenhas,1974).

It means that structure and dynamics of microtubules can be regulated by electric component of electromagnetic field, which accompanied the nerve excitation. In turn, dynamics of microtubules hollow core affects the properties of internal ordered water in state of mesoscopic Bose condensation (mBC).

For example, shift of the [open ⇔ closed] states equilibrium of cavity between $\alpha$ and $\beta$ tubulins to the open one in a course of excitation should lead to:

[I]. Increasing the dimensions and life-time of coherent clusters, represented by primary *lb* effectons (mBC)

[II]. Stimulation the distant interaction between MT of different neurons as a result of

increased frequency and amplitude/coherency of IR librational photons, radiated/absorbed by primary librational effectons of internal water;

[III] Turning the mesoscopic entanglement between water molecules in coherent clusters to nonuniform macroscopic entanglement.

Twisting of the centrioles of distant interacting cells and bending of MTs can occur after [gel→sol] transition. This tuning is necessary for enhancement of the number of MTs with the parallel orientation, most effective for their remote exchange interaction by means of 3D coherent IR photons and vibro-gravitational waves.

Reorganization of actin filaments and MTs system should be accompanied by corresponding changes of neuron's body and its dendrites shape and activity of certain ionic channels and synaptic contacts redistribution; This stage is responsible for long-term memory emergency.

At [sol]-state $Ca^{2+}$ - dependent $K^+$ channels turns to the open state and internal concentration of potassium decreases. The latter oppose the depolarization and decrease the response of neuron to external stimuli. Decay of neuron's response is termed "adaptation". This *response adaptation* is accompanied by *MTs-adaptation*, i.e. their reassembly in conditions, when concentration of $Ca^{2+}$ tends to minimum. The reverse [sol→gel] transition stabilize the new equilibrium state of given group of cells.

The described hierarchic sequence of stages: from mesoscopic Bose condensation to macroscopic one, providing entanglement of big number of cells, their simultaneous synaptic reorganization and synhronization of the excitation ⇌ relaxation cycles of nerve cells, are different stages of elementary act of consciousness.

**6.8 Possible mechanism of wave function collapsing**, following from the HMC

A huge number of superimposed possible quantum states of any quantum system always turn to "collapsed" or "reduced" single state as a result of measurement, i.e. interaction with detector.

In accordance to "Copenhagen interpretation", the collapsing of such system to one of possible states is unpredictable and purely random. Roger Penrose supposed (1989) that this process is due to quantum gravity, because the latter influences the quantum realm acting on space-time. After certain gravity threshold the system's wave function collapsed "under its own weight".

Penrose (1989, 1994) considered the possible role of quantum superposition and wave function collapsing in synaptic plasticity. He characterized the situation of learning and memory by synaptic plasticity in which neuronal connections are rapidly formed, activated or deactivated: "Thus not just one of the possible alternative arrangements is tried out, but vast numbers, all superposed in complex linear superposition". The collapse of many cytoskeleton configuration to single one is a nonlocal process, required for consciousness.

This idea is in-line with our model of elementary acts of consciousness as a result of transitions between nonuniform macroscopic and mesoscopic Bose condensation (BC) of big number of electromagnetically tuned neurons and corresponding oscillation between their entangled and non-entangled states.

Herbert (1993) estimated the mass threshold of wave function collapse roughly as $10^6$ daltons. Penrose and Hameroff (1995) calculated this threshold as

$$\Delta M_{col} \sim 10^{19} D \qquad 6.17$$

Non-computable self-collapse of a quantum coherent wave function within the brain may fulfill the role of non-deterministic free will after Penrose and Hameroff (1995).

For the other hand, in accordance with proposed in this author model, the induced coherency between coherent water clusters (primary librational effectons - mesoscopic

Bose condensate) in MTs, as a result of distant exchange of librational photons, emitted ⇌ absorbed by them, leads to formation of *macroscopic* BC in microtubules.

The increasing of the total mass of water, involved in macroscopic nonuniform BC in a big system of remote MTs and corresponding 'tuned' neuron ensembles, up to gravitational threshold may induce the wave function collapse in accordance to Penrose hypothesis.

In our approach we explain the selection of certain configurational space of huge number of 'tuned' neurons, not by structural changes of tubulins like in Hameroff-Penrose model, but by increasing of mass of water in state of macroscopic BC in brain in the process of condensation of spatially separated mesoscopic BC (coherent water clusters in MTs). The macroscopic BC is initiated by correlated shift of dynamic equilibrium ($a \rightleftharpoons b$) of nonpolar cavities, formed by pairs of tubulins, between the open (*b*) and closed (*a*) states to the open one, stabilizing water clusters. The time of development/evolution of coherence in remote neurons, accompanied by increasing of scale of macroscopic BC is much longer, than that of mesoscopic BC (about $10^{-6}$ *s*, equal to average period of pulsation of tubulin dimers cavity between open and close state) and can be comparable with time between axonal firing (about $1/40 = 2.5 \times 10^{-2}$ s). The time of coherence determines the period between corresponding wave function collapsing.

The corresponding structural rearrangements of tubulins and their pairs in the process of shift of open⇌ closed clefts to the right or left, do not change their mass and can not be a source of wave function collapsing "under its own weight" in contrast to increasing of mass of water in evolution of nonuniform macroscopic BC from mBC.

The dynamics of $[\,increasing \,\rightleftharpoons\, decreasing\,]$ of the entangled water mass in state of macroscopic BC is a result of correlated shift of dynamic equilibrium between primary *librational (lb) effectons* (coherent water clusters, mBC), stabilized by *open* inter-tubulins cavities and primary *translational (tr) effectons (independent water molecules)*, corresponding to closed cavities.

The correlated conversions between librational and translational effectons [$lb \rightleftharpoons tr$] of water in remote MTs, representing the association ⇌ dissociation of the entangled water clusters in state of mBC reflect, in fact, the reversible cycles of [coherence ⇌ decoherence] corresponding to cycles of mesoscopic wave function of these clusters collapsing. The relatively slow oscillations of dynamic equilibrium of [$lb \rightleftharpoons tr$] conversions are responsible for alternating contribution of macroscopic quantum entanglement in consciousness.

*Let's make some simple quantitative evaluations in proof of our interpretation of the wave function collapsing.* The mass of water in one microtubule in nerve cell body with most probable length $\sim 10^5$ Å and diameter 140 Å is about

$$m_{H_2O} \sim 10^8 D$$

In accordance with our calculations for bulk water, the fraction of molecules in composition of primary *tr* effectons is about 23% and that in composition of primary *librational* effectons (mBC) is about ten times less (Figure 53) or 2.5%. In MTs due to clusterphilic interaction, stabilizing water clusters, this fraction mBC can be few times bigger.

We assume, that in MTs at least 10% of the total water mass ($10^8 D$) can be converted to primary librational effectons (coherent clusters) as a result of IR photons exchange and entanglement between mBC of the same MTs. This corresponds to increasing of mass of these quasiparticles in each MT as:

$$\Delta m_{H_2O} \simeq 10^6 D \qquad \qquad 6.18$$

Such increasing of the coherent water fraction is accompanied by decreasing of water mass,

involved in other types of excitations in MT.

Based on known experimental data that each nerve cell contains about 50 microtubules, we assume that the maximum increasing of mass of primary librational effectons in one cell could be:

$$\Delta M_{H_2O} \sim 50\, \Delta m_{H_2O} = 5 \times 10^7 D \qquad 6.19$$

If the true value of mass threshold, responsible for wave function collapse, $\Delta M_{col}$ is known (for example $10^{16} D$), then the number ($N_{col}$) of neurons in assemblies, required for this process is

$$N_{col} \sim (\Delta M_{col}/\Delta M_{H_2O}) = 2 \times 10^8 \qquad 6.20$$

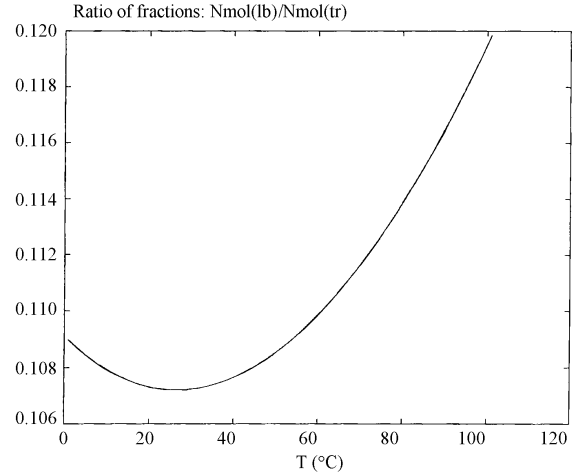

**Figure 6.5**. Calculated ratio of water fractions involved in primary [lb] effectons to that, involved in primary [tr] effectons for the bulk water.

MAP– microtubules associated proteins stabilize the overall structure of MTs. They prevent the disassembly of MTs in bundles of axons and *cilia* in a course of their coherent bending. In neuron's body the concentration of MAP and their role in stabilization of MTs is much lower than in cilia.

The total number of nerve cells in human brain is about: $N_{tot} \sim 10^{11}$. The critical fraction of cells population, participating in elementary act of consciousness, following from our model, can be calculated as:

$$f_c = (N_{col}/N_{tot}) \sim 0.05 \qquad 6.21$$

This value is dependent on correct evaluation of critical mass ($\Delta M_{col}$) of collapsing.

The [gel → sol] transition, induced by coherent "collapsing" of macroscopic brain wave function and dissociation of water clusters in MTs in state BC of huge number of tuned neurons, followed by synaptic contacts reorganization, represents the elementary act of consciousness.

Our approach agree with general idea of Marshall (1989) that Bose- condensation could be responsible for "unity of conscious experience". However, our model explains how this idea can work in detail and what kind of Bose condensation is necessary for its realization.

*We can resume now, that in accordance with our HMC, the sequence of following interrelated stages is necessary for elementary act of perception and memory:*

1. The change of the electric component of cell's electromagnetic field as a result of neuron depolarization;

2. Shift of $A \rightleftharpoons B$ equilibrium between the closed (A) and open to water (B) states of cleft, formed by $\alpha$ and $\beta$ tubulins in microtubules (MT) to the right due to the piezoelectric effect;

3. Increasing the life-time and dimensions of coherent "flickering" water clusters, representing the 3D superposition of de Broglie standing waves of $H_2O$ molecules with properties of Bose-condensate (*effectons*) in hollow core of MT. This process is stimulated by the open nonpolar clefts of tubulin dimers in MT with regular 80Å spacing;

4. Increasing the super-radiance of coherent IR photons induced by synchronization of quantum transitions of the *effectons* between *acoustic* and *optic* like states;

5. Opening the potential dependent $Ca^{2+}$ channels and increasing the concentration of these ions in cytoplasm;

6. Activation of $Ca^{2+}$ - dependent protein gelsolin, which induce fast disassembly of actin filaments and [gel-sol] transition, decreasing strongly the viscosity of cytoplasm and water activity;

7. Spatial "tuning" of quasi-parallel MTs of distant simultaneously excited neurons due to distant electromagnetic and vibro-gravitational interaction between them and centrioles twisting;

8. The coherent volume/shape pulsation of big group of depolarized cells as a consequence *of (actin filaments+MTs) system disassembly and [gel→sol] transition*. It happens as a result of filaments system reversible disassembly to huge number of subunits and increasing of water fraction in hydration shell of actin and tubulin subunits due to increasing of their surface. This should decrease the water activity in cytoplasm and increase the passive osmotic diffusion of water from the external volume to the cell.

*This stage should be accompanied by four effects:*

*(a)* Increasing the volume of the nerve cell body;

*(b)* Disrupting the (+) ends of MTs with cytoplasmic membranes, making MTs possible to bend in cell and to collective spatial tuning of huge number of MTs in the ensembles of even distant excited neurons;

(c) Origination of new MTs system switch on/off the ionic channels and change the number and properties of synaptic contacts;

(d) Decreasing the concentration of $Ca^{2+}$ to the limit one when its ability to disassembly of actin filaments and MT is stopped and [gel $\rightleftharpoons$ sol] equilibrium shifts to the left again, stabilizing a new MTs and synaptic configuration.

This cyclic consequence of quantum mechanical, physico-chemical and nonlinear classical events can be considered as elementary act of memory and consciousness realization. This act can be as long as 500 ms, *i.e.* half of second, like proposed in Hamroff-Penrose model.

The elementary act of consciousness include a stage of coherent electric firing in brain (Singer, 1993) of distant neurons groups with period of about 1/40 sec.

Probability of super-deformons and cavitational fluctuations increases after [gel→sol] transition. This process is accompanied by high-frequency (UV and visible) "biophotons" radiation due to recombination of part of water molecules, dissociated as a result of cavitational fluctuation.

The dimension of IR super-deformon edge is determined by the length of librational IR standing photon - about 10 microns. It is important that this dimension corresponds to the average microtubule length in cells confirming in such a way our idea. Another evidence in proof is that is that the resonance wave number of excitation of super-deformons, leading from our model is equal to 1200 (1/*cm*).

The experiments of Albreht-Buehner (1991) revealed that just around this frequency the response of surface extensions of 3T3 cells to weak IR irradiation is maximum. Our model predicts that IR irradiation of microtubules system *in vitro* with this frequency will dramatically increase the probability of gel →sol transition.

Except super-radiance, two other cooperative optic effects could be involved in supercatastrophe realization: self-induced bistability and the pike regime of IR photons radiation (Bates, 1978; Andreev et al.,1988).

The characteristic frequency of pike regime can be correlated with frequency of [gel-sol] transitions of neuronal groups in the head brain.

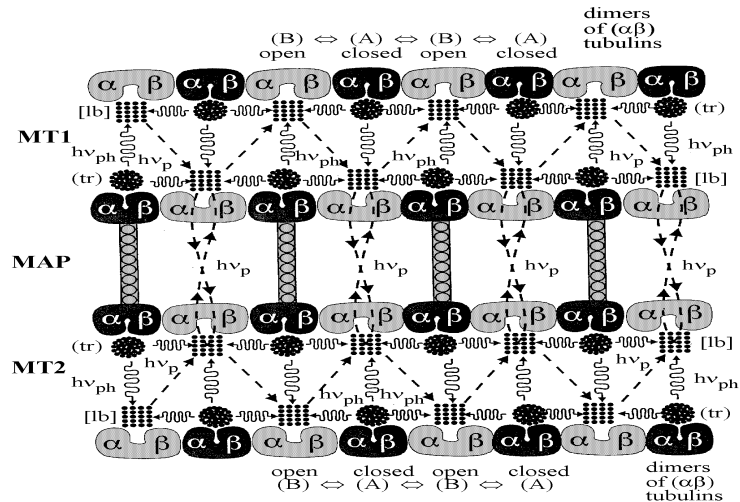

**Figure 6.6**. The correlation between local, conformational and distant - electromagnetic interactions between pairs of tubulins and microtubules (MT1 and MT2), connected by MAP by mean of librational IR photons exchange.
The dynamics of [ *increasing* ⇌ *decreasing* ] of the entangled water mass in state of macroscopic BC in the process of elementary act of consciousness is a result of correlated shift of dynamic equilibrium between open and closed cavities between alpha and beta tubulins. As a result of these cavities transition from the open to closed state the primary *librational (lb) effectons* (coherent water clusters in state of mesoscopic Bose condensation - mBC) disassembly to small primary *translational (tr) effectons* (independent water molecules), induced by transition of the open states of cavities to the closed one. The nonuniform macroscopic entanglement between the remote water clusters in state of mBC is stimulated by coherent IR photons exchange and vibro-gravitational interaction between these clusters.

MAP– microtubules associated proteins stabilize the overall structure of MTs. They prevent the disassembly of MTs in bundles of axons and *cilia* in a course of their coherent bending. In neuron's body the concentration of MAP and their role in stabilization of MTs is much lower than in cilia (Kaivarainen, 1995, 2003).

The distant electromagnetic and vibro-gravitational interactions between different MT are the consequence of IR photons and coherent vibro-gravitational waves exchange. The corresponding two types of waves are excited as a result of correlated ($a \Leftrightarrow b$) transitions of water primary librational effectons, localized in the open B- states of ($\alpha\beta$) clefts. Frequency of ($a \Leftrightarrow b$) transitions and corresponding superradiated IR photons is about $2\times10^{13}\ s^{-1}$. It is much higher, than frequency of transitions of clefts of $\alpha\beta$ tubulin dimers between open and closed states.

When the neighboring ($\alpha\beta$) clefts has the alternative open and closed states like on Fig 54, the general spatial structure remains straight. However, when [$A \Leftrightarrow B$] equilibrium of all the clefts from one side of MT are shifted to the left and that from the opposite side are shifted to the right, it leads to bending of MT. Coherent bending of MTs could be responsible for [volume/shape] vibrations of the nerve cells and the cilia bending.

Max Tegmark (2000) made evaluation of decoherence time of neurons and microtubules for analyzing the correctness of Hameroff-Penrose idea of wave function collapsing as a trigger of neurons ensembles axonal firing.

The following three sources of decoherence for the ions in the act of transition of neuron between 'resting' and 'firing' are most effective:

1. Collisions with other ions
2. Collisions with water molecules
3. Coulomb interactions with more distant ions.

The coherence time of such process, calculated from this simple model appears to very short: about $10^{-20}$ s.

The electrical excitations in tubulins of microtubules, which Penrose and others have suggested may be relevant to consciousness also where analyzed. Tegmark considered a simple model of two separated but superimposed (entangled) positions of kink, travelling along the MT with speed higher than 1 m/s, as it supposed in Hameroff-Penrose (H-P) model. The life-time of such quantum state was evaluated as a result of long-range electromagnetic interaction of nearest ion with kink.

His conclusion is that the role of quantum effects and wave function collapsing in H-P model is negligible because of very short time of coherence: $10^{-13}$ s for microtubules.

Hagan, Hameroff and Tuszynski (2002) responded to this criticism, using the same formalism and kink model. Using corrected parameters they get the increasing of life-time of coherent superposition in H-P model for many orders, up to $10^{-4}$ s. This fits the model much better.

Anyway the approach, used by Tegmark for evaluation of the time of coherence/entanglement is not applicable to our model.

### 6.9 Experimental data, confirming the model of elementary act of consciousness or Quantum of Mind

There are some experimental data, which support the role of microtubules in the information processing. Good correlation was found between the learning, memory peak and intensity of tubulin biosynthesis in the baby chick brain (Mileusnic *et al.*1980). When baby rats begin their visual learning after they first open eyes, neurons in the visual cortex start to produce vast quantities of tubulin (Cronley- Dillon et. al., 1974). Sensory stimulation of goldfish leads to structural changes in cytoskeleton of their brain neurons (Moshkov *et al.*, 1992).

There is evidence for interrelation between cytoskeleton properties and nerve membrane excitability and synaptic transmission (Matsumoto and Sakai, 1979; Hirokawa, 1991). It has been shown, that microtubules can transmit electromagnetic signals between membranes (Vassilev *et al.*, 1985).

Desmond and Levy (1988) found out the learning-associated change in dendritic spine shape due to reorganization of actin and microtubules containing, cytoskeleton system. After "learning" the number of receptors increases and cytoskeleton becomes more dense.

Other data suggest that cytoskeleton regulates the genome and that signaling along microtubules occurs as cascades of phosphorylation/dephosphorylation linked to calcium ion flux (Puck, 1987; Haag et al, 1994).

The frequency of super-deformons excitations in bulk water at physiological temperature ($37^0 C$) is around:

$$v_S = 3 \times 10^4 \ s^{-1}$$

The frequency of such cavitational fluctuations of water in MT, stimulating in accordance to our model cooperative disassembly of the actin and partly MT filament, accompanied by gel→ *sol* transition, could differ a bit from the above value for bulk water.

Our model predicts that if the neurons or other cells, containing MTs, will be treated by acoustic or electromagnetic field with resonance frequency of intra-MT water ($v_{res} \sim v_S^{MT} \geq 10^4 s^{-1}$), it can induce simultaneous disassembly of th actin filaments and destabilization of MTs system, responsible for maintaining the specific cell volume and geometry. As a result, it activates the neuron's body volume/shape pulsation.

Such external stimulation of gel⇌ *sol* oscillations has two important consequences:

-*The first one* is generation of oscillating high-frequency nerve impulse, propagating via axons and exciting huge number of other nerve cells, i.e. distant nerve signal transmission in living organism;

-*The second one* is stimulation the leaning process as far long-term memory in accordance to HMC, is related to synaptic contacts reorganization, accompanied the neuron volume/shape pulsation.

Lehardt *et al.* (1989) supposed that ultrasonic vibrations are perceptive by tiny gland in the inner ear, known as the *Saccule*. It looks that *Saccule may* have a dual functions of detection *gravity and auditory signals. Cohlea could be a result of Saccule evolution in mammals*.

Lenhardt and colleagues constructed the an amplitude modulated by audio-frequencies ultrasonic transmitter that operated at frequencies: (28-90) kHz. The output signal from their device was attached to the deaf people heads by means of piezoelectric ceramic vibrator. All people "heard" the modulated signal with clarity.

Our approach allows to predict the important consequence. Excitation of water super-deformons in cells, leading to gel→ *sol* transition, cell's volume/shape pulsation and generation of high-frequency nerve impulse - could be stimulated by hypersound, electromagnetic waves and coherent IR photons with frequency, corresponding to excitation energy of super-deformons.

The calculated from this assumption frequency is equal to

$$v_p^S = c \times \widetilde{v}_p^S = (3 \times 10^{10}) \times 1200 = 3.6 \times 10^{13} \ c^{-1} \qquad 6.22$$

the corresponding photons wave length:

$$\lambda_p^S = c/(n_{H_2O} \times v_p^S) \simeq 6.3 \times 10^{-4} \ cm = 6.3 \mu \qquad 6.23$$

where $\widetilde{v}_p^S = 1200 \ cm^{-1}$ is wave number, corresponding to energy of super-deformons excitation;

$n_{H_2O} \simeq 1.33$ is refraction index of water.

*18.9.1 The additional experimental verification of the Hierarchic Model of Consciousness "in vitro"*

It is possible to suggest some experimental ways of verification of HMC using model systems. The important point of HMC is stabilization of highly ordered water clusters (primary librational effectons) in the hollow core of microtubules. One can predict that in this case the IR librational bands of water in the oscillatory spectra of model system, containing sufficiently high concentration of MTs, must differ from IR spectra of bulk water as follows:

- the shape of librational band of water in the former case must contain 2 components: the first one, big and broad, like in bulk water and the second one small and sharp, due to increasing coherent fraction of librational effectons. The second peak should disappeared after disassembly of MTs with specific reagents;

- the velocity of sound in the system of microtubules must be bigger, than that in disassembled system of MTs and bulk pure water due to bigger fraction of ordered ice-like water;

- all the above mentioned parameters must be dependent on the applied electric potential, due to piezoelectric properties of MT;

- the irradiation of MTs system in vitro by ultrasonic or electromagnetic fields with frequency of super-deformons excitation of the internal water of MTs at physiological temperatures $(25 - 40^0 C)$ :

$$\nu_s = (2 - 4) \times 10^4 \; Hz$$

have to lead to increasing the probability of disassembly of MTs, induced by cavitational fluctuations. The corresponding effect of decreasing turbidity of MT-containing system could be registered by light scattering method.

Another consequence of super-deformons stimulation by external fields could be the increasing of intensity of radiation in visible and UV region due to emission of corresponding "biophotons" as a result of recombination reaction of water molecules:

$$HO^- + H^+ \stackrel{h\nu}{\rightleftharpoons} H_2O \qquad 6.24$$

Cavitational fluctuations of water, representing in accordance to our theory super-deformons excitations, are responsible for dissociation of water molecules, *i.e.* elevation of protons and hydroxyls concentration. These processes are directly related to sonoluminiscence phenomena.

The coherent transitions of $(\alpha\beta)$ dimers, composing MTs, between "closed" (A) and "open" (B) conformers with frequency $(\nu_{mc} \sim 10^7 \; s^{-1})$ are determined by frequency of water macroconvertons (flickering clusters) excitation, localized in cavity between $\alpha$ and $\beta$ tubulins. If the charges of (A) and (B) conformers differ from each other, then the coherent $(A \rightleftharpoons B)$ transitions generate the vibro-gravitational and electromagnetic field with the same radio-frequency. The latter component of biofield could be detected by corresponding radio waves receiver.

We can conclude that the Hierarchic Theory of condensed matter and its application to water and biosystems provide reliable models of informational exchange between different cells and correlation of their functions. Hierarchic Model of consciousness is based on proposed quantum exchange mechanism of interactions between neurons, based on very special properties of microtubules, [gel-sol] transitions and interrelation between spatial distribution of MTs in neurons body and synaptic contacts.

The described mechanism of IR photons - mediated conversion of mesoscopic Bose condensation to macroscopic one with corresponding change of wave function spatial scale and its collapsing time, as a precondition of 'Quantum of Mind' can be exploit in the construction of artificial quantum computers, using inorganic microtubules, supplied by nanotechnology